\documentclass[traditabstract]{aa} 
\usepackage{graphicx,natbib,longtable,txfonts,lscape,color}

\newcommand{\kms}{$\rm{\,km \,s}^{-1}$}

\newcommand{\Ha} {H$\alpha$}  
\newcommand{\Hb} {H$\beta$}  

\newcommand{\FRo}{FR0{\sl{CAT}}}

\newcommand{\WHz}{\>{\rm W}\,{\rm Hz}^{-1}}
\DeclareUnicodeCharacter{2212}{-}
\begin{document}

\title{The LOFAR view of giant, early-type galaxies:\\ Radio emission
  from active nuclei and star formation} \author{A. Capetti\inst{1}
  \and M. Brienza\inst{2,3} \and B. Balmaverde\inst{1} \and
  P.N. Best\inst{4} \and R.~D. Baldi\inst{2} \and A. Drabent\inst{5}
  \and G. G\"urkan\inst{5} \and H.J.A. Rottgering\inst{6} \and
  C. Tasse\inst{7,8} \and B. Webster\inst{9}}

\institute{INAF - Osservatorio Astrofisico di Torino, Strada
  Osservatorio 20, I-10025 Pino Torinese, Italy \and Dipartimento di
  Fisica e Astronomia, Universit\`a di Bologna, Via P. Gobetti 93/2,
  I-40129, Bologna, Italy \and INAF - Istituto di Radio Astronomia,
  Via P. Gobetti 101, I-40129 Bologna, Italy \and SUPA, Institute for
  Astronomy, Royal Observatory, Blackford Hill, Edinburgh, EH9 3HJ, UK
  \and Th\"uringer Landessternwarte, Sternwarte 5, D-07778 Tautenburg,
  Germany \and Leiden Observatory, Leiden University, PO Box 9513,
  2300 RA Leiden, The Netherlands \and GEPI \& USN, Observatoire de
  Paris, Université PSL, CNRS, 5 Place Jules Janssen, 92190 Meudon,
  France \and Department of Physics \& Electronics, Rhodes University,
  PO Box 94, Grahamstown, 6140, South Africa \and The Open University,
  Walton Hall, Milton Keynes, MK7 6AA, UK} \date{}

\abstract{We studied the properties and the origin of the radio
  emission in the most luminous, early-type galaxies (ETGs) in the
  nearby Universe (M$_{K} \le-25$, recession velocity $ \le 7,500$
  \kms), as seen by the 150 MHz Low-Frequency ARray (LOFAR)
  observations. LOFAR images are available for 188 of these giant ETGs
  (gETGs), and 146 (78 \%) of them are detected above a typical
  luminosity of $\sim 10^{21} \WHz$. They show a large spread in
  power, reaching up to $\sim 10^{26} \WHz$. We confirm a positive
  link between the stellar luminosity of gETGs and their median radio
  power, the detection rate, and the fraction of extended
  sources. About two-thirds (91) of the detected gETGs are unresolved,
  with sizes $\lesssim 4$ kpc, confirming the prevalence of compact
  radio sources in local sources. Forty-six gETGs show extended
  emission on scales ranging from 4 to 340 kpc, at least 80\% of which
  have a FR~I class morphology. Based on the morphology and spectral
  index of the extended sources, $\sim$30\% of them might be remnant
  or restarted sources, but further studies are needed to confirm
  this. Optical spectroscopy (available for 44 gETGs) indicates that
  for seven gETGs the nuclear gas is ionized by young stars suggesting
  a contribution to their radio emission from star forming
  regions. Their radio luminosities correspond to a star formation
  rate (SFR) in the range 0.1 - 8 M$_\odot$yr$^{-1}$ and a median
  specific SFR of $0.8 \times 10^{-12}$ yr$^{-1}$. The gas flowing
  toward the center of gETGs can accrete onto the supermassive black
  hole but also stall at larger radii and form new stars, an
  indication that feedback does not completely quench star formation.
  The most luminous gETGs (25 galaxies with M$_K < -25.8$) are all
  detected at 150 MHz; however, they are not all currently turned on:
  at least four of them are remnant sources and at least one is likely
  powered by star formation.}

\maketitle

\section{Introduction}
\label{intro}

Feedback from active galactic nuclei (AGNs) is an important ingredient
in the evolution of galaxies. For example, the transfer of energy and
matter from relativistic jets to the external medium, the so-called
radio mode feedback (e.g., \citealt{fabian12}), is thought to be
able to quench star formation and produce the exponential cut-off at
the bright end of the galaxy luminosity function
\citep{croton06}. Exploring the properties and the origin of the radio
emission in the most massive galaxies is an essential step to
understanding how this process operates. In particular, it is necessary to
separate sources in which the radio emission is produced by an AGN
from those powered by star formation in order to study their morphology and
to explore their duty-cycle.

The study of the radio emission in early-type galaxies (ETGs) has been
the subject of many studies in the past (e.g.,
\citealt{ekers73,colla75,fanti78,sadler89,wrobel91a,wrobel91b}). The
general conclusions have been the large fraction of ETGs associated
with radio sources, the positive dependence of radio power with the
luminosity of the host, and the large spread of radio luminosity at a
given host mass.

A significant issue for these studies is the ability to separate radio
emission produced by AGN and star forming regions, a process that
becomes dominant at low power \citep{condon02}. This is best obtained
from the optical spectra of these sources and the large area surveys
obtained in the last decades provided us with required information on
both the radio and the optical.

\citet{best05b} studied a sample of 2215 radio-loud AGN with $0.03 < z
< 0.3$ obtained by combining data from the Faint Images of
the Radio Sky at Twenty centimeters survey (FIRST,
\citealt{becker95,helfand15}), the National Radio Astronomy
Observatory Very Large Array Sky Survey (NVSS; \citealt{condon98}), and
spectra of galaxies included in the main galaxy spectroscopic sample
of the Sloan Digital Survey (SDSS, \citealt{york00}). They confirmed
the link between radio source prevalence and host luminosity derived
from previous studies: the integral radio luminosity function is well
described with a broken power law whose normalization grows with the
stellar mass as $M_*^{2.5}$ and the AGN fraction is as high as $30\%$
in the most massive galaxies.

\citet{mauch07} identified 7824 radio sources in the NVSS in the six-degree Field Galaxy Survey \citep{jones04} associated with galaxies
brighter than K = 12.75 mag, spanning the $0.003 < z <
0.3 $ range. They separated radio sources powered by star formation and AGNs
and found that radio-loud AGNs are preferentially hosted by the most
massive galaxies.

\citet{shabala08} found, by combining SDSS, FIRST, and NVSS data, that
the length of the active phase has a strong dependence on the host
mass. The active phase length is connected with the gas cooling rate,
suggesting the quiescent phase is due to fuel depletion.

The analysis of a heterogeneous set of radio measurements of an
optically selected sample led \citet{brown11} to the conclusion that
all ETGs with an absolute magnitude M$_K < -25.5$ have radio flux
densities greater than zero. The analysis of the Low-Frequency ARray
(LOFAR) DR1 images of galaxies included in the main galaxy
spectroscopic sample of the Sloan Digital Survey (SDSS) led
\citet{sabater19} to the conclusion that all galaxies with a mass M
$>10^{11}$ M$_\odot$ display radio-AGN activity with a power $>10^{21}
\WHz$, that is, they are always switched on.  A similar result
  was found by \citet{grossova22} from their study of 42 nearby and
  X-ray-bright, early-type galaxies, all of them being detected at
  radio frequencies.

In order to explore the radio emission in the most
massive ETGs as seen by the LOFAR surveys in greater detail, we selected a complete
volume-limited sample (recession velocity $v< 7500$ \kms) of bright
galaxies based on infrared surveys, which were less subjected to internal and
Galactic absorption. The resulting sample is sufficiently large to
obtain robust statistical results, and, thanks to the depth of the
LOFAR data, it is possible to reach unprecedented low levels of radio
luminosity. With respect to previous studies, we also investigated the
radio morphology, an important clue for the nature of RGs, which,
thanks to the proximity of these sources can be studied in detail (the
scale of the most distant sources is $\sim$ 1 kpc per arcsecond). In
addition, we probed the radio spectra by comparing the data at 150
MHz with those of the publicly available surveys at 1.4 GHz.

The paper is organized as follows. In Sect. 2, we describe the sample
of the selected sources and the available radio observations with
LOFAR and from surveys at 1.4 GHz. In Sect. 3, we present the main
results, including a description of the radio morphology and spectral
shape of the sources of the sample. The information available from
optical spectroscopy is described in Sect. 4. In Sect. 5, we discuss
the results, which we then summarize in Sect. 6 and draw our
conclusions.

\section{Sample selection and the LOFAR observations}
\label{sample}

\begin{figure*}
\includegraphics[scale=0.5]{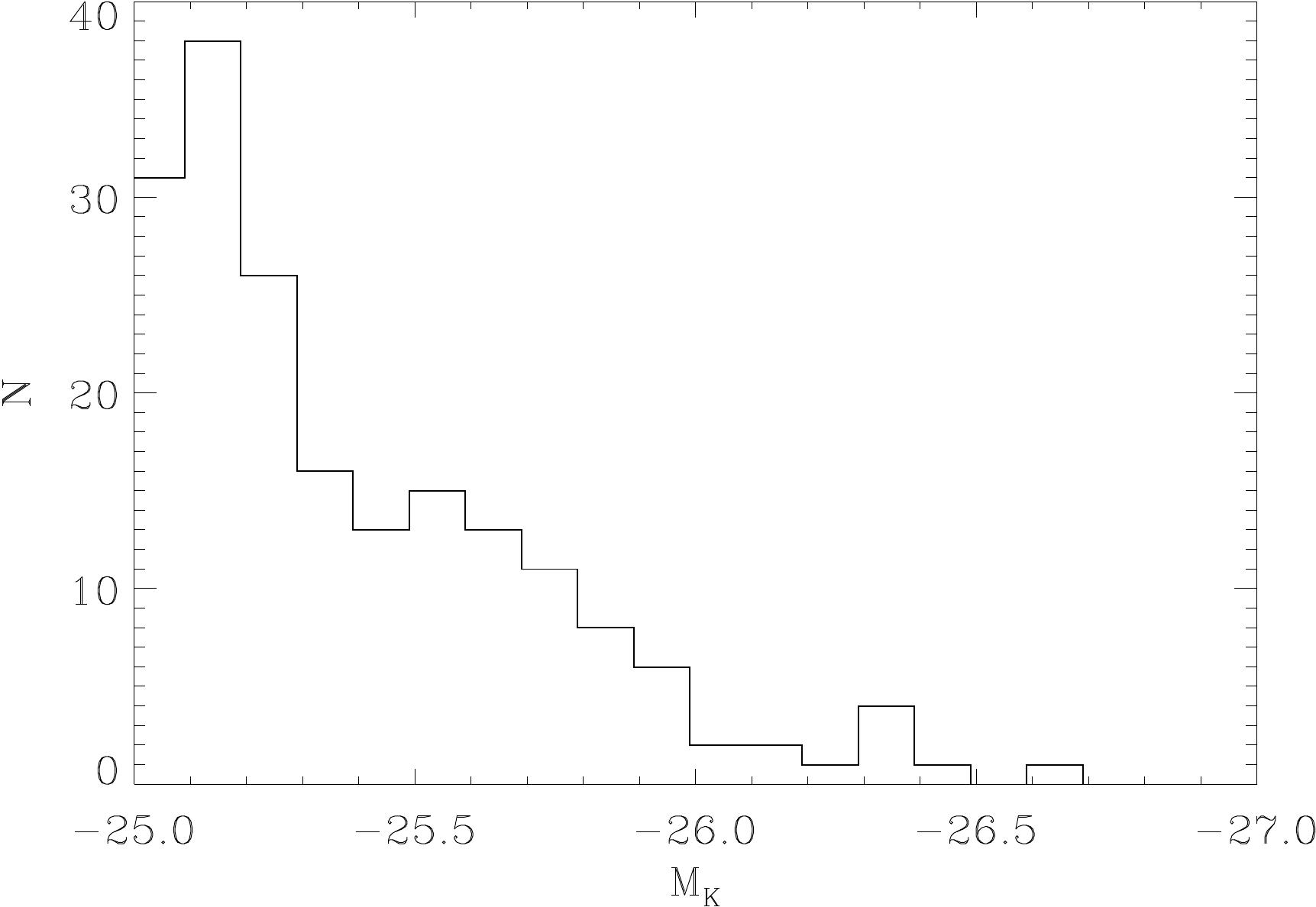}
\includegraphics[scale=0.5]{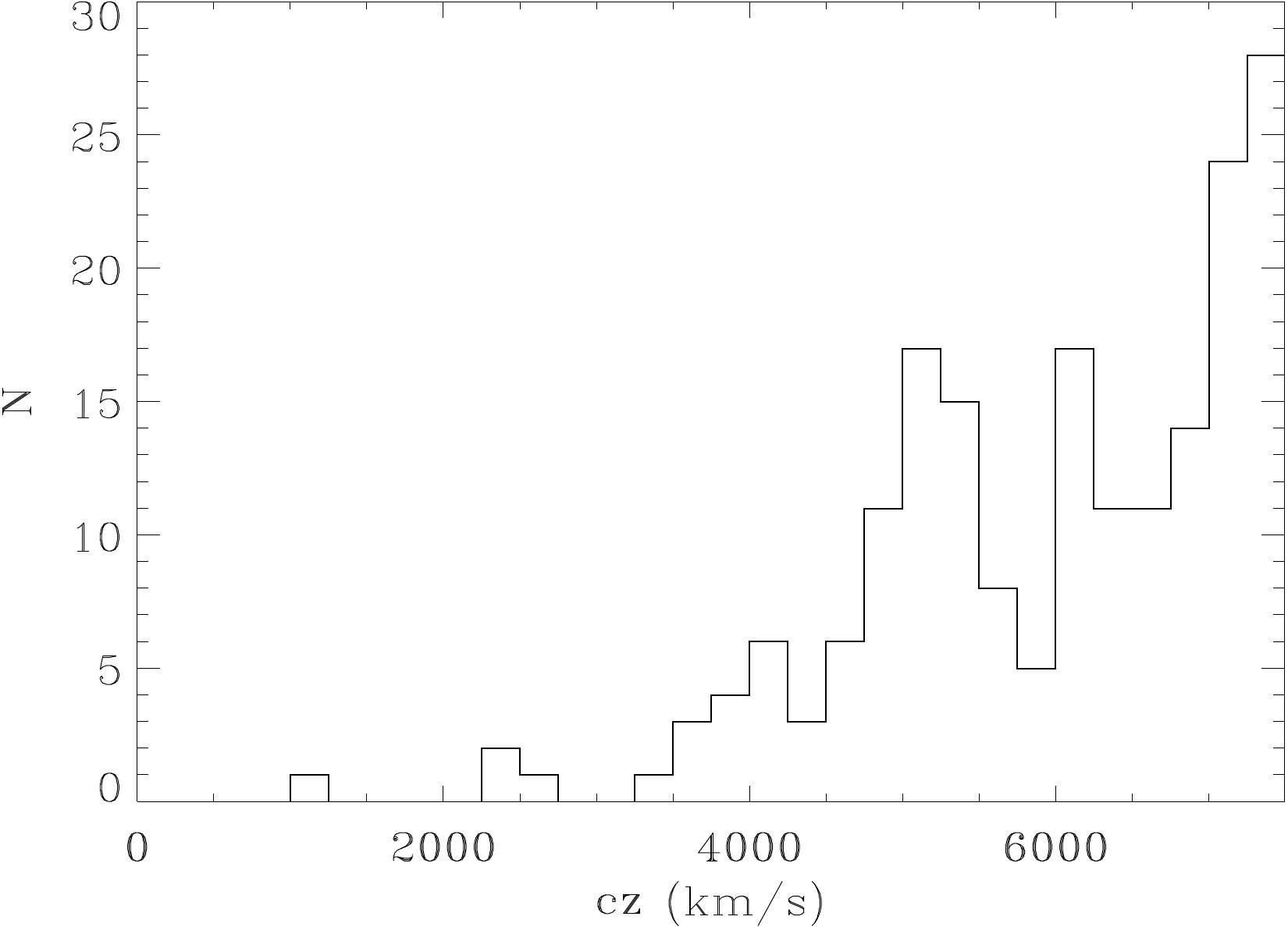}
\caption{Distributions of absolute K-band magnitude (left) and
  recession velocity (right) of the 188 sources of
  the sample.}
\label{kzhist}
\end{figure*}

The LOFAR Two-metre Sky Survey (LoTSS, \citealt{shimwell17}) will
cover the whole northern sky with 3168 pointings of at least eight
hours of dwell time each in the frequency range between 120 and 168
MHz.  The LoTSS first data release (DR1) \citep{shimwell19} presented
the results obtained from observations of 424 square degrees in the
HETDEX Spring Field. The final release images were obtained by
combining the images from individual pointings of the survey,
producing mosaics covering the region of interest at a
6\arcsec\ resolution and with a median noise of 71 $\mu$Jy/beam. The
flux density scale was adjusted to ensure consistency with previous
surveys (see \citealt{hardcastle16} for further details).  The second
LoTSS data release (DR2) will consist of two contiguous fields at high
Galactic latitude centered around 0h and 13h and covering
approximatively 5,700 square degrees (Shimwell et al., in prep.). The
DR2 provides fully calibrated mosaics at a resolution of $\sim
6\arcsec$, catalogs, and pipeline products. With respect to the DR1,
the DR2 products were obtained with an upgraded pipeline
\citep{tasse21}. LOFAR images can also be obtained from individual
LoTSS pointings, outside the DR2 area.

We selected the galaxies included in the 2MASS Redshift Survey
\citep{huchra12} requiring a declination DEC $>0^\circ$, a Hubble type
$T \le -1$, a total absolute magnitude $M_K \le -25$, and a recession
velocity (corrected for the effects of the Virgo Cluster, the Great
Attractor, and the Shapley Supercluster, \citealt{mould00}) $v \le 7500$
\kms, corresponding to a distance of $\lesssim 100$ Mpc. As of 31 May
2021, LOFAR observations are available for 188 out of 489 selected
galaxies, 111 covered by the DR2 and 77 obtained from individual LoTSS
survey pointings (we refer to these as ``S'' images).

We estimated the r.m.s. of each image in various regions, usually
centered 45$\arcmin$ away from the source of interest. For the
galaxies falling into the DR2 the median r.m.s. is 90$\mu$Jy/beam,
while this is 200$\mu$Jy/beam for the ``S'' images. The higher noise
of the ``S'' images is due to the fact that they are single pointings
and they do not benefit from the combination of adjacent pointings in
the mosaicking process.

The flux density of the sources included in the DR2 is available from
the internally released catalog, while for the sources included in the
``S'' group we measured their flux density from the LOFAR images after
correction for the preliminary scaling factor. The flux density errors
are dominated by the uncertainties in the absolute calibration and are
typically $\sim$10\%. Several galaxies in the DR2 area have large
scale and complex radio structures, not always fully included in the
catalog measurement. For these objects, we measured the flux densities
on the sky area including the whole source emission within the
3$\sigma$ contour. We obtained a detection at $>5\sigma$ significance
for 146 (78\%) of the sources of the sample.

The measurements at 1.4 GHz were obtained from the Faint Images of the
Radio Sky at Twenty centimeters survey (FIRST,
\citealt{becker95,helfand15}) and the National Radio Astronomy
Observatory Very Large Array Sky Survey (NVSS; \citealt{condon98}).
Eighty-six galaxies are included in the FIRST area and 40 of them are
detected by this survey having adopted a search radius of
3$\arcsec$. For the undetected sources, we estimated upper limits at
five times the local noise, typically $\sim 0.14$ mJy beam$^{-1}$. For
the 102 galaxies outside the FIRST area, we collected the NVSS
measurements with a search radius of 5$\arcsec$; 37 of these sources
have an NVSS detection, while for the remaining 65 we generally adopted a
limit of 2 mJy. For the extended galaxies, we measured the FIRST flux
densities on the same region used for the LOFAR images. In some cases,
the LOFAR radio structures have large angular sizes (exceeding $\sim
1^\prime$) and their counterparts at higher frequencies might be
resolved out in the FIRST images, due to the poor coverage of the
shortest baselines. For these objects, we obtained their flux densities
integrating the NVSS images. We followed the same method to
  estimate the flux density upper limits in the few cases of extended
  sources not detected by the NVSS.

The list of giant early-type galaxies (gETGs) covered by LOFAR images
is presented in Appendix A, Table \ref{tab}, where we list their main
properties; that is, name, coordinates, recession velocity, absolute
K-band magnitude, a code that indicates the origin of the image
(``DR2'' = within the DR2 area, ``S'' = individual survey's pointing),
the local r.m.s. of the LOFAR image, the flux density at 150 MHz, and
the size of the central component of the 150 MHz source. For the
extended sources, we also give the largest angular size at the
3$\sigma$ level and a morphological description, the luminosity at 150
MHz, the 1.4 GHz flux density (from either FIRST or NVSS), and the
spectral index between these frequencies\footnote{Spectral indices
  $\alpha$ are defined as $F_{\nu}\propto\,\nu^{-\alpha}$.}.

\begin{figure}
\includegraphics[scale=0.5]{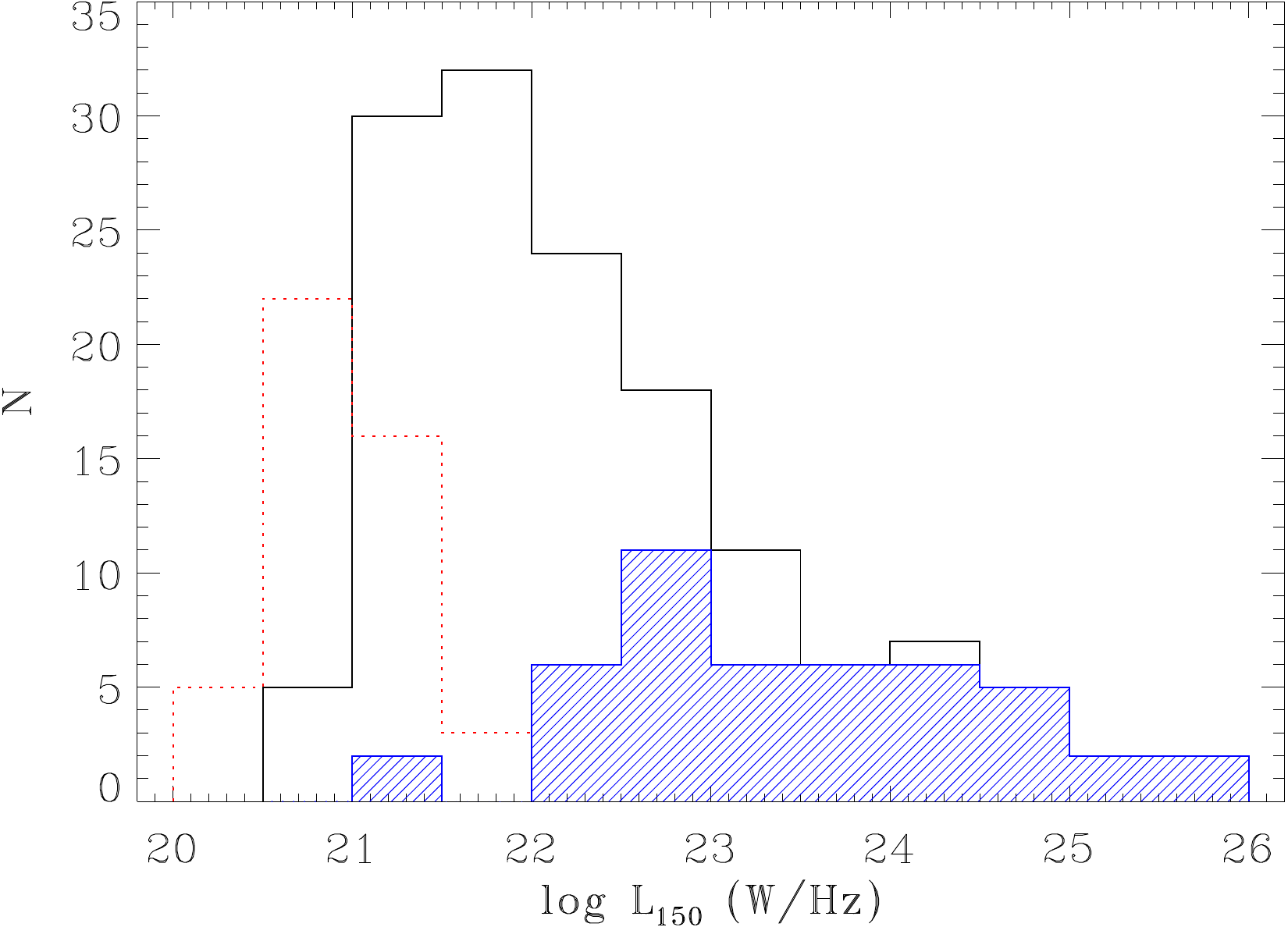}
\caption{Distribution of 150 MHz luminosity of the gETGs
  sample. The blue histogram represents the contribution of the 46
  extended sources, the dominant population at high luminosities, while
  the red dotted histogram corresponds to the upper limits of the
  undetected sources.}
\label{lum}
\end{figure}

\begin{figure*}
\includegraphics[scale=0.13]{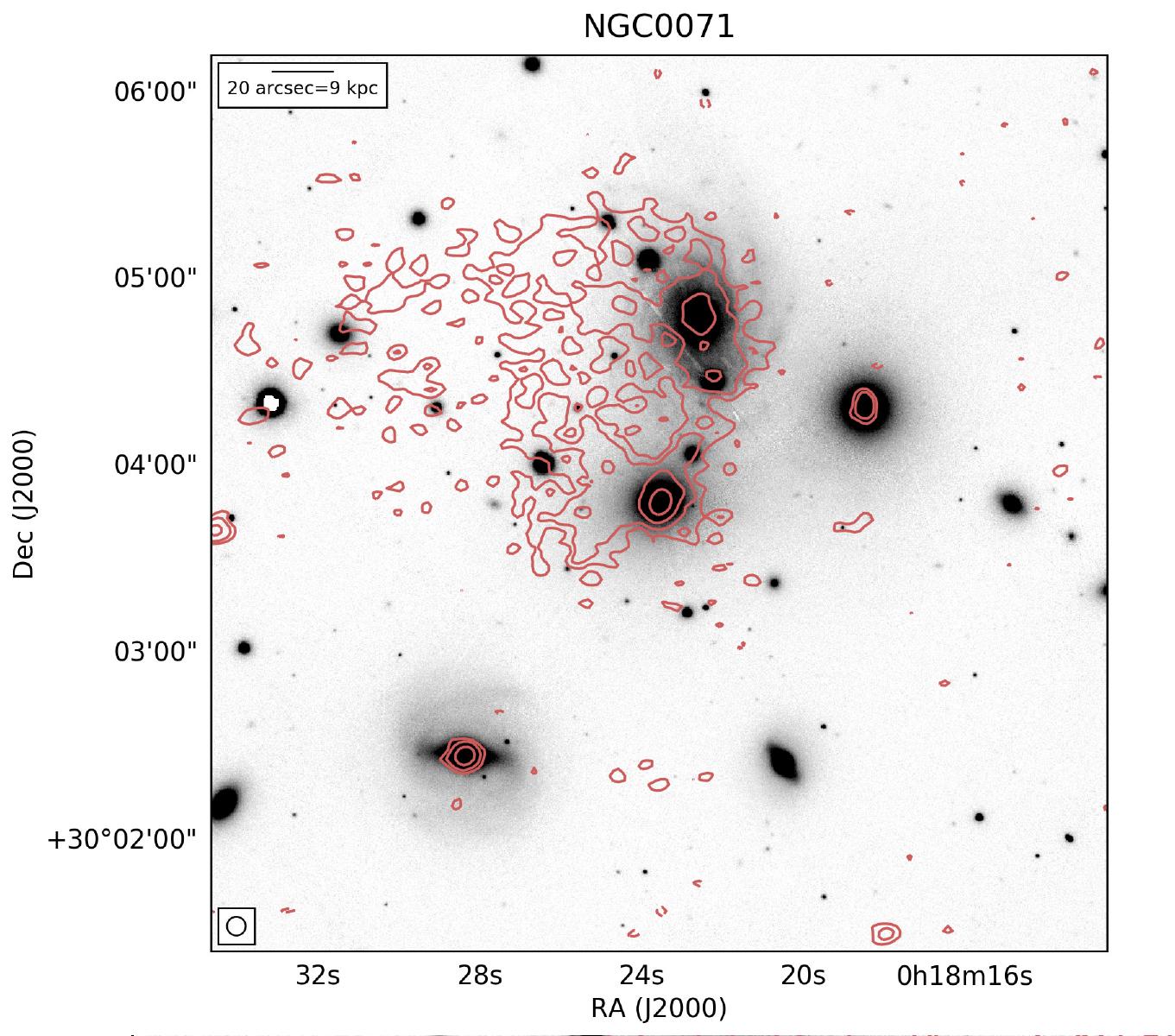}
\includegraphics[scale=0.13]{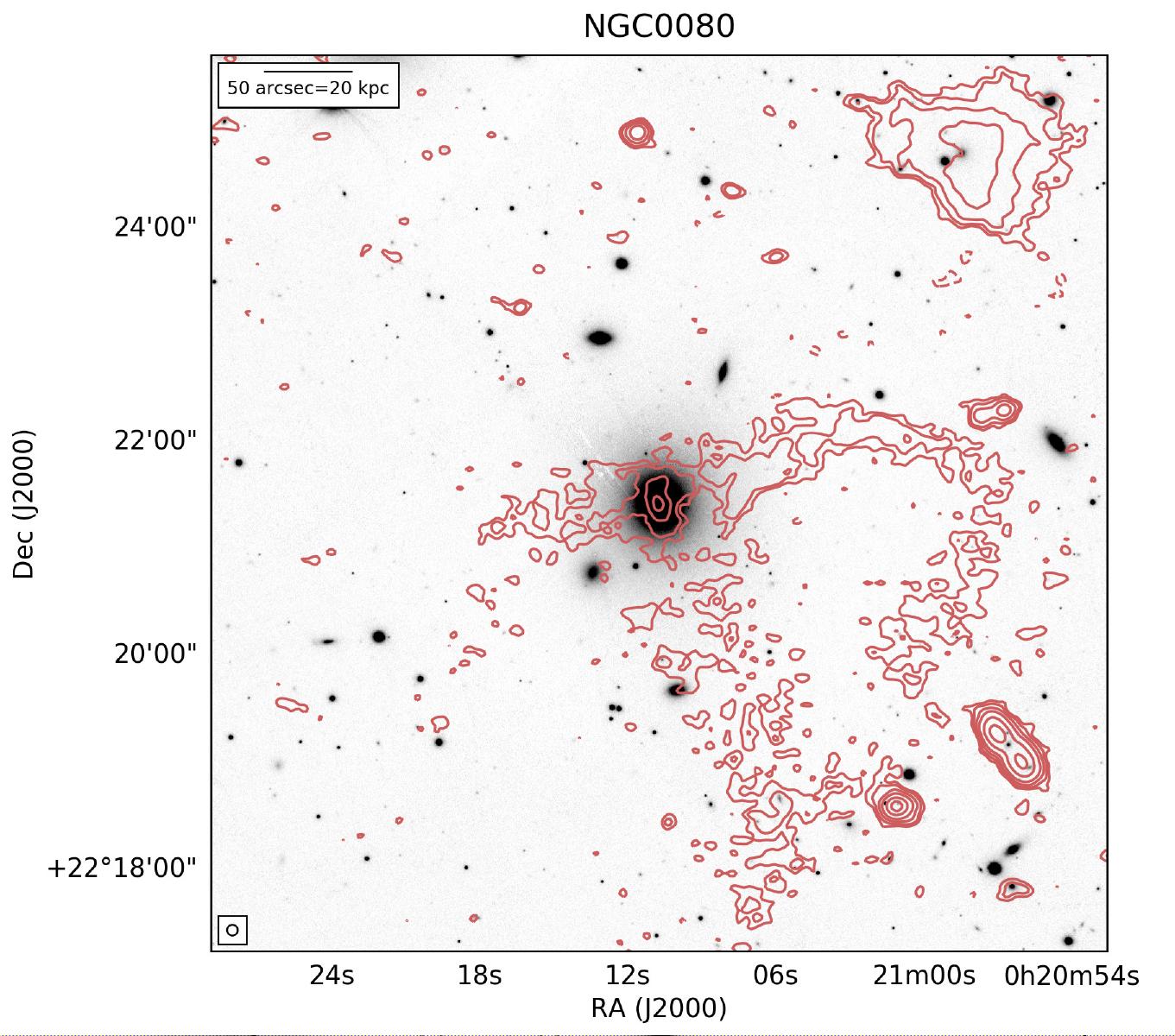}
\includegraphics[scale=0.13]{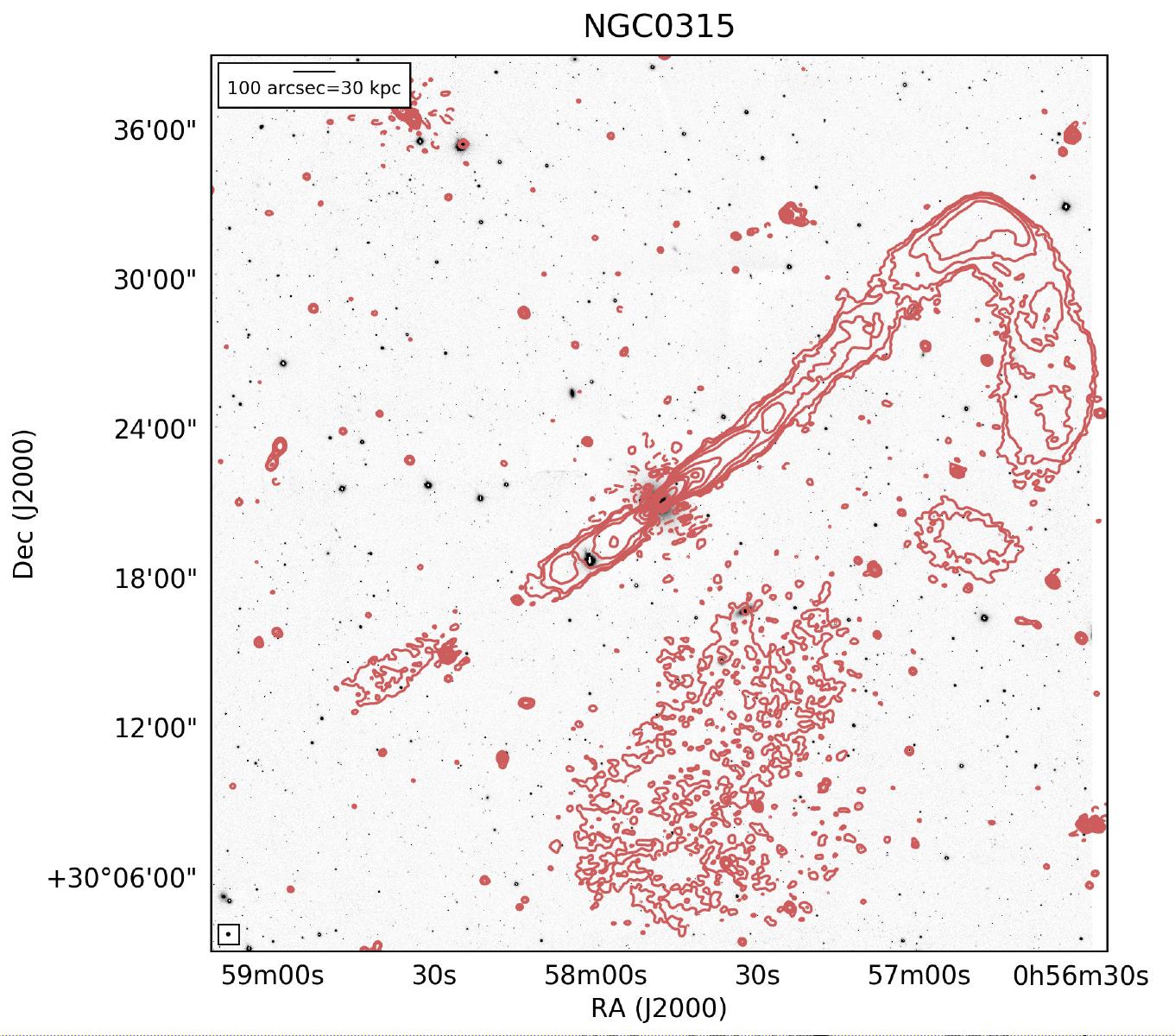}
\caption{Three examples of LOFAR images at 150
  MHz with extended emission superposed to the optical images from
  Pan-STARRS. Images of all extended sources are shown in the
  appendix B. The lowest contour is drawn at three times the local
  r.m.s., as reported in Table \ref{tab}. The following contours
  increase with a common ratio of 2.}
\label{estese}
\end{figure*}

\begin{figure}
\includegraphics[scale=0.5]{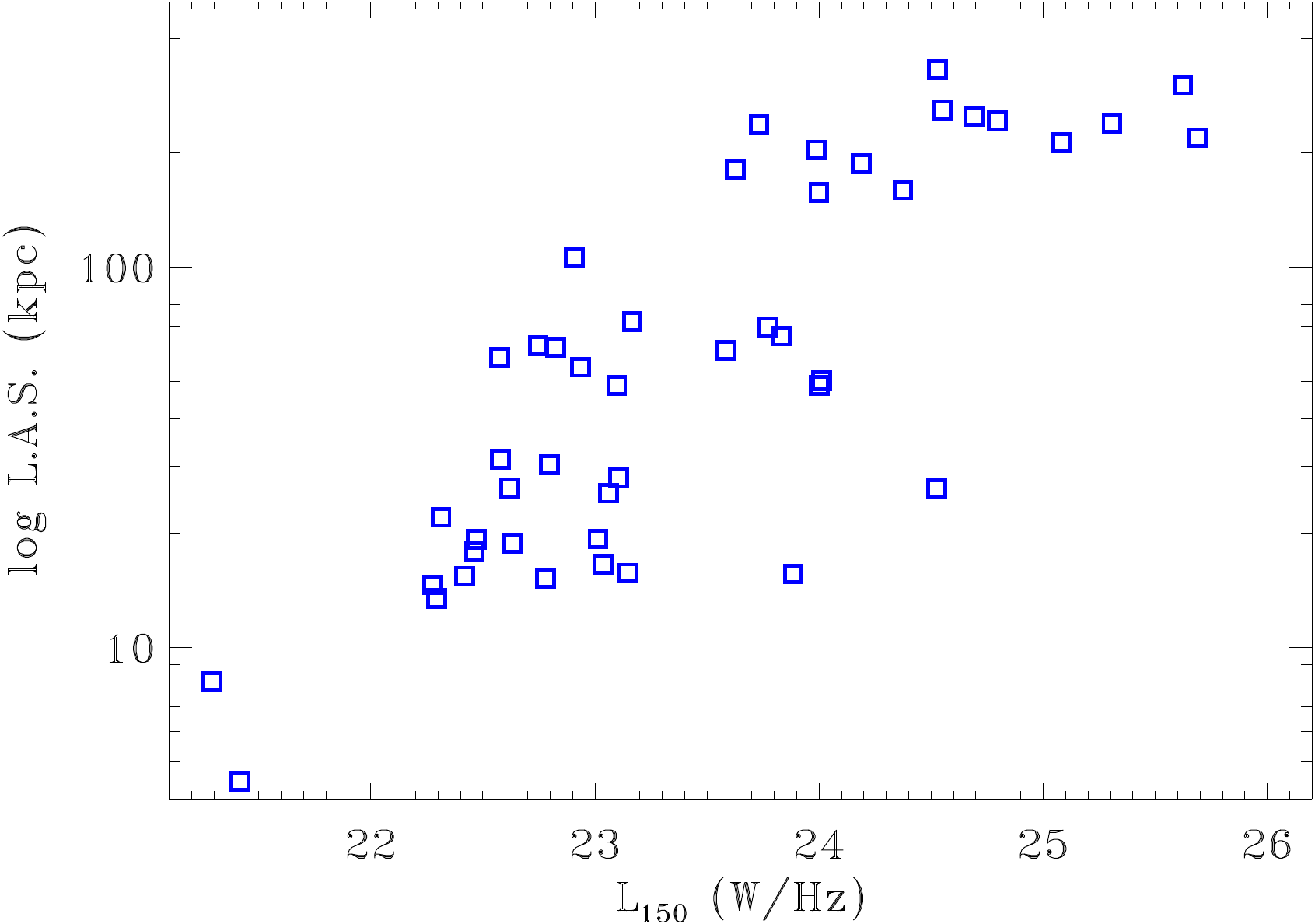}
\caption{Radio power at 150 MHz versus the largest angular size of the
  46 extended sources. A positive trend between radio power and size
  is present, with behavior consistent with that found for other
  samples of radio galaxies.}
\label{las}
\end{figure}

\begin{figure*}
\includegraphics[scale=0.13]{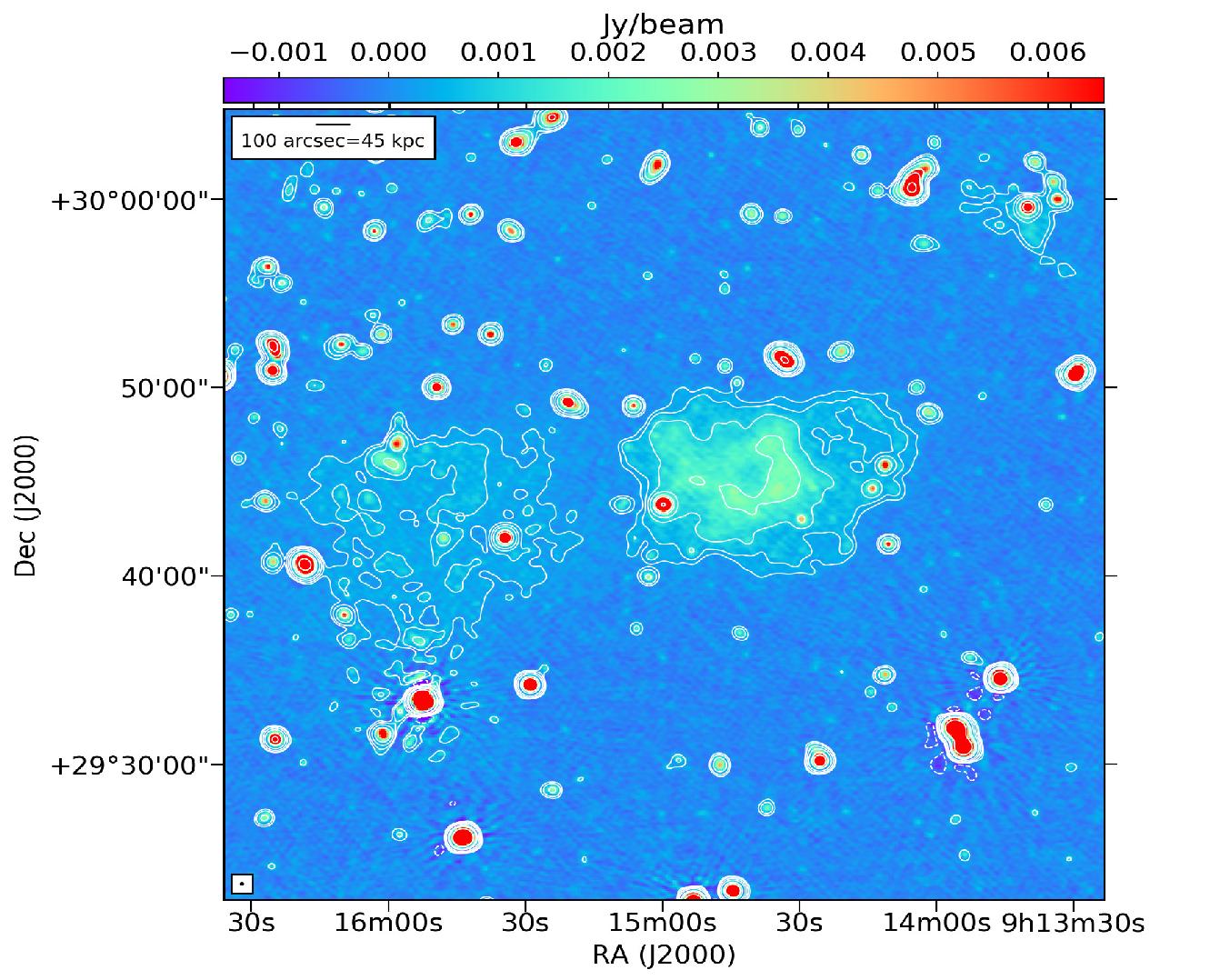}
\includegraphics[scale=0.13]{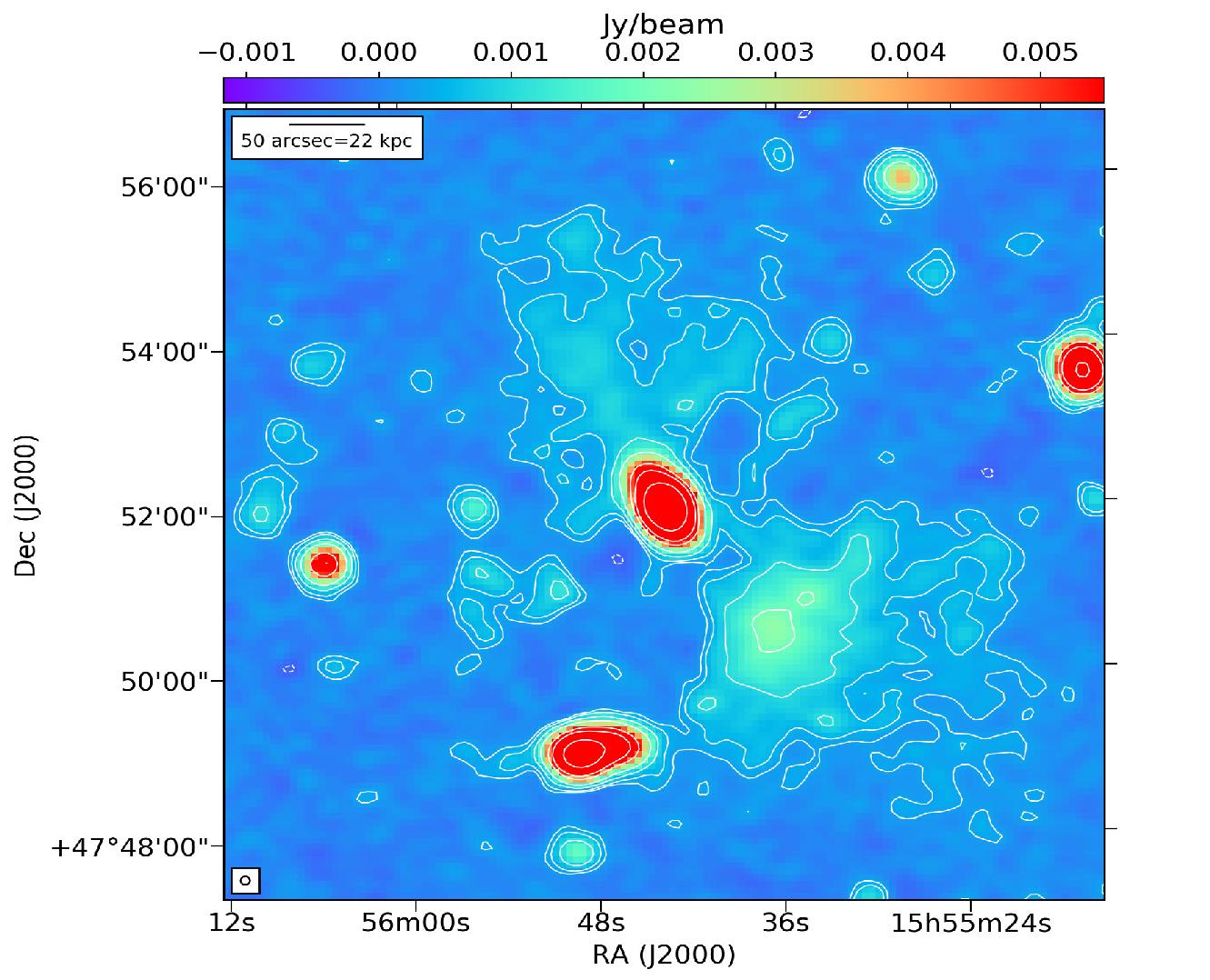}
\includegraphics[scale=0.13]{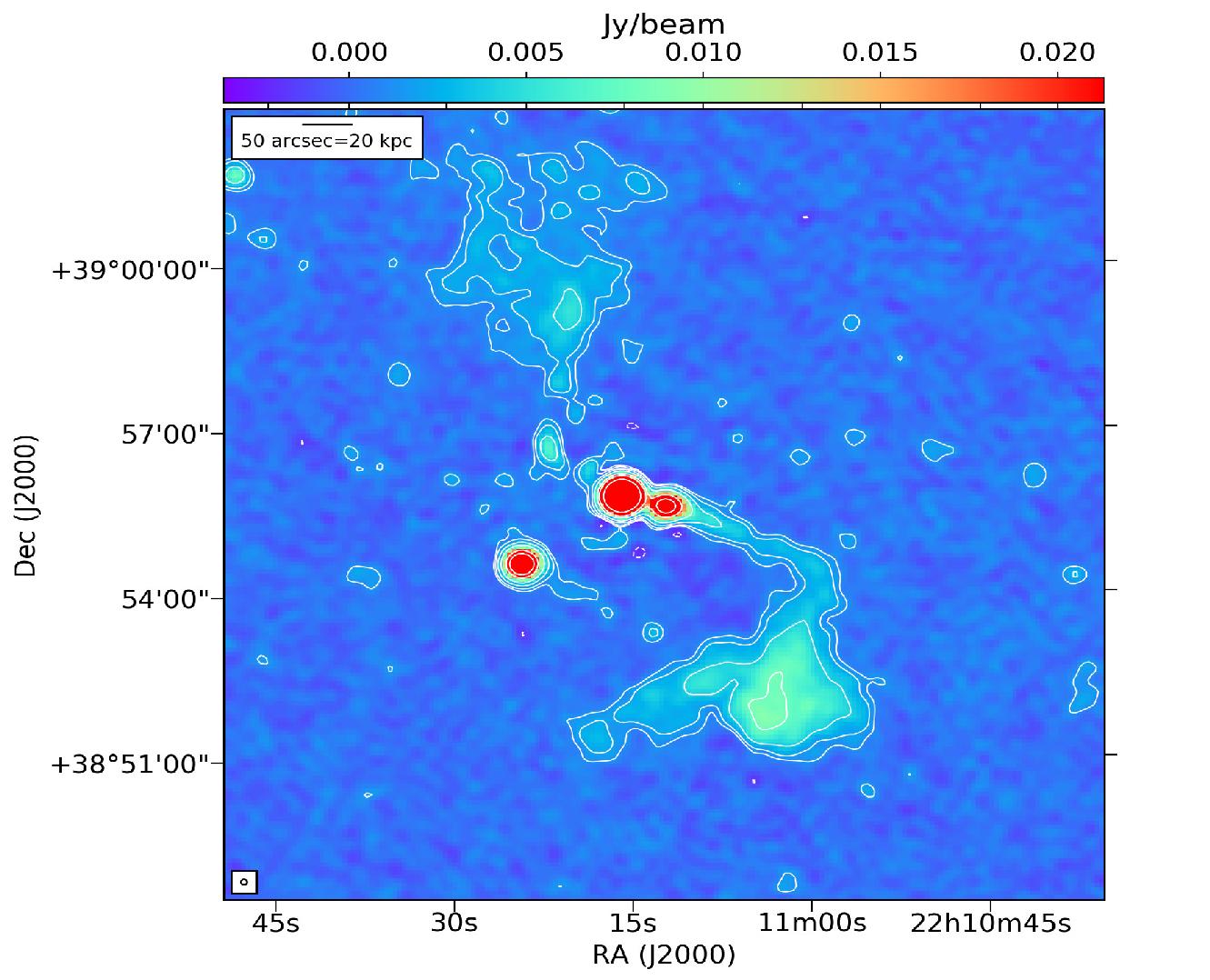}
\caption{Low-resolution images of the three extended sources showing
  large-scale low-brightness structures, not (or barely) visible in
the full-resolution images. From left to right: NGC~2789, UGC~10097, and
  IC~5180.}
\label{low}
\end{figure*}

\begin{figure}
\includegraphics[scale=0.5]{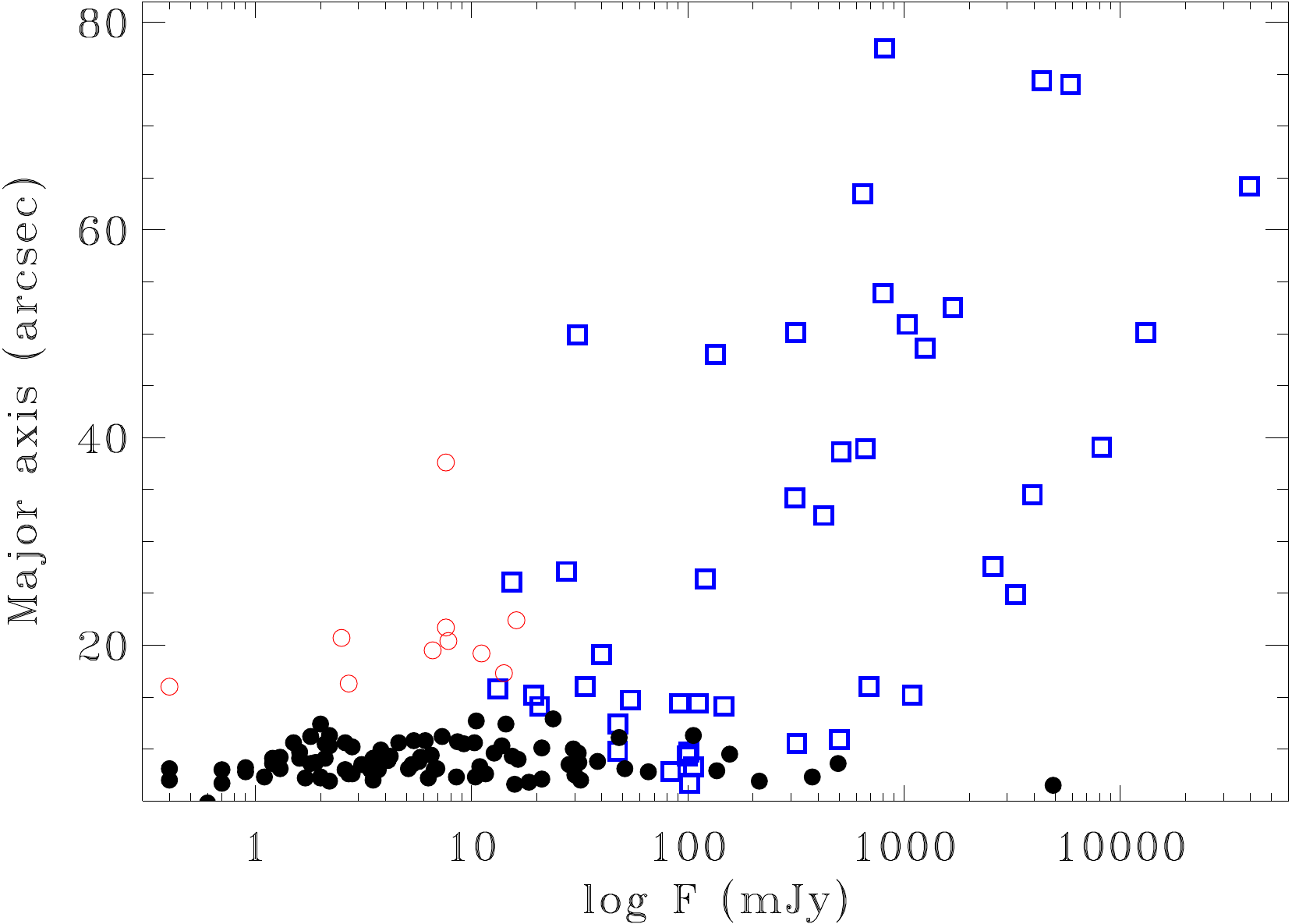}
\caption{Major axis length of the central component, obtained through
  Gaussian fitting, versus flux density. The extended sources are
  marked as empty blue squares, the black dots are those with a
   full width at half maximum (FWHM) $<15\arcsec$. The empty red circles are the low-brightness
  objects, that is, ones whose 3$\sigma$ level contours do not reach a
  radius of 15$\arcsec$ but whose FWHM exceeds 15$\arcsec$.}
\label{fwhm}
\end{figure}

\begin{figure*}
\includegraphics[scale=0.13]{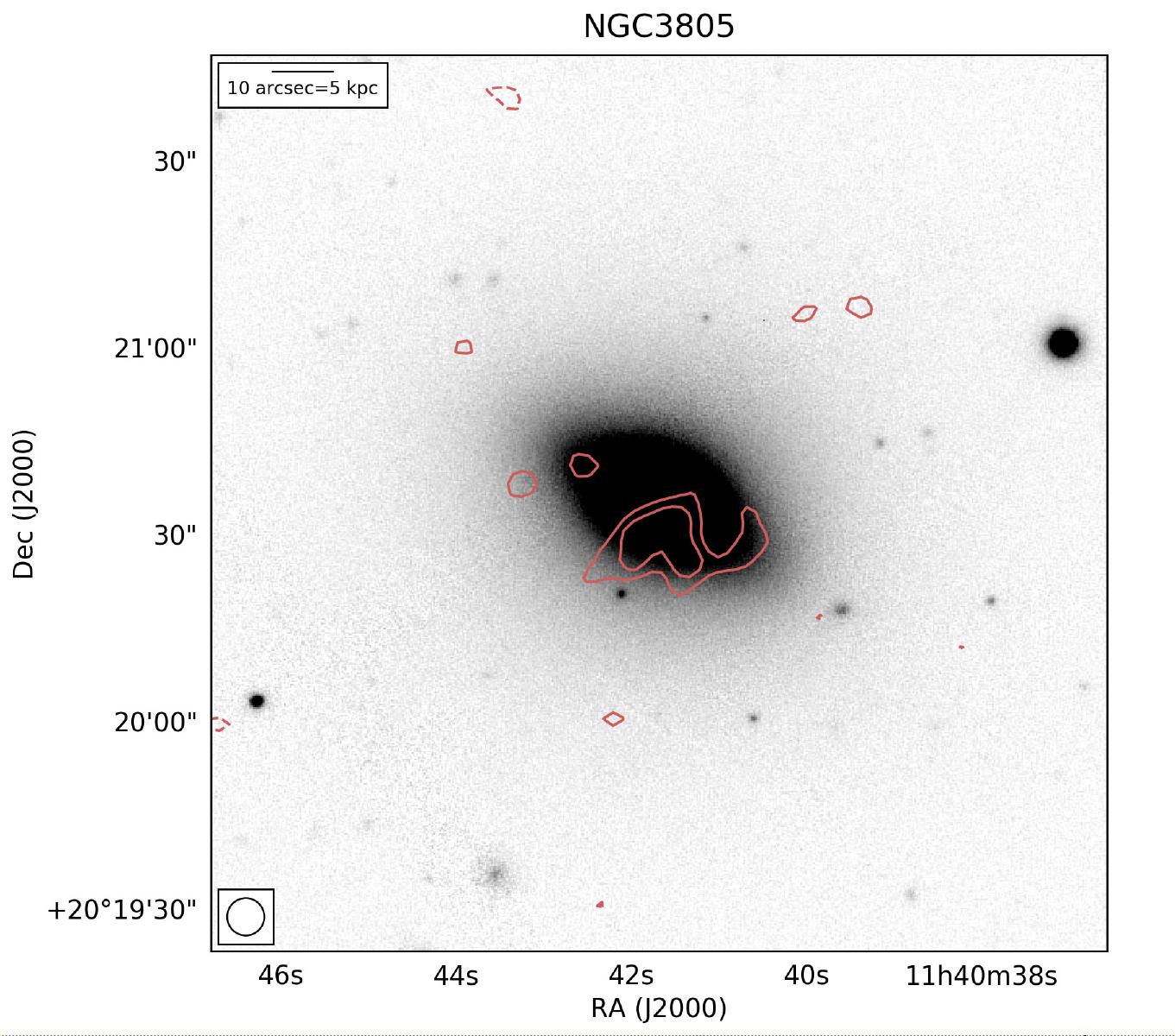}
\includegraphics[scale=0.13]{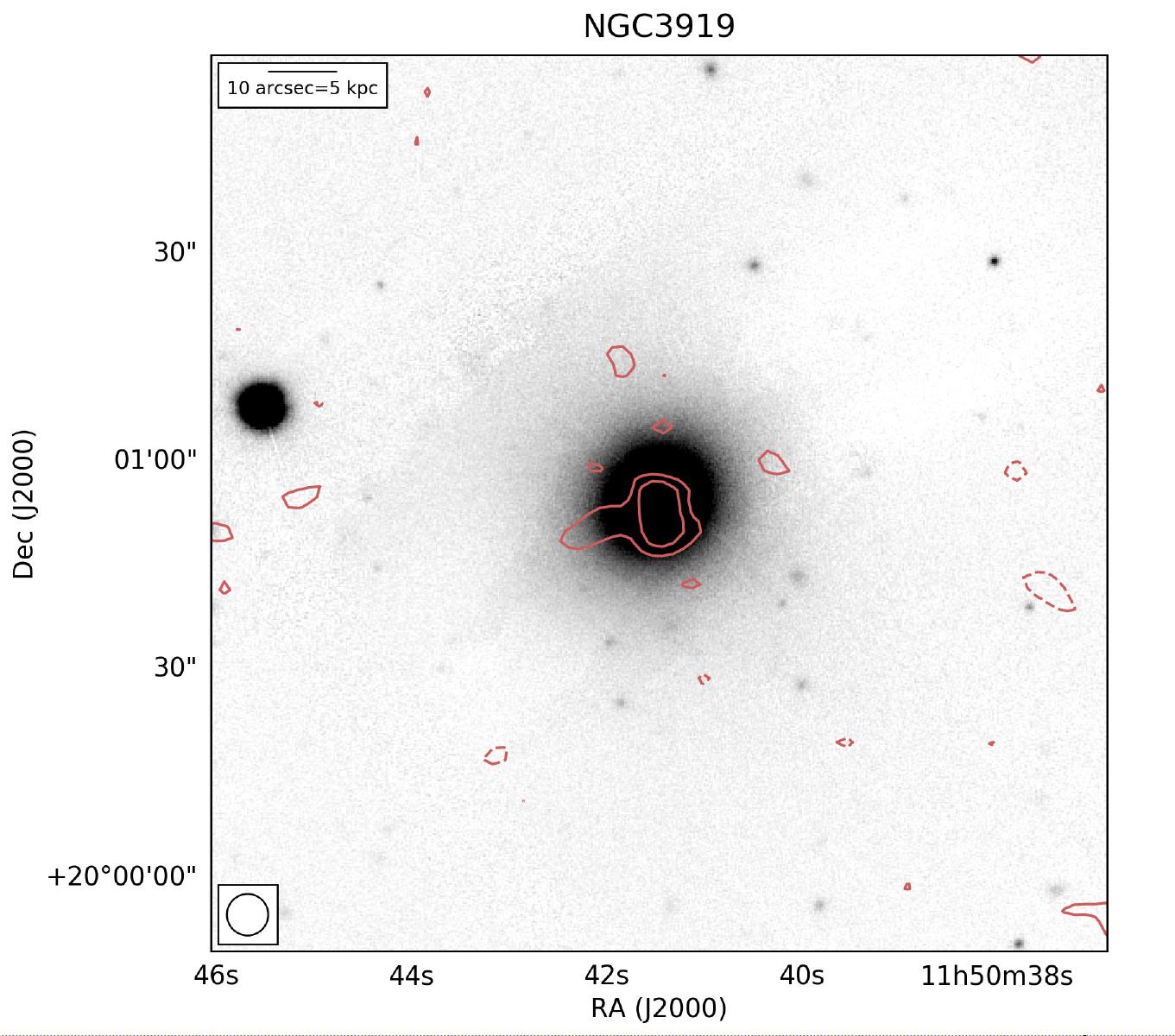}
\includegraphics[scale=0.13]{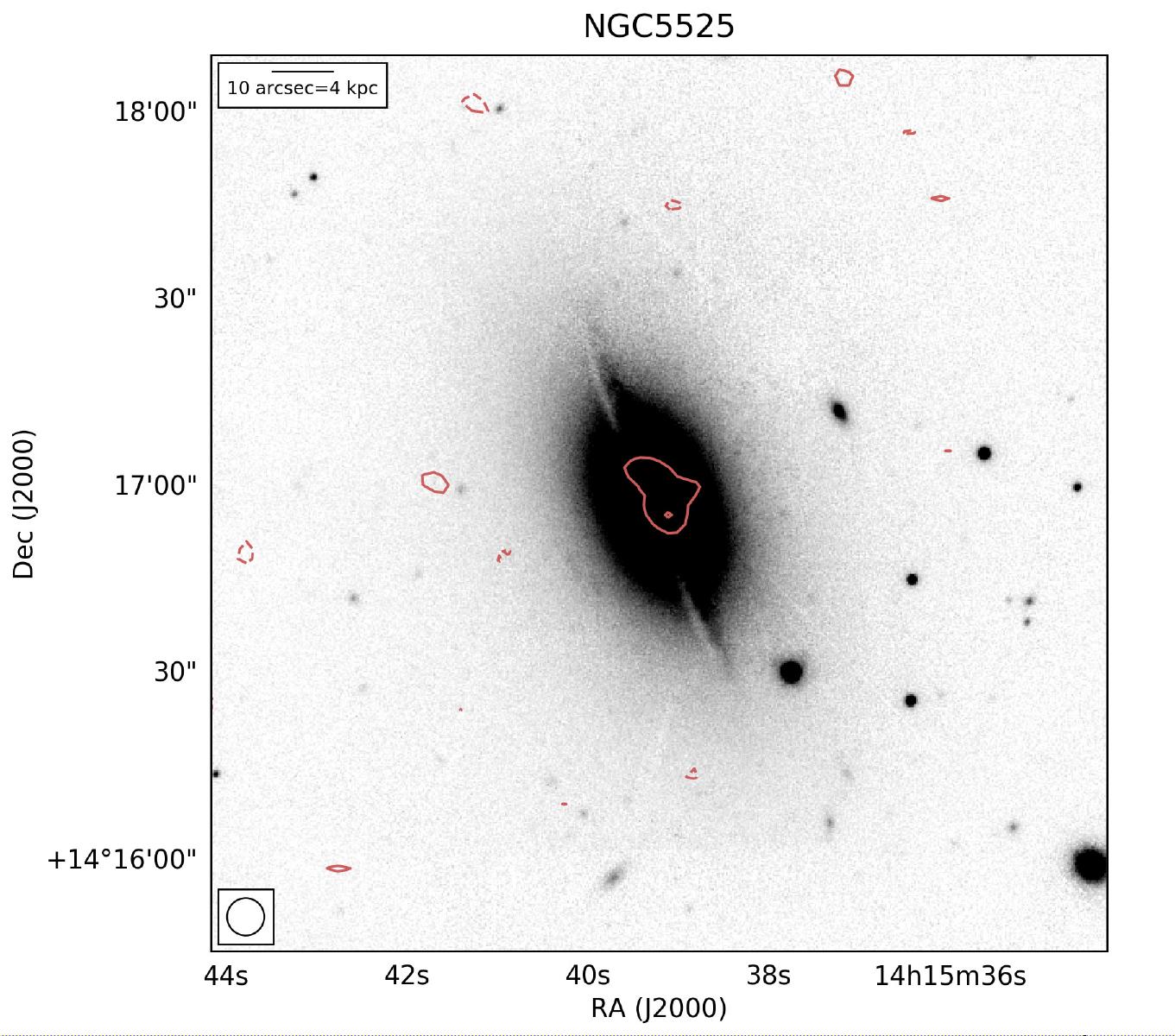}
\caption{Three examples of LOFAR images at 150 MHz of galaxies not
  classified as extended sources (that is, whose 3$\sigma$ level
  contours do not reach a radius of 15$\arcsec$ but whose FWHM,
  measured by fitting a 2D Gaussian to the central portion of the
  images, exceeds 15$\arcsec$.}
\label{diffuse}
\end{figure*}

\section{Results}
\label{results}

The distributions in redshift and absolute K-band magnitude (see
Fig. \ref{kzhist}) show the expected peaks close to the thresholds of
the sample selection. The median recession velocity is 6,200 \kms, and
the host galaxies have a median magnitude of M$_{\rm K}=-25.3$ and
cover a range of a factor of $\sim$ 5 in near-infrared luminosity. Using
the relation between the dynamical mass and K-band absolute magnitude,
log $M_* = 10.58-0.44\times(M_{Ks}+ 23)$ (derived by
\citealt{cappellari13} from a sample of ETGs with $-21.5 < M_K < -26$), and
the corresponding stellar masses of our sample are in the $2.9
\times10^{11} - 1.5 \times10^{12} {\rm M}_\odot$ range (median $\sim4
\times10^{11} {\rm M}_\odot$).

The distribution of luminosity at 150 MHz for the sources of the
sample (Fig. \ref{lum}) shows a large spread in power, from $\lesssim
10^{21} \WHz$ to $\sim 10^{26} \WHz$.  Forty-two sources
of the sample are not detected by the LOFAR observations at a
5$\sigma$ flux density limit ranging from 0.4 to 7.4 mJy, with a
median value of 1.0 mJy. The corresponding limits to the radio
luminosity range from 1.6$\times10^{20}$ to 4.8 $\times10^{21} \WHz$,
with a median value of 1.1 $\times10^{21} \WHz$.

\subsection{Radio morphologies of gETGs}

We defined the 46 sources in which the 3$\sigma$
radio contour extends to a radius of at least 15\arcsec, about twice
the beam FWHM, as extended objects. The extended sources cover
a large range of luminosities ($10^{21} \lesssim L_{150} \lesssim
10^{26} \WHz$) and represent the majority (82\%) of the objects with
$L_{150} > 10^{23} \WHz$ (see Fig. \ref{lum}).

Three examples of their radio images are presented in
Fig. \ref{estese} and all are collected in Fig. \ref{esteseall} in the
appendix B. The sizes of their radio structures (measured as the largest
distance at which the radio emission is detected above the 3$\sigma$
limit) range from $\sim$ 4 to $\sim$ 340 kpc. A clear positive trend
between radio power and size is present (see Fig. \ref{las}), with a
behavior consistent with that found for other samples of
radio-galaxies (such as the B2 and 3C, e.g., \citealt{deruiter90} and
more recent samples, e.g., \citealt{hardcastle19}). Most of these
extended radio sources (34, $\sim$ 76\%) have a morphology clearly
indicating the presence of jets, usually with an edge-darkened
morphology typical of \citet{fanaroff74} FR~I class (well defined in
30 cases), with only one example of an FR~II (NGC~2892). A few more
sources are barely resolved but suggest the presence of a jet
structure. Several gETGs show a complex morphology, in at least two
cases (namely, NGC~3842 and UGC~12482) presenting a central edge
brightened structure and large diffuse tails, suggestive of restarted
activity. In two objects (NGC~0687 and NGC~2672) the radio
emission, extending over $\sim$ 40\arcsec ($\sim$ 15 kpc) and $\sim$
105\arcsec ($\sim$ 30 kpc), respectively, is diffuse and lacks of any
well defined radio core. For three gETGs (namely IC~5180, UGC~10097,
and NGC~2789) the inspection of the low-resolution ($\sim 25\arcsec
\times 25\arcsec$, rms $\sim$250 $\mu$Jy/beam) LOFAR images (see
Fig. \ref{low}) reveals the presence of low-brightness lobes not (or
barely) visible in the full resolution images. The radio morphology of
the gETGs is indicated in Table. \ref{tab}. In Appendix B, we also
give a brief description of the most interesting or complex cases.

For all detected sources in the DR2 area we retrieved from the
internally released catalog the estimate of their major axis length
$R_{\rm maj}$, obtained using Gaussian fitting as described in
\citet{shimwell19}, while for those observed in single pointings we
measured this parameter by fitting a 2D Gaussian to the central
portion of the images. These sizes, reported in Table. \ref{tab}, refer
effectively to the central component of the radio source; while it can
be considered as a robust size estimate for the compact sources, it
should be treated with caution for the more extended ones.

In Fig. \ref{fwhm}, we show the distribution of $R_{\rm maj}$ against
their flux density. There is a good correspondence between the
sources classified as extended based on the manual size measure and
those showing R$_{\rm maj} >$15\arcsec\ in the catalog based on the
automatic fitting. There are, however, some exceptions; in particular,
there are nine objects not included in the list of the extended
sources but with a manually measured size $\gtrsim 15 \arcsec$.  We
mark them with red circles in Fig. \ref{fwhm} and as P* in
Table. \ref{tab}.
Inspection of their images (see Fig. \ref{diffuse} and Fig.
\ref{diffuseall}) shows that at least two of them (namely NGC~1508 and
NGC~2493) are possibly jetted sources, while at least another two
(namely NGC~0750 and NGC~7722) are diffuse sources, similar to
NGC~0687 but of even lower surface brightness.

However, most of the radio sources associated with the gETGs are
compact. Out of the 146 detected objects, more than 60\% of the
sources has a major axis $\lesssim 10 \arcsec$, corresponding to a
size of $\lesssim 4$ kpc.

\subsection{Spectral properties of gETGs}
\label{spixsec}
\begin{figure}
\includegraphics[scale=0.5]{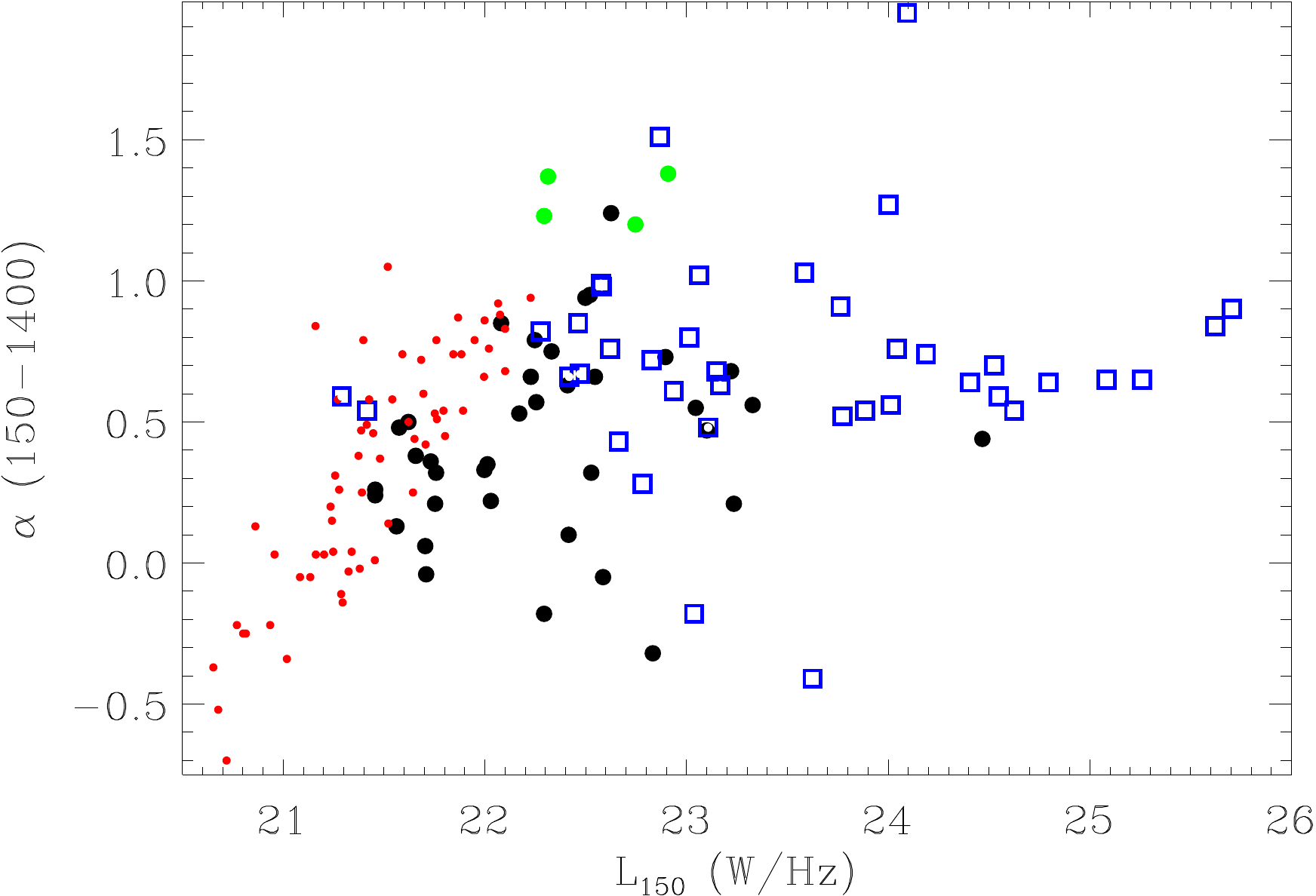}
\caption{Radio spectral index between 150 MHz and 1.4 GHz defined as
  $F_{\nu}\propto\,\nu^{-\alpha}$ versus source luminosity at 150 MHz. The
  green circles are the lower limits derived for the five extended
  sources not detected at 1.4 GHz. The remaining extended sources in
  the LOFAR images are marked with blue squares, while the black dots
  are the unresolved ones. The smaller red dots represent lower
  limits due to the lack of a detection at 1.4 GHz. }
\label{spix}
\end{figure}

In Fig. \ref{spix}, we compare the LOFAR luminosity with the spectral
index between 150 MHz and 1.4 GHz. As already mentioned, the
measurements at 1.4 GHz were obtained when available, and reliable,
from the FIRST otherwise from the NVSS. All sources detected at 1.4
GHz are also detected in the LOFAR images.

The point-like objects show a large spread in their spectral slopes,
$-0.4 < \alpha_{150}^{1400} < 1.2$, but the range could be even larger
considering the large fraction of lower limits. Only one compact
source detected at both frequencies, NGC~4555, has a spectral index of
$>$1 ($\alpha_{150}^{1400}=1.24$). 

Most extended sources have $\alpha_{150}^{1400}$ in the 0.5 - 1 range,
which is typical of sources with active jets, but there are a few notable
exceptions. Seven of them show a very steep spectral slope (
$\alpha_{150}^{1400} > 1.2$, namely NGC~0080, NGC~0507,
NGC~0687, NGC~0910, NGC~2672, NGC~2832, and NGC~3842), only three of
them being detected (NGC~0507, NGC~2672, and NGC~3842) by the surveys
at 1.4 GHz. NGC~507 is already reported to be a remnant source by
\citealt{murgia11}). All of them are characterized by rather weak
cores, surrounded large-scale emission, often showing a diffuse
structure.  On the opposite extreme, there is NGC~3894 showing an
inverted spectrum, $\alpha_{150}^{1400} = -0.18$: this source has two
symmetric faint jets and one bright core, which dominates the flux
density of the source and drives its integrated spectral shape in this
frequency range.

\begin{figure}
\includegraphics[scale=0.5]{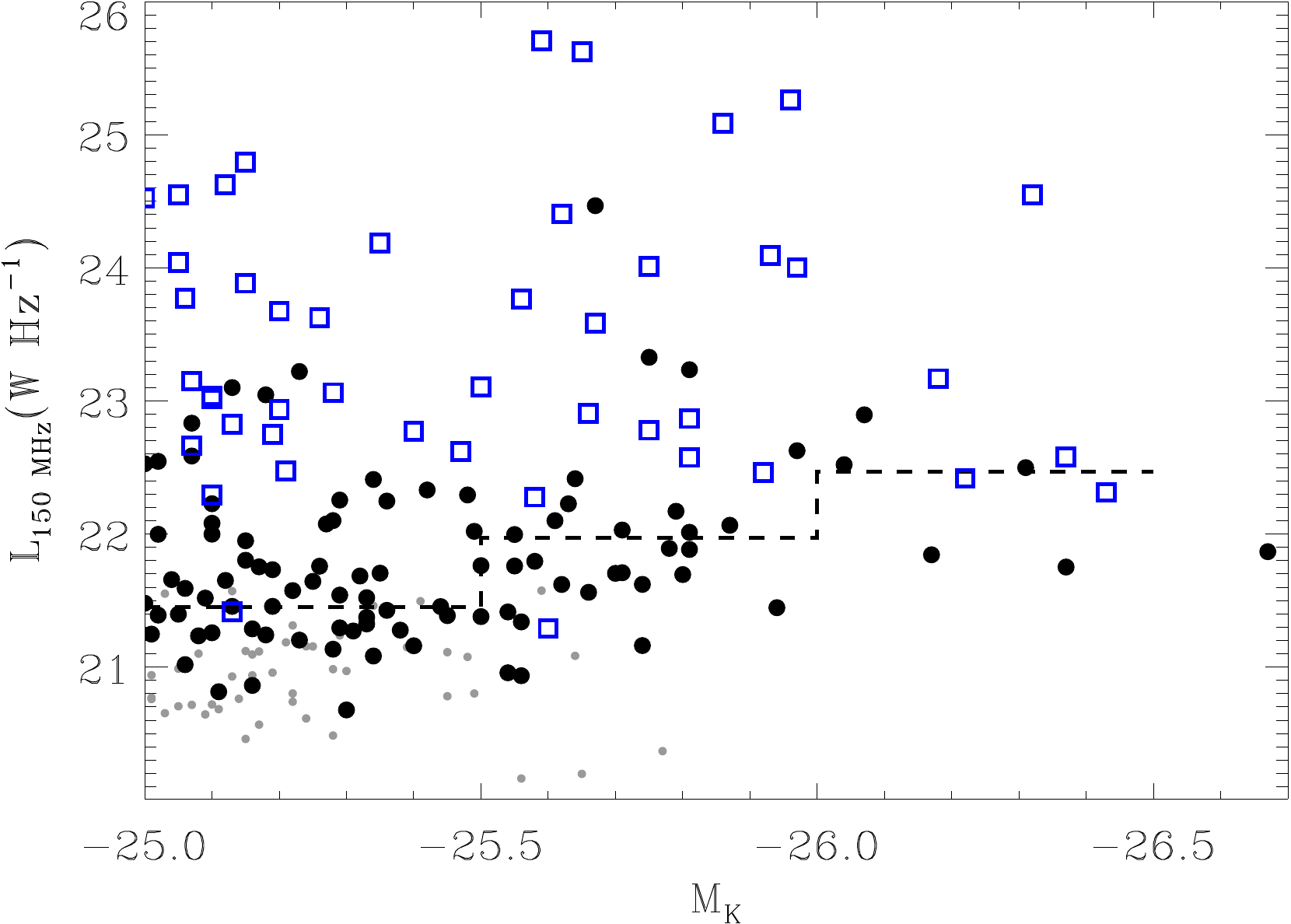}
\caption{Comparison between absolute K-band magnitude and the
  luminosity at 150 MHz. The gETGs show a positive link between the
  stellar luminosity and median radio power but with a very large
  spread. The blue squares are the extended sources, the black circles
  the unresolved ones, and the small gray dots are the undetected
  objects. The dashed histogram represents the median radio luminosity
  in three bins of absolute magnitude. \label{mlum}}
\end{figure}

\begin{figure}
\includegraphics[scale=0.5]{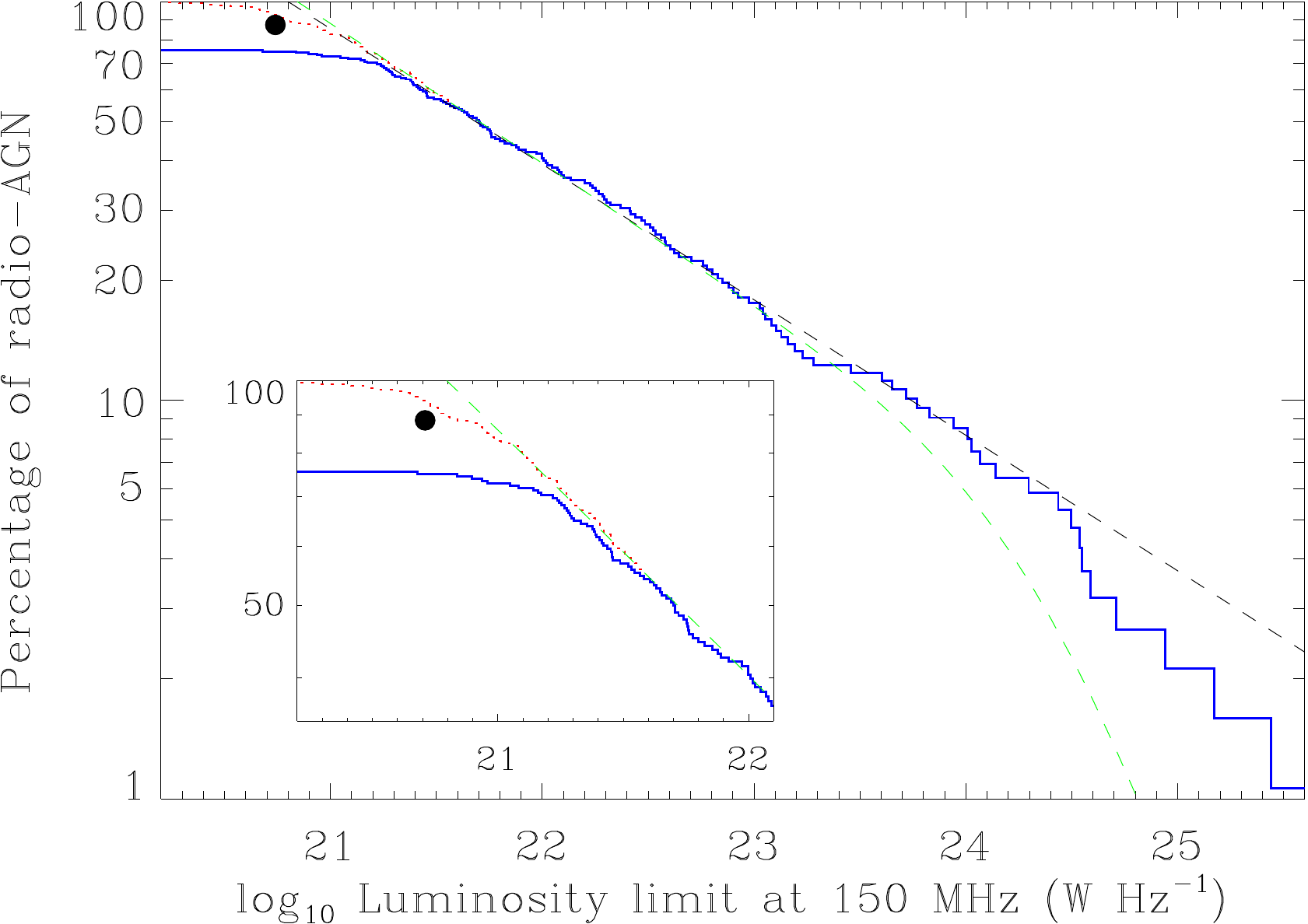}
\caption{Fraction of galaxies that host a radio source
  above a given luminosity limit. The blue curve corresponds to the
  detected sources, the red curve includes the upper limits and
  represents the upper envelope of the distribution. The black dot
  represents the result of the stacking of the 42 undetected
  galaxies. The black dashed line is a power law with a slope of
  0.34. The green dashed line is instead the fit obtained by
  \citet{best05a}. The inset shows a zoomed-in look at the lowest
  luminosities. }
\label{flum}
\end{figure}

\begin{figure}
\includegraphics[scale=0.5]{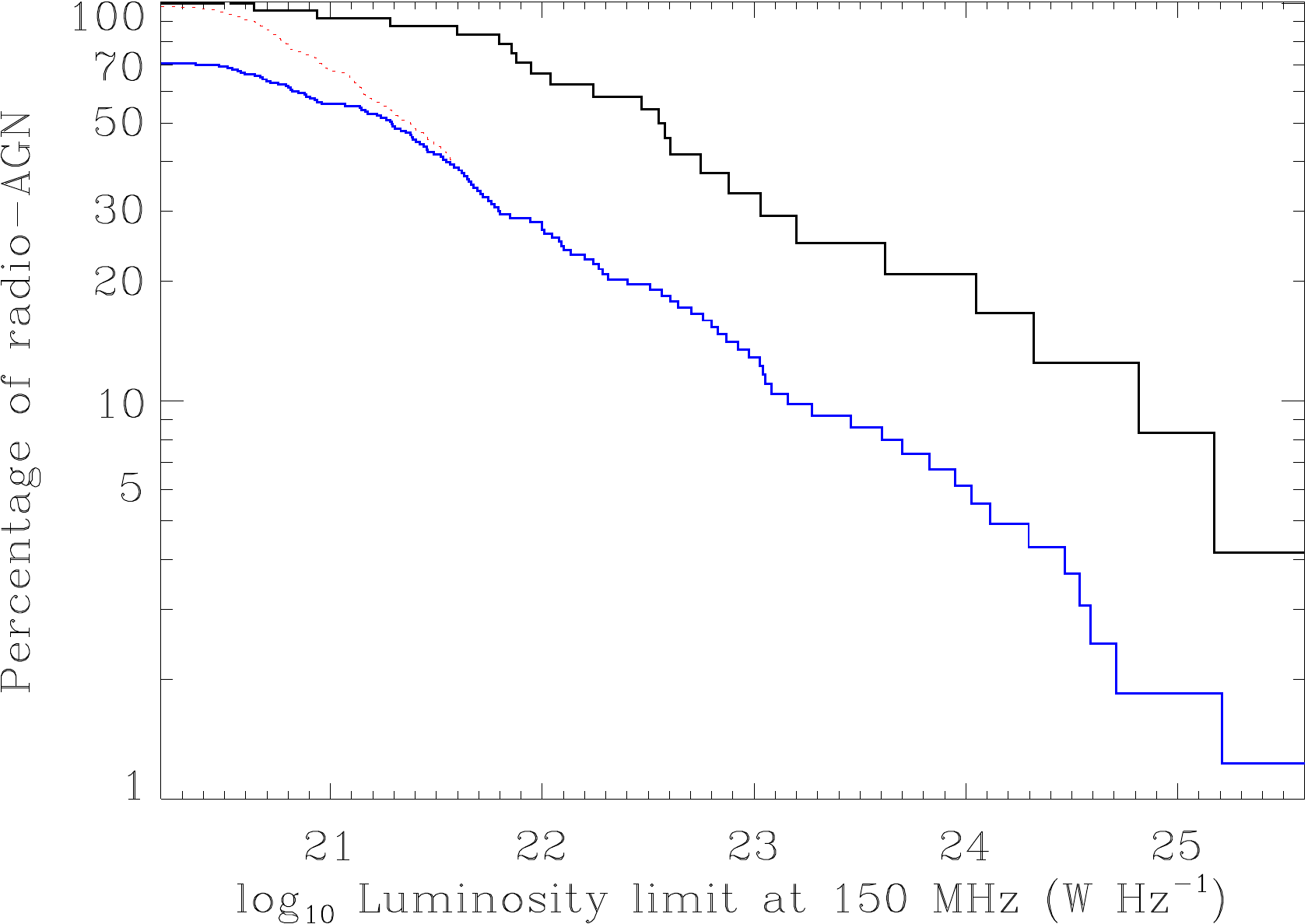}
\caption{Same as Fig. \ref{flum}, but splitting the sample into two
  bins of host luminosity. The black represents the most massive gETGs
  (K$<$-25.8), the blue curve corresponds to the less massive gETGs
  detected with LOFAR, while the red curve includes the upper limits
  and represents the upper envelope of the distribution for the less
  massive gETGs.}
\label{flum2}
\end{figure}

\subsection{Radio and near-infrared host luminosity}
  
In Fig. \ref{mlum}, we compare the near infrared and radio luminosity
of the gETGs. By splitting the sample into three bins, 0.5 magnitudes
wide (that is, separated by a factor of $\sim 1.6$ in luminosity), we
find that the median radio luminosity increases by a factor of $\sim$3 in
each subsequent bin of host luminosity with a dependence of $L_{150}
\sim L_K^{2.4}$, in good agreement with previous estimates
(\citealt{best05a}). Nonetheless, the spread in $L_{150}$ within each
bin is extremely large, reaching six orders of magnitude for the
galaxies of intermediate values of $L_K$. The fraction of detected
sources also increases with $L_K$: all gETGs with $M_K <-26.0$ (and
actually down to $M_K = -25.8$) are detected in the LOFAR images, only
five (that is, $\sim9$\%) are undetected in the bin $-25.5 < M_K
<-26.0$, while the fraction of undetected sources increases to $\sim
35\%$ for the least luminous galaxies. Similarly, the fraction of
gETGs with extended radio structures decreases with decreasing
infrared luminosity from $\sim 45\%$ to $\sim 34\%$, and, finally, to
$\sim 18\%$ for the same three bins as above.  The connection between
$L_{150}$ and $L_K$ is present also considering the
extended and point-like (or undetected) sources separately.

\subsection{The fractional radio luminosity function}

Figure \ref{flum} reports the fraction of gETGs in which we detected a
radio source with a power larger than a given threshold. This function
is fully determined for $\log L_{150} > 21.6$ (where $f=55\%)$, the
luminosity where we encounter the first upper limit. Below this value,
the presence of undetected sources produces two branches, representing
the lower and upper envelope of the true distribution. For values
$\log L_{150} > 21.15$, however, they differ by less than 5\% (71\%
and 76\%, respectively).

This analysis can be extended to even lower radio luminosities by
taking advantage of the stacking technique (e.g.,
\citealt{white07}). By performing a median filtering of the images of
the 42 undetected sources, we obtain a source with a flux of 0.55 mJy
at a 8$\sigma$ significance. By stacking in the luminosities domain we
estimate a median power of $5.1\times10^{20} \WHz$ corresponding to
the 89th percentile for the whole sample.\footnote{The median
luminosity refers to the subsample of 42 upper limits, that is, to
the 21st most luminous source among them, to be compared to the full
sample size of 188 objects.}  The small offset between the upper
envelope of the luminosity distribution from the results of the
stacking is rather small ($\sim 5\%$), indicating that at least half
of the undetected sources have luminosities very close to the
detection threshold.

The dependence of the radio luminosity on the host mass can be clearly
seen by splitting the sample into two bins of magnitude, considering
separately the galaxies more luminous than M$_K<$-25.8 (see
Fig. \ref{flum2}). The shape of the luminosity functions for the two
groups (whose median magnitudes are M$_K<$ -25.97 and M$_K<$ -25.25,
respectively) are quite similar, but with an offset of a factor 20 in
radio luminosity or, looking from an orthogonal perspective, the more
massive hosts are $\sim$3 times more likely to have a
radio luminosity above a given threshold.

\section{Optical spectroscopic properties of gETGs}

\begin{figure*}
\includegraphics[scale=1.0]{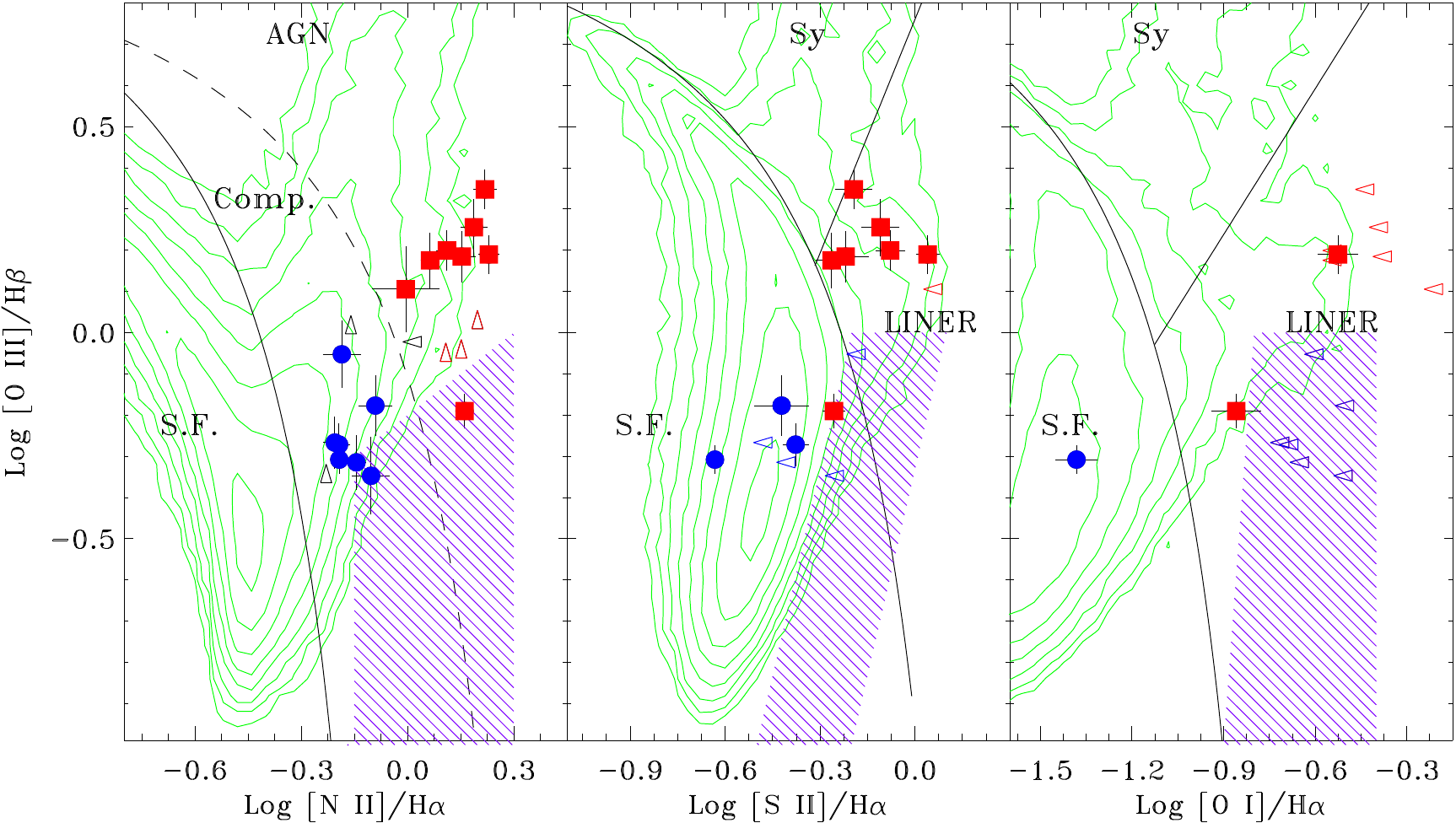}
\caption{Optical spectroscopic diagnostic diagrams for gETGs with
  available SDSS spectra adopting a conservative threshold of S/N$>$5
  for the lines detection. The solid lines are from \citet{kewley06}
  and separate star forming galaxies, LINER, and Seyfert; in the first
  panel, the region between the two curves is populated by the
  composite galaxies. The blue circles (red squares) represent gETGs
  located within the region populated by the composite galaxies
  (LINERs). A black symbol corresponds to a source of uncertain
  classification. The green contours represent the iso-densities of
  all DR7 emission line galaxies. Purple dashed regions mark the
    location of ionized gas filaments in galaxy clusters as measured
    by \citet{mcdonald12}.}
\label{diag}
\end{figure*}

\begin{table*}
\caption{Optical emission line properties from the SDSS spectra}
\begin{tabular}{l| r r r r | r r | l | l}
\label{tab2}
Name & [O~III]/\Hb & [N~II]/\Ha & [S~II]/\Ha & [O~I]/\Ha &  L$_{150}$ & L$_{\rm H\alpha}$ & Class & Radio morph.\\ 
\hline  
NGC2492  &        $>$-0.29  &           ---      &           ---    &           ---    &    21.39 &  $<$4.40&        &  P  \\       
 IC2393  &             ---  &           ---      &           ---    &           ---    & $<$20.76 &  $<$4.78&        &     \\       
UGC04767 &             ---  &           ---      &           ---    &           ---    &    22.47 &  $<$4.71&        & FRI \\       
NGC2759  &        $>$-0.34  & -0.23$\pm$ 0.10    &        $<$-0.04  &        $<$-0.33  &    21.32 &     5.19&        &  P  \\       
NGC2783  &  0.19$\pm$ 0.05  &  0.23$\pm$ 0.03    &  0.04$\pm$ 0.04  & -0.53$\pm$ 0.07  &    22.78 &     5.54&  LINER & E ? \\ 
NGC2789  & -0.31$\pm$ 0.04  & -0.19$\pm$ 0.01    & -0.63$\pm$ 0.02  & -1.38$\pm$ 0.07  &    23.01 &     6.30&  SF    & E ? \\    
UGC04972 &        $>$-0.05  &  0.11$\pm$ 0.08    &        $<$-0.01  &        $<$-0.13  &    23.77 &     5.06&  LINER & FRI \\ 
UGC04974 & -0.11$\pm$ 0.10  &           ---      &           ---    &           ---    & $<$20.71 &  $<$4.71&        &     \\       
NGC2918  & -0.05$\pm$ 0.08  & -0.19$\pm$ 0.05    &        $<$-0.19  &        $<$-0.60  &    20.96 &     5.39&  SF    &  P  \\    
NGC3158  &        $<$ 0.19  &        $<$-0.42    &        $<$-0.03  &        $<$-0.15  &    22.58 &     5.25&        & FRI \\       
NGC3615  &             ---  &           ---      &           ---    &           ---    &    21.57 &  $<$4.89&        &  P  \\       
NGC3710  &        $>$-0.24  &           ---      &           ---    &           ---    & $<$21.55 &  $<$4.53&        &     \\       
NGC3713  &        $<$ 0.29  &           ---      &           ---    &           ---    & $<$21.15 &     4.69&        &     \\       
NGC3805  & -0.27$\pm$ 0.06  & -0.21$\pm$ 0.03    &        $<$-0.48  &        $<$-0.72  &    21.89 &     5.63&  SF    &  P* \\    
NGC3816  &             ---  &           ---      &           ---    &           ---    & $<$21.16 &  $<$4.56&        &     \\       
NGC3837  & -0.31$\pm$ 0.07  & -0.14$\pm$ 0.03    &        $<$-0.41  &        $<$-0.65  &    21.49 &     5.56&  SF    &  P  \\    
NGC3842  &        $<$ 0.13  &           ---      &           ---    &           ---    &    23.99 &  $<$4.76&        & E ? \\       
NGC3862  & -0.19$\pm$ 0.04  &  0.16$\pm$ 0.02    & -0.26$\pm$ 0.04  & -0.86$\pm$ 0.08  &    25.71 &     5.61&  LINER & FRI \\ 
NGC3886  &        $<$ 0.15  &           ---      &           ---    &           ---    & $<$20.94 &  $<$4.89&        &     \\       
NGC3919  & -0.18$\pm$ 0.07  & -0.09$\pm$ 0.05    & -0.42$\pm$ 0.09  &        $<$-0.50  &    21.63 &     5.56&  SF    &  P* \\    
NGC3937  &        $>$-0.22  &           ---      &           ---    &           ---    &    21.68 &  $<$5.05&        &  P  \\       
NGC3971  &        $>$-0.04  &  0.15$\pm$ 0.06    &        $<$ 0.14  &        $<$-0.28  &    21.26 &     5.06&  LINER &  P  \\ 
NGC4065  &  0.03$\pm$ 0.09  &           ---      &           ---    &           ---    &    22.43 &  $<$4.67&        &  P  \\       
UGC07115 & -0.27$\pm$ 0.05  & -0.19$\pm$ 0.03    & -0.38$\pm$ 0.04  &        $<$-0.69  &    24.46 &     5.58&  SF    & FRI \\    
UGC07132 &        $>$ 0.02  & -0.16$\pm$ 0.07    &        $<$-0.06  &        $<$-0.39  &    21.74 &     5.44&        &  P  \\       
NGC4213  & -0.02$\pm$ 0.09  &        $<$-0.50    &        $<$ 0.03  &        $<$-0.12  &    21.73 &     5.11&        &  P  \\       
NGC4229  &        $>$ 0.03  &  0.20$\pm$ 0.07    &        $<$ 0.18  &        $<$-0.17  &    20.77 &     4.98&  LINER &  P  \\ 
 IC0780  &        $<$ 0.10  &        $<$-0.68    &        $<$ 0.10  &           ---    &    22.69 &     5.00&        & FRI \\       
NGC4555  & -0.02$\pm$ 0.12  &           ---      &           ---    &           ---    &    22.59 &  $<$4.84&        &  P  \\       
NGC4583  &  0.20$\pm$ 0.05  &  0.11$\pm$ 0.03    & -0.08$\pm$ 0.05  &        $<$-0.55  &    21.57 &     5.14&  LINER &  P  \\ 
NGC4715  &             ---  &           ---      &           ---    &           ---    & $<$20.93 &  $<$4.55&        &     \\       
NGC4841A &        $>$-0.22  &           ---      &           ---    &           ---    &    21.45 &  $<$4.79&        &  P  \\       
NGC4886  &             ---  &           ---      &           ---    &           ---    &    21.87 &  $<$4.05&        &  P  \\       
NGC4952  & -0.02$\pm$ 0.12  &           ---      &           ---    &           ---    &    20.86 &     4.99&        &  P  \\       
NGC4957  &             ---  &           ---      &           ---    &           ---    & $<$20.80 &  $<$4.82&        &     \\       
NGC4978  & -0.04$\pm$ 0.07  &           ---      &           ---    &           ---    &    21.44 &     5.02&        &  P  \\       
 IC0885  &  0.11$\pm$ 0.10  & -0.00$\pm$ 0.10    &        $<$ 0.06  &        $<$-0.21  & $<$21.57 &     5.11&  LINER &     \\ 
NGC5127  &  0.26$\pm$ 0.07  &  0.19$\pm$ 0.04    & -0.11$\pm$ 0.06  &        $<$-0.39  &    24.80 &     4.93&  LINER & FRI \\ 
NGC5141  &  0.35$\pm$ 0.05  &  0.22$\pm$ 0.03    & -0.19$\pm$ 0.06  &        $<$-0.44  &    24.53 &     5.29&  LINER & FRII\\ 
NGC5322  &             ---  &           ---      &           ---    &           ---    &    22.62 &     3.90&        & FRI \\       
NGC5513  &             ---  &           ---      &           ---    &           ---    & $<$21.19 &  $<$4.41&        &     \\       
NGC5525  & -0.35$\pm$ 0.09  & -0.10$\pm$ 0.05    &        $<$-0.25  &        $<$-0.51  &    21.36 &     5.01&  SF    &  P* \\    
 IC1153  &  0.18$\pm$ 0.06  &  0.15$\pm$ 0.04    & -0.22$\pm$ 0.07  &        $<$-0.38  &    21.73 &     5.24&  LINER &  P  \\ 
NGC7722  &  0.18$\pm$ 0.07  &  0.06$\pm$ 0.03    & -0.26$\pm$ 0.05  &        $<$-0.55  &    21.66 &     4.92&  LINER &  P* \\ 
  \hline
\end{tabular}
\label{tab1}

\smallskip
\small{Column description: (1) Name, (2-5) emission line ratios, (6)
  radio luminosity at 150 MHz (in $\WHz$), (7) \Ha\ line luminosity in
  solar luminosities, (8) spectral classification, and (9) radio morphology.}
\end{table*}

The spectroscopic diagnostic diagrams are commonly used to constrain the
gas ionization mechanism
(e.g., \citealt{heckman80,baldwin81,veilleux87,kewley06}) by measuring
ratios of selected emission lines.  From the main sample of
galaxies with spectra available from the SDSS, Data Release 7 (DR7), we
used the MPA-JHU DR7 release of spectral
measurements,\footnote{Available at {\sl
  http://www.mpa-garching.mpg.de/SDSS/DR7/}} to obtain fluxes of the
diagnostic optical emission lines. SDSS spectra are available for 44
of the gETGs of our sample. However, only in a minority of these
spectra it is possible to measure a sufficient number of emission
lines, adopting a conservative threshold of a 5$\sigma$ detection, to
locate the gETGs in the diagnostic diagrams (see Fig.~\ref{diag}). In
Table \ref{tab2}, we list the key diagnostic line ratios and the
\Ha\ luminosity, in units of solar luminosity.

Eleven gETGs fall, in the [O~III]/\Hb\ versus [N~II]/\Ha\ diagram, into
the AGN region, while seven of them are located into the region of
``composite galaxies''. None of the gETGs are located in the Seyfert
region. For three sources, the limits on the line ratios do not allow
us to derive a secure classification. The study presented by
\citet{brinchmann04} indicates that in 70\% of the composite galaxies
the AGN contribution accounts for less than 20\% of the
\Ha\ luminosity, a fraction that rarely exceeds 50\%. We consider
these seven gETGs as star forming (SF) candidates for which the radio
emission may be associated with star-formation. The dominance of star
formation to the gas ionization is confirmed based on their
[S~II]/\Ha\ ratios, while [O~I]/\Ha\ ratio can be measured only in one
source. These galaxies span the luminosity range -25.0 $<$ M$_K<$-25.8
with a median value of M$_K=$-25.4, similar to the median value of the
whole gETGs sample.

Alternative explanations for the origin of ionized gas in massive
galaxies, often located at the center of groups or cluster of
galaxies, have been proposed and include weak shocks (e.g.,
\citealt{sparks89}), ionization due to cooling of the hot
intra-cluster medium \citep{voit94}, thermal conduction
\citep{mcdonald10}, and reconnection diffusion \citep{fabian11}. The
location in the diagnostic diagrams of the ionized gas filaments
measured by \citet{mcdonald12} is reported as the dashed purple
regions in Fig. \ref{diag}, and it is generally inconsistent with the
line ratios measured in the SF candidates, further supporting our
conclusion that the gas ionization in these sources is due to young
stars.

The seven SF candidates are all detected by LOFAR. They display
different radio morphologies: two (NGC~2918 and NGC~3837) are compact
sources, with measured sizes of 8.0\arcsec and 10.6\arcsec,
respectively (deconvolved sizes $\sim$3 kpc), three (NGC~3805,
NGC~3919, and NGC~5525) are diffuse, low-brightness sources with sizes
between 6 and 10 kpc (see Fig. \ref{diffuse}), while UGC~07115 is an
extended ($\sim 300$ kpc) edge-darkened source.  The last SF
candidate, NGC~2789, is a compact radio source ($\sim$6 kpc)
associated with a pair of large-scale ($\sim$800 kpc), low-brightness
diffuse lobes, detached from the central component (see
Fig. \ref{low}). From the point of view of the radio spectra, four SF
candidates are detected at 1.4 GHz, with $\alpha_{150}^{1400}$ ranging
from 0.22 to 0.88, consistent with the low-frequency radio slope of
star forming galaxies ($\alpha=0.59$, \citealt{klein18}), while the
lower limits for those undetected at 1.4 GHz are in the 0.03 -
0.51 range.

\begin{figure}
\includegraphics[scale=0.60]{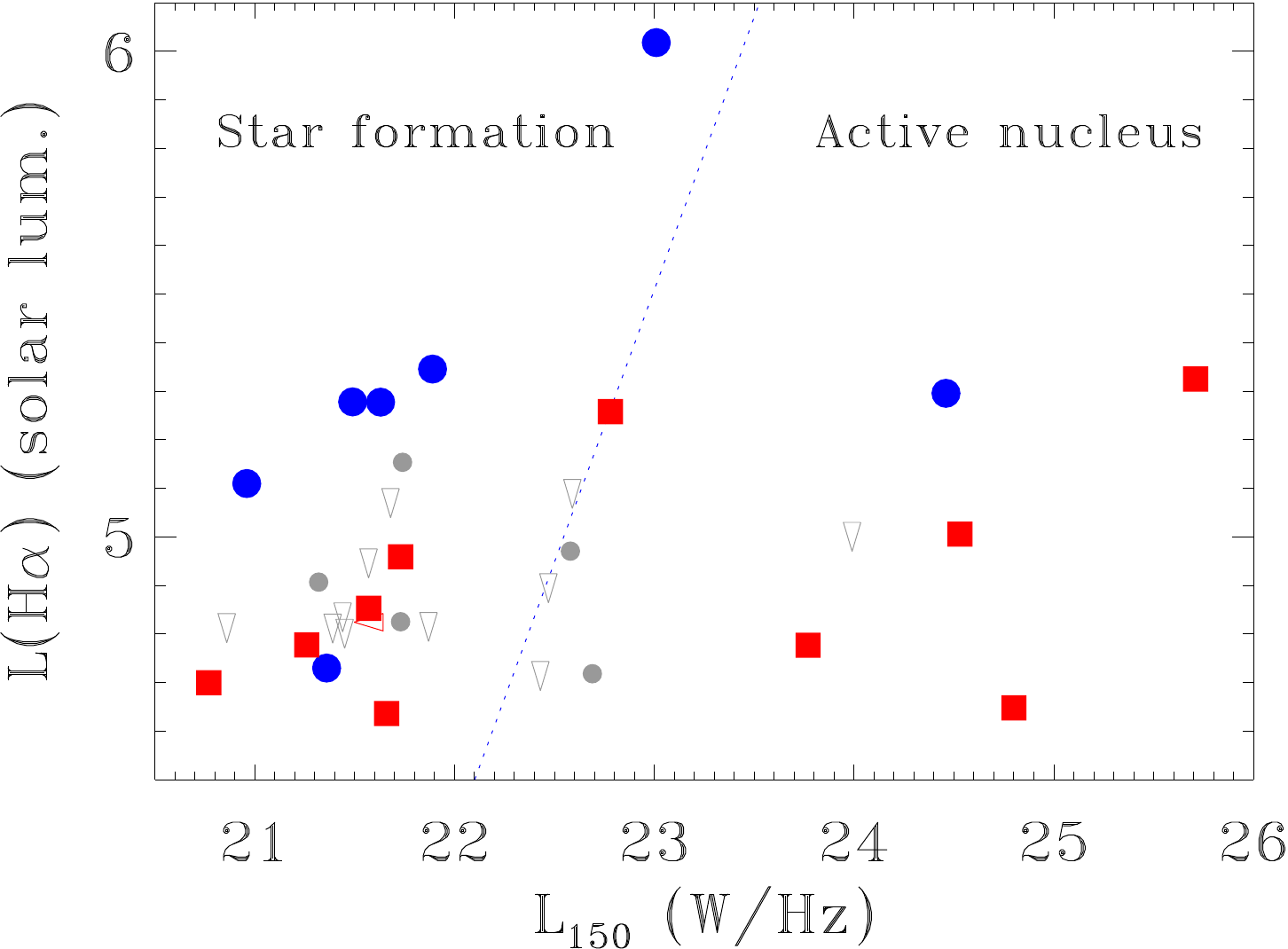}
\caption{Comparison of radio and \Ha\ luminosities for the gETGs
  with available SDSS spectra. The diagonal line marks the empirical
  separation derived by \citet{best12} between sources in which the
  radio emission is dominated by star forming regions (to the left)
  from those in which it is powered by an AGN (to the right). The blue
  circles correspond to the sources located in the region of star
  forming galaxies in Fig. \ref{diag}, the red squares are those
  located in the LINERs region, the gray ones are those for which no
  optical spectral classification is possible, the gray downward
  pointing triangles are upper limits.}
\label{lum2}
\end{figure}

By adopting the relation from \citet{gurkan18}, the radio luminosities
at 150 MHz of the star forming candidates can be translated into a
star formation rate (SFR). By excluding UGC~07115, whose radio
emission is dominated by a large-scale structure powered by jets, we
obtain a range between 0.1 and 0.5 M$_\odot$yr$^{-1}$, with the
exception of NGC~2789, for which this estimate is $\sim$8
M$_\odot$yr$^{-1}$. These values compare favorably with those derived
by \citet{brinchmann04} based on the analysis of their SDSS spectra,
with an SFR in the 0.05 - 4 M$_\odot$yr$^{-1}$ range.

The origin of the radio emission of the gETGs with a LINER spectrum (or
unclassified) on the left side of Fig. \ref{lum2} is uncertain,
because they can be at least in part powered by a low-luminosity
AGN. Consequently, we can only derive upper limits to their star
formation rates, $\lesssim0.2-1.2$ M$_\odot$yr$^{-1}$. For the gETGs
not detected by LOFAR, the SFR limits are in the
$\lesssim0.1-0.3$ M$_\odot$yr$^{-1 }$range.

These results suggest that star formation might be the dominant
mechanism for the radio emission in a significant fraction (6 out 44;
that is, $\sim$ 14\%) of gETGs. This conclusion is also supported by the
diffuse radio morphology observed in three of them.

A different approach to separate radio sources powered by star
formation and by an AGN is based on the ratio between \Ha\ and radio
luminosity, see Fig. \ref{lum2}. The dividing line is based on the
empirical relation derived by \citet{best12}, representing a rather
conservative boundary to recognize radio AGNs. We scaled this relation
to 150 MHz luminosities by adopting the mean slope of the
low-frequency spectrum of star forming galaxies ($\alpha=0.59$,
\citealt{klein18}). Nine gETGs, not detected by LOFAR and in the
\Ha\ line, cannot be located in this diagram. UGC~07115, the SF
candidate with the large-scale radio structure, is located to the
right of the boundary between SF- and AGN-powered sources (by
considering only its core emission, log$L_{150}=22.85$, this galaxy
falls on the boundary between SF and AGN). The remaining six SF
candidates are in the region populated by star forming galaxies,
providing further support to the conclusion that their line emission
is associated with the presence of young stars. It must also be noted
that the measurements of the \Ha\ luminosity refer only to the region
covered by the SDSS fiber (3\arcsec\ of diameter, $\sim 1.5$ kpc) and
are likely to underestimate the total line luminosity of the galaxy.

The presence of sources with a LINER spectrum falling into the region
of star forming galaxies in this diagram is, at first sight,
puzzling. However, \citet{capetti11b} and \citet{cid11} have shown
that, following earlier analysis (e.g.,
\citealt{trinchieri91,binette94}), the LINERs region in the diagnostic
diagram is actually mainly populated by objects in which the ionizing
radiation is produced by old stars, such as post-asymptotic giant
branch stars and white dwarves. \citet{capetti11b} noted that in
massive quiescent ETGs, the [O~III] line EW is strongly clustered
around 0.75\AA, while genuine AGNs are associated with larger EW
values. The six gETGs located on the left side of the dividing line
in Fig. \ref{lum2} have EW values between 0.5 and 1.0 \AA. They are
consistent with their being sources in which the ionizing field is dominated
by their old stellar population rather than by young stars.

We conclude that at very low levels of luminosity, the comparison
between radio and line emission does not provide a robust tool to
understand the origin of radio emission due to the possible
contribution to the line emission from different processes, which is not
related to star formation and active nuclei.

\section{Discussion}

\subsection{The radio morphology of gETGs}

The morphology of radio sources powered by an AGN is determined by the
interplay of its age, environment, and jet properties. The analysis
presented provides us with an unbiased census of the radio morphology
of the most massive early-type galaxies in the local Universe,
exploring a complete sample selected only on the basis of their
stellar luminosity and regardless of the radio properties.

The size distribution of the radio sources associated with the gETGs
shows that 91 (62\% of the detected sources) are compact (with a size
limit of $\sim$ 4 kpc), nine are ``diffuse'', and 46 are extended; within
the latter class, 17 (12\%) have an L.A.S. between 10 and 30 kpc, eight (5\%)
extend to 60 kpc, 4 (3\%) to 100 kpc, and finally 15 (10\%) gETGs
have a size exceeding 100 kpc. In \citet{baldi19}, we found that in the
\citet{best12} sample of radio AGNs selected at 1.4 GHz, 63\% are the
compact FR~0s, 23\% have an L.A.S. between 10 and 30 kpc, 13\% have 30
$<$L.A.S.$<$60 kpc, and only one source reaches 70 kpc. The lack of
very extended sources among those considered by \citeauthor{baldi19}
is due to the combination of the higher depth of the LOFAR images with
respect to the FIRST ones (particularly important for the extended
steep spectrum structures) and to the fact that structures larger than
$\sim1^\prime$ might be resolved out due to the lack of short
baselines in the u−v coverage of the FIRST data \citep{helfand15}.

The main result is that the majority of the radio sources associated
with gETGs have a size smaller than $\sim$ 4 kpc. As already discussed
by \citet{baldi18,baldi19}, the large fraction of compact radio
sources rules out the possibility that these are all young sources
that will evolve into more extended radio galaxies. In fact, by
assuming a constant expansion speed, the relative space densities of
radio sources are expected to be proportional to the range of sizes
covered by each class, contrarily to our findings of a large fraction
of compact sources. This argument relies on the assumption that the
source luminosity does not undergo a temporal evolution. It is well
known that the luminosity of powerful RGs decreases with time (e.g.,
\citealt{maciel14}). However, for the low-power RGs the evolution is
likely to produce an increase in their radio luminosity due to the
development of an extended structure. Thus, the size distribution of
the radio sources in gETGs cannot be ascribed solely to an effect of age,
but, most likely, compact and extended radio sources differ in their
jet properties. \citet{baldi19} suggested that the compact sources are
characterized by a lower jet Lorentz factor, an interpretation
supported by proper motion studies \citep{cheng21}, related to a
lower spin of the central black hole.

We now discuss how these sources compare with the compact FR~0s radio
galaxies selected by \citet{baldi18} and forming the
\FRo\ sample. \FRo\ objects were selected from the sample of radio AGNs
\citep{best12} imposing a redshift $z<0.05$, a flux density $>$5 mJy,
a limit to their sizes of 5 kpc, and requiring an optical spectral
classification as low-excitation galaxies (LEGs). The size and optical
spectral requirements (at least for those with available spectra) are
met by the point-like gETGs. From the point of view of the radio
spectral shape, the compact radio sources in gETGs show a range of
slopes similar to that observed in \FRo\ \citep{capetti20}.

The contribution of compact sources to the overall population of radio
galaxies is similar in the two studies (60\% and 63\%, respectively);
by removing the contamination of sources powered by star forming
regions, the fraction of compact gETGs is reduced to $\sim$48\%. The
resulting slightly lower contribution of compact fraction in gETGs is
probably accounted for, noting that 1) as shown above, compact sources
are more common in less luminous hosts, and 2) the host luminosity of
the \FRo\ sources is lower than those of the gETG, with only half of them
having $M_K < -25$.

The luminosity of the compact radio sources associated with the gETGs
is generally lower than the range covered by FR~0s ($\sim 10^{22} -
10^{24} \WHz$). This is due to the combination of two effects: the
smaller median distance of the gETGs and the lower flux threshold
reached by the LOFAR images. Considering the increasing fraction of
compact radio sources for lower stellar luminosities, we expect that
they represent an even more dominant population in less massive
hosts. These results indicate that the FR~0s population selected from
the FIRST images is just the tip of the iceberg of a much larger
population of compact radio galaxies.

Considering the extended sources, at least 80\% of them have a jetted,
and usually FR~I-class, morphology. This is expected based on their
radio luminosities, which are below separation between the two FR
classes, $\sim 2\times 10^{26} \WHz$ \citep{fanaroff74}. Only a
minority show a complex radio structure in several cases suggestive of
restarted activity.

\subsection{Restarted and remnant sources.}

The analysis of a complete sample of giant early-type galaxies, selected only
on the basis of their stellar luminosity, can be used to explore the
duty cycle of their nuclear activity. In particular, by combining
spectral and morphological information it is possible to look for
remnant and restarted radio sources (RRSs).

From the point of view of the radio morphology, among the 46 extended
sources, there are eight galaxies (NGC~2789, NGC~3665, NGC~3842,
UGC~10097, IC~5180, NGC7436B, UGC~12482, and MCG+05-10-007; see
\citet{brienza21} for a detailed study of this source) in which a
bright, small-scale structure is accompanied by large-scale diffuse
emission. These gETGs can be considered as possible restarted sources
with the structures observed on different scales (and of different
surface brightnesses) being produced by subsequent phases of
activity. In addition, as discussed in Section \ref{spixsec}, there
are seven extended sources with an overall radio spectral index of
$\alpha_{150}^{1400}>1.2$, often showing a diffuse radio emission and
(with the exception of NGC~3842 as discussed) lacking a prominent
central component. These are likely galaxies in which the jets have
switched of,f and the integrated spectrum of the lobes is therefore
undergoing strong radiative losses \citep{pacholczyk70}. By combining
these two groups, the overall fraction of candidate RRSs reaches $\sim
7\%$ of all gETGs and $\sim 28\%$ of those with extended emission.

Remnant and restarted sources has been the subject of several studies,
spurred in particular by the advent of the deep low frequency images
provided by LOFAR. \citet{brienza17} and \citet{jurlin20,jurlin21}
study the properties of 158 extended radio sources in the Lockman Hole
and by combining several indicators (e.g., core prominence,
low-surface brightness of the extended emission, and steep spectrum of
the central region) find a fraction between 13\% and 15\% of candidate
restarted radio galaxies and 7\% of remnant radio
galaxies. \citet{quici21} explored 104 radio sources in the GAMA 23
field with data covering the frequency range 0.1-9 GHz; by adopting a
morphological criterion based on the absence of a radio core, they
found ten candidate remnant sources, three of them also have a steep
spectrum and a diffuse radio morphology. \citet{mahatma18} used the
core-detection method and found 9\% of remnants in the Herschel-ATLAS
field. \citet{morganti21} combined the images at 1.4 GHz produced by
Apertif with those from LOFAR in an area within the Lockman Hole.
Based on the spectral index maps, they found 7\% (3/46) of restarted
sources and 9\% (4/46) remnants. The precise number of RRSs clearly
depends on the defining criterion and on the sample selection; but,
overall, all these works point to an RRS fraction of $\sim$10-20\%.

The fraction of remnant and restarted sources among the gETGs with
extended radio structure appears to be larger, $\sim 30$\%, although
this is somewhat uncertain due to the small number statistics. The presence,
at the low radio luminosity probed in these very nearby sources of
radio emission powered by star formation does not affect this fraction,
because none of the SF candidates is included in the RRSs list. A more
likely possibility is that gETGs are often found in very dense
environments. The high density of the external medium slows the
adiabatic expansion of the radio source, increasing the time over
which the remnant emission can be detected \citep{murgia11}.

Considering instead the gETGs associated with compact radio sources,
we clearly lack any morphological information, and the data on their
radio spectral index is generally very limited. Only one of those
detected at both frequencies, NGC~4555, has a steep spectrum ($\alpha_{150}^{1400}
= 1.24$), suggesting that the observed compact emission might be
originated by compact jets which have switched off and are now in the
process of fading away. For many of them, we are only able to set a
lower limit to their spectral slope but these limits never require a
steep spectrum. Based on these data, we are unable to draw any
conclusion on their duty cycles, which must await deeper high-frequency radio observations. We note, however, that no source among
the 66 compact sources in the \FRo\ sample with available LOFAR
observations has a spectral index larger than 1 \citep{capetti20}. This
suggests that remnants among the compact sources are rare, possibly
due to a shorter duration of this phase of their life cycle. 

\subsection{The fractional radio luminosity function of gETGs.}

The fractional radio luminosity function of gETGs is well described by
a power-law with an index of 0.34$\pm$0.02 in the $3\times10^{21} <
L_{150} < 3\times10^{24} \WHz$ range, and it then steepens at higher
luminosities. This behavior is consistent with the results of
\citet{best05a} (they found an index of 0.35$\pm$0.03) up to a
luminosity of $\sim 2\times10^{23} \WHz,$ while for higher power we
found a substantial excess of sources. Albeit within the limited
statistical significance of our curve at the highest luminosities (only 10
sources have $ L_{150} > 3\times10^{24} \WHz$), it appears that the
transition to a steeper slope occurs at a power higher by a factor of  $\sim$ 10 than that derived by \citeauthor{best05a} This is expected
considering the link between host mass and radio power combined with
our selection of the more luminous ETGs; that is, our sample is biased
in favor of galaxies with a larger radio power.

We also found sources in which the radio emission is likely powered by
star formation. Leaving aside the UGC~07115, the SF source with an
FR~I morphology, five of the SF candidates have $10^{21} < L_{150} <
10^{22} \WHz,$ and only one has$L_{150} \sim 10^{23} \WHz$. Taking
into account the fraction of gETGs for which SDSS spectra are
available (44/188, about one quarter of the whole sample), the
contribution of SF to the luminosity function at low power may be
substantial.

\subsection{Star formation in gETGs}

As described in Section 4, we found seven gETGs whose optical spectral
properties suggest that the main source of gas ionization is due to
the presence of young stars. In at least three cases, the diffuse
morphology of their radio emission supports this interpretation.  The
detection of gas ionized by young stars in two gETGs characterized by
large scale radio emission (namely, UGC~7115 and NGC~2789) indicates
that star formation can co-exist with an active nucleus.

The mass of the six SF candidates (leaving aside the extended FR~I
radio source) are in the $2.9 \times10^{11} - 6.6 \times10^{11}
{\rm M}_\odot$ range. The corresponding values for the sSFR are $0.2 - 10
\times 10^{-12} {\rm yr}^{-1}$ with a median value of $0.8 \times
10^{-12} {\rm yr}^{-1}$. The median values of the upper limits to the
sSFR for the gETGs located in the star forming region in Fig.
\ref{lum2} and for the nine galaxies not detected by LOFAR are
$\lesssim 0.6 \times 10^{-12} {\rm yr}^{-1}$ and $\lesssim 0.4 \times
10^{-12} {\rm yr}^{-1}$, respectively.

\citet{salvador20} found that star formation is ubiquitous in massive
early-type galaxies with a redshift of 0.35 - 0.6, confirming previous
claims from \citet{Vazdekis16}. They selected the above redshift range
(corresponding to a look back time of $\sim$ 4-6 Gyrs) motivated by
the need to include the key UV absorption
index strengths in the optical spectra. The estimate of the fraction of stars formed in the
last 2 Gyr is $\sim$0.5\%, corresponding to a specific star formation
rate (sSFR) of $\sim 2.6 \times 10^{-12} {\rm yr}^{-1}$, only slightly
higher than measured in the candidate SF gETGs at low redshift of our
sample. This implies that star formation is a common process in
elliptical galaxies over a large range of time across their evolution.

\subsection{Comparison with previous studies.}

Several studies, already cited in the introduction, indicate that
massive ETGs are all associated with a radio source
\citep{brown11,sabater19,grossova22}, suggesting that their active
nuclei are always switched on. Our results confirm this observational
evidence because the 25 galaxies in the present sample more luminous
than M$_K = -25.8$ are all detected in the LOFAR images. However, the
additional information derived from their morphology, spectral
indices, and optical spectra that we used for our analysis indicate that
four of them are remnant sources and at least one is likely powered by
star formation, but this fraction could be higher considering the
limited coverage of the optical spectral data of our sample. The
detection of a radio source does not necessarily imply that the
active nucleus is currently active.

A significant difference with respect to the study in
\citet{grossova22} is related to the radio morphology; these authors
found that 67\% of their galaxies are associated with extended radio
sources, a higher fraction than obtained in our sample (24\%). This
discrepancy is reduced by considering the different criterion adopted
for the definition of extended sources: in our analysis, we considered
extended sources to be those in which the 3$\sigma$ radio contour extends
to a radius of at least 15\arcsec, $\sim 6$ kpc at the median distance
of the sample; \citeauthor{grossova22}, thanks to the higher spatial
resolution of their radio images, considered all sources
with a size exceeding twice the beam size as extended, which is typically $\sim
2\arcsec$. To perform an appropriate comparison, we must include in the
list of extended sources in the \citeauthor{grossova22} sample only
those of the ``D'' (diffuse) and ``J/L'' (jet/lobe) classes, excluding
the compact sources (defined as objects with a size smaller than 5
kpc) for a total of 16/42 (38\%) galaxies.

The residual difference in the fraction of extended sources might be
related to the selection criteria of the sample, because
\citeauthor{grossova22} only considered X-ray bright sources. More
specifically, they adopted a threshold of the X-ray luminosity within a
radius of $<$ 10 kpc (but excluding the nuclear source) measured by
\citet{lakhchaura18} in the 0.5 $–-$ 7.0 keV band of 10$^{40}$ erg
s$^{-1}$. \citet{dunn10} found a positive trend between the radio and
X-ray luminosities; the selection of X-ray-bright galaxies biases the
sample toward higher radio power and, consequently, to a higher
fraction of extended sources, considering the link between radio
luminosity and size discussed in Sect. 3.

\section{Summary and conclusions}

We explored the radio properties of the most massive early type
galaxies (gETGs) in the local Universe by selecting 489 sources with
an absolute magnitude of $M_K \le -25$ and a recession velocity of $v \le
7500$ \kms. LOFAR observations at 150 MHz are available for 188 of
them and 146 gETGs are detected above a median threshold of $\sim 1$
mJy. They span a very large range in radio power (from $\sim 10^{21}$
to $\sim 10^{26} \WHz$) and sizes (from unresolved objects,
i.e., smaller than $\sim$4 kpc, to $ \sim 340$ kpc). We confirm the
positive link between the median stellar luminosity of the host and
the radio power. A positive connection of $L_{K}$ with the fraction of
extended galaxies is also found. The RLF of gETGs is in agreement with
previous findings, being described by a power-law over a large range
of $L_{150}$ and with a steepening for the most luminous sources. We
also found that the most luminous gETGs (25 galaxies with M$_K <
-25.8$) are all detected at 150 MHz, but they are not all currently
active: at least four of them are remnant sources, and at least one
is likely powered by star formation.

However, the luminosity function only describes a statistical trend;
while we found, in line with several previous studies, a very large
spread ($\sim$ 5 orders of magnitude) of radio luminosity at any given
galaxy's mass. The analysis of the morphology of sources associated
with the gETGs reveals that a connection between the size of radio
source and its luminosity is present, the less luminous objects being
usually associated with compact sources, with size limits of only a
few kpc. We argued that compact and extended radio sources differ in
their jet properties, the compact ones possibly being produced by
slower jets. When compact and extended sources are considered
separately, the spread in radio power is reduced, but it is still
quite large. Other factors, such as accretion rate, environment, and
the source age, must play a role.

Most sources (62\%) are unresolved with a limit to their extension of
$\sim$ 4 kpc, confirming previous results showing that compact radio sources
represent the dominant population of radio galaxies at low
luminosities. The study of gETGs indicates that the class of FR~0
sources, selected at higher flux densities and higher frequencies,
represents just the tip of the iceberg of a much larger population of
compact sources. The high relative fraction of compact sources cannot
be ascribed to their youth, but it is more likely related to different
jet properties between extended and compact sources; for example, the latter
class being characterized by a lower jet Lorentz factor.

Within the gETGs, there are 46 extended sources, with sizes ranging
from $\sim4$ to $\sim$340 kpc. At least 3/4 of the extended sources
have a morphology clearly indicating that they are powered by jets, and
most of them are edge-darkened FR~I sources, as expected considering
their radio luminosity.

Among the extended sources, we found sources whose radio morphology, a
large-scale diffuse emission accompanied to a bright small-scale
structure, suggests that they are restarted AGNs. In addition, there are
galaxies with a steep radio spectral slope ($\alpha_{150}^{1400} >$
1.2), an indication that these are remnant sources. The fraction of
remnant/restarted sources represents $\sim 30$\% of those with
extended radio emission. This fraction, although somewhat uncertain
due to the small number of statistics, is higher than what was found in
other studies. However, we stress that the criteria used by the
various authors to include an object in the RRS class differ
significantly, and we must be cautious in drawing strong
conclusions. It is possible that the environment plays a
major role by slowing the expansion of the relativistic plasma. In
order to explore the nature of the candidate RRSs among the gETGs in
more detail, dedicated studies at different radio frequencies are
required. In particular, the (general) non-detection at 1.4 GHz
prevents us from measuring their spectral index distribution and deriving information concerning their ages and evolution.

The combination of deep, low-frequency radio observations and optical
spectra enabled us to isolate a subpopulation of gETGs in which the
radio emission is likely powered by star formation. This is suggested
by both the optical emission line ratios and by their diffuse radio
morphology. The sSFR in these sources is quite
low, $\sim 10^{-12} {\rm yr}^{-1}$, similarly to the estimates based on
studies of optical absorption lines of elliptical galaxies at $z\sim
0.35 - 0.6$. The star formation in ETGs detected at higher redshift is
also present in, at least, a subset of the local galaxies at a
comparable level. This implies that the gas flowing toward the center
of these galaxies in part accretes onto the supermassive black hole,
powering the AGN, but it can also stall at larger radii and form new
stars. Clearly, such a level of star formation does not have a
significant impact on the evolution of the host but it is an
indication that AGN feedback does not completely quench star
formation.

\bigskip

M. Brienza acknowledges financial support from the ERC-Stg 674
DRANOEL, no 714245 and the ERC Starting Grant ``MAGCOW", no. 714196.
PNB is grateful for support from the UK STFC, via grant
ST/V000594/1. AD acknowledges support by the BMBF Verbundforschung
under the grant 05A20STA. The J\"ulich LOFAR Long Term Archive and the
German LOFAR network are both coordinated and operated by the J\"ulich
Supercomputing Centre (JSC), and computing resources on the
supercomputer JUWELS at JSC were provided by the Gauss Centre for
supercomputing e.V. (grant CHTB00) through the John von Neumann
Institute for Computing (NIC).  BW acknowledges a studentship from the
UK Science and Technology Facilities Council (STFC).

LOFAR, the Low Frequency Array designed and constructed by ASTRON, has
facilities in several countries, which are owned by various parties
(each with their own funding sources), and are collectively operated
by the International LOFAR Telescope (ILT) foundation under a joint
scientific policy. The ILT resources have benefited from the following
recent major funding sources: CNRS-INSU, Observatoire de Paris and
Universit\'e d'Orl\'eans, France; BMBF, MIWF-NRW, MPG, Germany;
Science Foundation Ireland (SFI), Department of Business, Enterprise
and Innovation (DBEI), Ireland; NWO, The Netherlands; the Science and
Technology Facilities Council, UK; Ministry of Science and Higher
Education, Poland; The Istituto Nazionale di Astrofisica (INAF),
Italy.

Part of this work was carried out on the Dutch national
e-infrastructure with the support of the SURF Cooperative through
grant e-infra 160022 \& 160152. The LOFAR software and dedicated
reduction packages on {\sl https://github.com/apmechev/GRID\_LRT} were
deployed on the e-infrastructure by the LOFAR e-infragrop, consisting
of J.\ B.\ R.\  (ASTRON \& Leiden Observatory), A.\ P.\ Mechev
(Leiden Observatory) and T. Shimwell (ASTRON) with support from
N.\ Danezi (SURFsara) and C.\ Schrijvers (SURFsara). The J\"ulich
LOFAR Long Term Archive and the German LOFAR network are both
coordinated and operated by the J\"ulich Supercomputing Centre (JSC),
and computing resources on the supercomputer JUWELS at JSC were
provided by the Gauss Centre for supercomputing e.V. (grant CHTB00)
through the John von Neumann Institute for Computing (NIC).

This research made use of the University of Hertfordshire
high-performance computing facility and the LOFAR-UK computing
facility located at the University of Hertfordshire and supported by
STFC (ST/P000096/1), and of the Italian LOFAR IT computing
infrastructure supported and operated by INAF, and by the Physics
Department of Turin University (under an agreement with Consorzio
Interuniversitario per la Fisica Spaziale) at the C3S Supercomputing
Centre, Italy.

The Pan-STARRS1 Surveys (PS1) and the PS1 public science archive have
been made possible through contributions by the Institute for
Astronomy, the University of Hawaii, the Pan-STARRS Project Office,
the Max-Planck Society and its participating institutes, the Max
Planck Institute for Astronomy, Heidelberg and the Max Planck
Institute for Extraterrestrial Physics, Garching, The Johns Hopkins
University, Durham University, the University of Edinburgh, the
Queen's University Belfast, the Harvard-Smithsonian Center for
Astrophysics, the Las Cumbres Observatory Global Telescope Network
Incorporated, the National Central University of Taiwan, the Space
Telescope Science Institute, the National Aeronautics and Space
Administration under Grant No. NNX08AR22G issued through the Planetary
Science Division of the NASA Science Mission Directorate, the National
Science Foundation Grant No. AST-1238877, the University of Maryland,
Eotvos Lorand University (ELTE), the Los Alamos National Laboratory,
and the Gordon and Betty Moore Foundation.

\bibliographystyle{./aa}

\begin{appendix}
\clearpage
\newpage

\section{Properties of the sample}

\onecolumn
\begin{landscape}
 \begin{longtable}{l r r c c c r r l r r r r r}

\caption{Properties of the sample}
\label{tab} \\                                   

\hline 
Name & R.A. & DEC. & v & M$_K$ & Image & rms  & F$_{150}$ & Morph. & R$_{\rm maj}$ & L.A.S. & L$_{150}$ & F$_{1.4}$ & $\alpha_{150}^{1400}$   \\ 
\hline  
\endfirsthead

\multicolumn{3}{c}{{\tablename} \thetable{} -- Continued} \\                                   [0.5ex]
\hline 
Name & R.A. & DEC. & v & M$_K$ & Image & rms & F$_{150}$ & Morph. &
R$_{\rm maj}$ & L.A.S.  & L$_{150}$ & F$_{1.4}$ & $\alpha$ \\
\hline
\endhead

%This is the footer for all pages except the last page of the table...
\hline
  \multicolumn{10}{c}{{Continued on Next Page}} \\                                   
\endfoot

%This is the footer for the last page of the table...
  \\                                   [-1.8ex] 
\endlastfoot

                  NGC0057 &     3.8787 &    17.3284 & 5339 & -25.81 & DR2 &  0.16 &     10.5  &  P    & 12.7 $\pm$ 0.8   &          &  21.88     &     ---   &   $>$  0.74\\
                  NGC0068 &     4.5771 &    30.0718 & 5647 & -25.01 & DR2 &  0.10 &      2.2  &  P    & 10.3 $\pm$ 2.1   &          &  21.25     &     ---   &   $>$  0.04\\
                  NGC0071 &     4.5982 &    30.0632 & 6588 & -25.40 & DR2 &  0.09 &     53.0  &Compl. & 14.4 $\pm$ 1.0   &  90 ( 40)&  23.10     &     $^a$  &            \\
                  NGC0076 &     4.9075 &    29.9339 & 7205 & -25.07 & DR2 &  0.08 &     51.1  &  P    &  8.1 $\pm$ 0.1   &          &  22.83     &  104.0$^N$&       -0.32\\
                  NGC0080 &     5.2952 &    22.3572 & 5592 & -25.66 & DR2 &  0.09 &    101.0  &Compl. &  9.7 $\pm$ 1.3   & 280 (106)&  22.91     & $<$4.6$^N$&   $>$  1.38\\
                  NGC0083 &     5.3433 &    22.4336 & 6109 & -25.49 & DR2 &  0.10 &     10.9  &  P    &  8.3 $\pm$ 0.2   &          &  22.02     &     ---   &   $>$  0.76\\
                  NGC0128 &     7.3127 &     2.8640 & 4153 & -25.41 &   S &  1.41 &   $<$7.1  &       &                  &          & $<$  21.50 &  $<$0.5   &            \\
                  NGC0194 &     9.8268 &     3.0375 & 5105 & -25.17 &   S &  0.39 &   $<$2.0  &       &                  &          & $<$  21.12 &  $<$0.6   &            \\
                  NGC0315 &    14.4538 &    30.3524 & 4844 & -26.32 & DR2 &  0.14 &   5880.0  & FRI   & 74.0 $\pm$ 0.1   & 790 (260)&  24.55     & 1590.0$^N$&        0.59\\
                  NGC0379 &    16.8154 &    32.5204 & 5464 & -25.15 & DR2 &  0.14 &     11.6  &  P    &  7.6 $\pm$ 0.3   &          &  21.95     &     ---   &   $>$  0.79\\
                  NGC0383 &    16.8540 &    32.4126 & 4996 & -25.86 & DR2 &  0.16 &  19000.0  & FRI   &120.1 $\pm$ 1.9   & 630 (215)&  25.08     & 4430.0$^N$&        0.65\\
                  NGC0393 &    17.1540 &    39.6443 & 6000 & -25.50 & DR2 &  0.09 &      6.3  &  P    &  7.2 $\pm$ 0.1   &          &  21.76     &     ---   &   $>$  0.51\\
                  NGC0410 &    17.7453 &    33.1520 & 5187 & -26.04 & DR2 &  0.10 &     48.1  &  P    & 11.1 $\pm$ 0.1   &          &  22.52     &    5.8$^N$&        0.95\\
                  NGC0467 &    19.7922 &     3.3008 & 5308 & -25.45 &   S &  0.36 &   $<$1.8  &       &                  &          & $<$  21.11 &  $<$0.6   &            \\
                  NGC0499 &    20.7978 &    33.4601 & 4312 & -25.29 & DR2 &  0.20 &      7.3  &  P    & 11.2 $\pm$ 0.9   &          &  21.54     &     ---   &   $>$  0.58\\
                  NGC0507 &    20.9164 &    33.2561 & 4831 & -25.97 & DR2 &  0.21 &   1680.0  & FRI   & 52.5 $\pm$ 0.1   & 150 ( 49)&  24.00     &   98.8$^N$&        1.27\\
                  NGC0524 &    21.1988 &     9.5388 & 2403 & -25.60 &   S &  0.29 &     13.2  &Diff.  & 15.8 $\pm$ 2.3   &  50 (  8)&  21.29     &    3.5$^N$&        0.59\\
                  NGC0533 &    21.3808 &     1.7590 & 5385 & -26.07 &   S &  0.29 &    106.0  &  P    & 11.3 $\pm$ 0.2   &          &  22.90     &   20.5    &        0.73\\
                  NGC0529 &    21.4179 &    34.7130 & 4726 & -25.11 & DR2 &  0.14 &      1.1  &  P    &  7.3 $\pm$ 1.3   &          &  20.81     &     ---   &   $>$ -0.25\\
                  NGC0679 &    27.4324 &    35.7856 & 4942 & -25.18 & DR2 &  0.11 &      2.8  &  P    &  7.6 $\pm$ 0.5   &          &  21.24     &     ---   &   $>$  0.15\\
                  NGC0687 &    27.6385 &    36.3708 & 4984 & -25.10 & DR2 &  0.08 &     30.9  &Diff.  & 49.9 $\pm$ 3.6   &  40 ( 13)&  22.29     &     ---   &   $>$  1.23\\
                 UGC01308 &    27.7134 &    36.2758 & 5062 & -25.13 & DR2 &  0.08 &    102.0  & FRI   &  6.7 $\pm$ 0.1   & 180 ( 62)&  22.83     &   20.2$^N$&        0.72\\
                 UGC01332 &    28.0755 &    48.0878 & 6687 & -25.55 &   S &  0.09 &      8.6  &  P    & 10.7 $\pm$ 0.1   &          &  22.00     &     ---   &   $>$  0.66\\
                  NGC0708 &    28.1936 &    36.1518 & 4754 & -25.67 & DR2 &  0.11 &    662.0  & FRI   & 38.9 $\pm$ 0.2   & 188 ( 61)&  23.58     &   65.7$^N$&        1.03\\
                  NGC0712 &    28.2852 &    36.8199 & 5218 & -25.02 & DR2 &  0.09 &      3.5  &  P    &  9.1 $\pm$ 1.9   &          &  21.39     &     ---   &   $>$  0.25\\
                   IC0171 &    28.7925 &    35.2819 & 5240 & -25.56 & DR2 &  0.08 &      1.2  &  P    &  8.6 $\pm$ 1.3   &          &  20.94     &     ---   &   $>$ -0.22\\
                 UGC01389 &    28.8778 &    47.9550 & 7026 & -25.50 &   S &  0.09 &      1.9  &  P    &  8.7 $\pm$ 1.3   &          &  21.38     &     ---   &   $>$ -0.02\\
                 UGC01418 &    29.2375 &    40.3414 & 5307 & -25.10 & DR2 &  0.08 &     13.8  &  P    & 10.3 $\pm$ 0.2   &          &  22.00     &     ---   &   $>$  0.86\\
                  NGC0750 &    29.3864 &    33.2093 & 5049 & -25.80 & DR2 &  0.08 &      7.6  &  P*   & 37.6 $\pm$ 5.7   &          &  21.70     &     ---   &   $>$  0.60\\
                  NGC0759 &    29.4597 &    36.3431 & 4570 & -25.02 & DR2 &  0.09 &     65.6  &  P    &  7.8 $\pm$ 0.0   &          &  22.55     &   15.1$^N$&        0.66\\
                  NGC0777 &    30.0622 &    31.4294 & 4891 & -25.92 & DR2 &  0.11 &     47.3  & FRI   &  9.8 $\pm$ 0.1   &  54 ( 18)&  22.46     &    7.0$^N$&        0.85\\
                  NGC0890 &    35.5042 &    33.2661 & 3889 & -25.56 & DR2 &  0.08 &   $<$0.4  &       &                  &          & $<$  20.16 &     ---   &            \\
                 UGC01841 &    35.8041 &    42.9878 & 6476 & -25.65 & DR2 &  0.46 &  39000.0  & FRI   &216.4 $\pm$ 0.1   & 690 (304)&  25.62     & 5960.0$^N$&        0.84\\
                  NGC0910 &    36.3616 &    41.8243 & 5114 & -25.19 & DR2 &  0.10 &     83.2  &Compl. &  7.8 $\pm$ 0.3   & 180 ( 63)&  22.75     & $<$5.7$^N$&   $>$  1.20\\
                 UGC01930 &    37.1251 &    51.4464 & 4877 & -25.01 &   S &  0.28 &   $<$1.4  &       &                  &          & $<$  20.94 &     ---   &            \\
                  NGC0940 &    37.3646 &    31.6409 & 5667 & -25.27 & DR2 &  0.09 &     14.4  &  P    & 12.4 $\pm$ 0.4   &          &  22.08     &     ---   &   $>$  0.88\\
                 NGC0978A &    38.6957 &    32.8462 & 4631 & -25.08 & DR2 &  0.09 &      3.1  &  P    &  8.5 $\pm$ 0.6   &          &  21.23     &     ---   &   $>$  0.20\\
                  NGC1004 &    39.4242 &     1.9753 & 6271 & -25.05 &   S &  0.25 &   1090.0  & FRI   & 15.2 $\pm$ 0.1   & 480 (205)&  24.04     &  200.0$^N$&        0.76\\
                  NGC1016 &    39.5815 &     2.1193 & 6453 & -26.31 &   S &  0.25 &     29.5  &  P    & 10.0 $\pm$ 0.2   &          &  22.50     &    3.6    &        0.94\\
                  NGC1044 &    40.2757 &     8.7380 & 6157 & -25.12 &   S &  0.31 &   4330.0  & FRI   & 74.4 $\pm$ 0.1   & 600 (247)&  24.62     & 1300.0$^N$&        0.54\\
                  NGC1060 &    40.8127 &    32.4250 & 5061 & -26.22 & DR2 &  0.13 &     39.9  & FRI   & 19.1 $\pm$ 0.2   &  45 ( 15)&  22.42     &    9.2$^N$&        0.66\\
                  NGC1066 &    40.9579 &    32.4749 & 4242 & -25.15 & DR2 &  0.12 &   $<$0.6  &       &                  &          & $<$  20.46 &     ---   &            \\
                 UGC02249 &    41.8490 &    45.5316 & 7174 & -25.15 &   S &  0.20 &   $<$1.0  &       &                  &          & $<$  21.12 &     ---   &            \\
                 UGC02309 &    42.4148 &    21.2079 & 6083 & -25.08 &   S &  0.27 &   $<$1.3  &       &                  &          & $<$  21.10 &     ---   &            \\
                  NGC1129 &    43.6141 &    41.5796 & 5102 & -26.17 &   S &  0.09 &     10.4  &  P    &  7.3 $\pm$ 0.1   &          &  21.84     &     ---   &   $>$  0.74\\
                  NGC1167 &    45.4265 &    35.2056 & 4835 & -25.67 &   S &  0.15 &   4890.0  &  P    &  6.5 $\pm$ 0.1   &          &  24.47     & 1840.0$^N$&        0.44\\
                  NGC1175 &    46.1348 &    42.3393 & 5428 & -25.30 &   S &  0.17 &      0.6  &  P    &  4.8 $\pm$ 3.7   &          &  20.68     &     ---   &   $>$ -0.52\\
                 UGC02528 &    46.4796 &    41.5920 & 6180 & -25.16 &   S &  0.25 &   $<$1.3  &       &                  &          & $<$  21.09 &     ---   &            \\
                  NGC1226 &    47.7723 &    35.3868 & 6012 & -25.58 &   S &  0.19 &      6.7  &  P    &  8.0 $\pm$ 0.3   &          &  21.79     &     ---   &   $>$  0.54\\
              WISE0322+09 &    50.5608 &     9.5682 & 6948 & -25.07 &   S &  0.18 &     31.1  &  P    &  9.6 $\pm$ 0.1   &          &  22.59     &   35.0$^N$&       -0.05\\
              CGCG416-007 &    50.8640 &     8.9124 & 6909 & -25.05 &   S &  0.16 &   $<$0.8  &       &                  &          & $<$  20.99 &     ---   &            \\
              WISE0348+33 &    57.0032 &    33.1116 & 4146 & -25.22 &   S &  0.25 &   $<$1.2  &       &                  &          & $<$  20.74 &     ---   &            \\
                 UGC02902 &    59.2879 &    34.1609 & 5454 & -25.28 &   S &  0.17 &      1.8  &  P    & 11.2 $\pm$ 0.9   &          &  21.13     &     ---   &   $>$ -0.05\\
                  NGC1508 &    61.4486 &    25.4084 & 7001 & -25.36 &   S &  0.98 &     14.1  &  P*   & 17.3 $\pm$ 0.6   &          &  22.25     &    2.4$^N$&        0.79\\
            MCG+05-10-007 &    61.6572 &    30.3762 & 5273 & -25.56 &   S &  0.18 &    814.0  & FRI   & 77.5 $\pm$ 0.1   & 185 ( 64)&  23.76     &  108.0$^N$&        0.91\\
                   IC0359 &    63.1181 &    27.7019 & 3978 & -25.62 &   S &  0.37 &     10.3  &  P    & 10.6 $\pm$ 0.1   &          &  21.62     &    3.4$^N$&        0.50\\
                  NGC1539 &    64.7582 &    26.8275 & 5442 & -25.10 &   S &  0.22 &     15.8  &  P    &  6.6 $\pm$ 0.1   &          &  22.08     &    2.4$^N$&        0.85\\
                  NGC1587 &    67.6664 &     0.6617 & 3613 & -25.13 &   S &  0.55 &    376.0  &  P    &  7.3 $\pm$ 0.3   &          &  23.10     &  132.0$^N$&        0.47\\
                  NGC1671 &    72.3918 &     0.2528 & 6204 & -25.35 &   S &  0.30 &      5.1  &  P    &                  &          &  21.71     &     ---   &   $>$  0.42\\
                  NGC2256 &   101.8082 &    74.2365 & 5418 & -25.87 &   S &  0.10 &     15.4  &  P    &  9.3 $\pm$ 0.1   &          &  22.07     &     ---   &   $>$  0.92\\
                  NGC2274 &   101.8224 &    33.5672 & 5153 & -25.74 &   S &  0.17 &      6.1  &  P    & 10.8 $\pm$ 0.4   &          &  21.62     &     ---   &   $>$  0.50\\
                  NGC2258 &   101.9425 &    74.4818 & 4257 & -25.79 &   S &  0.10 &     31.9  &  P    &  7.0 $\pm$ 0.1   &          &  22.17     &    9.7$^N$&        0.53\\
                 UGC03894 &   113.2695 &    65.0791 & 7019 & -25.71 &   S &  0.12 &      8.5  &  P    &  7.3 $\pm$ 0.5   &          &  22.03     &    5.2$^N$&        0.22\\
                   IC2196 &   113.5406 &    31.4058 & 4897 & -25.05 & DR2 &  0.14 &      4.1  &  P    &  8.9 $\pm$ 0.7   &          &  21.40     &  $<$0.7   &   $>$  0.79\\
              CGCG286-070 &   117.5346 &    55.3841 & 6598 & -25.15 & DR2 &  0.13 &    687.0  & FRI?  & 16.0 $\pm$ 0.1   &  35 ( 15)&  23.88     &  207.0    &        0.54\\
                 UGC04051 &   117.8235 &    50.1794 & 6427 & -25.06 &   S &  0.10 &      3.7  &  P    &  8.0 $\pm$ 0.9   &          &  21.59     &  $<$0.7   &   $>$  0.74\\
                 UGC04052 &   117.8281 &    50.2355 & 6897 & -25.45 &   S &  0.10 &   $<$0.5  &       &                  &          & $<$  20.78 &  $<$0.7   &            \\
              CGCG310-022 &   118.5444 &    55.4953 & 7483 & -25.38 & DR2 &  0.07 &      1.3  &  P    &  8.1 $\pm$ 0.8   &          &  21.28     &  $<$0.7   &   $>$  0.26\\
                  NGC2475 &   119.5018 &    52.8618 & 5764 & -25.29 & DR2 &  0.19 &     21.1  &  P    &  7.1 $\pm$ 0.0   &          &  22.25     &    5.9    &        0.57\\
                  NGC2492 &   119.8738 &    27.0264 & 6947 & -25.45 & DR2 &  0.08 &      2.0  &  P    &  7.3 $\pm$ 0.5   &          &  21.39     &  $<$0.7   &   $>$  0.47\\
                  NGC2493 &   120.0986 &    39.8304 & 4124 & -25.09 & DR2 &  0.09 &      7.6  &  P*   & 21.7 $\pm$ 2.8   &          &  21.52     &  $<$0.7   &   $>$  1.05\\
                 UGC04420 &   127.2956 &    63.3379 & 7107 & -25.23 & DR2 &  0.05 &      1.2  &  P    &  9.1 $\pm$ 1.0   &          &  21.20     &  $<$1.2   &   $>$  0.03\\
                  NGC2639 &   130.9087 &    50.2055 & 3619 & -25.23 & DR2 &  0.07 &    495.0  &  P    &  8.6 $\pm$ 0.0   &          &  23.22     &  107.0    &        0.68\\
                   IC2393 &   131.7049 &    28.1713 & 6567 & -25.01 & DR2 &  0.10 &   $<$0.5  &       &                  &          & $<$  20.76 &  $<$0.7   &            \\
                  NGC2672 &   132.3412 &    19.0750 & 4628 & -25.81 &   S &  0.11 &    134.0  &Diff.  & 48.0 $\pm$ 5.2   & 105 ( 30)&  22.87     &    4.6$^N$&        1.51\\
                  NGC2693 &   134.2469 &    51.3474 & 5205 & -25.81 & DR2 &  0.06 &     54.1  & FRI   & 14.7 $\pm$ 0.1   & 165 ( 57)&  22.58     &    5.9    &        0.99\\
                  NGC2749 &   136.3389 &    18.3131 & 4503 & -25.18 &   S &  0.18 &    214.0  &  P    &  6.9 $\pm$ 0.1   &          &  23.05     &   63.2    &        0.55\\
                 UGC04767 &   136.4386 &    36.3546 & 7491 & -25.21 & DR2 &  0.08 &     20.6  & FRI   & 14.1 $\pm$ 0.3   &  38 ( 19)&  22.47     &    4.6    &        0.67\\
                 UGC04775 &   136.9115 &    66.5747 & 7134 & -25.24 & DR2 &  0.06 &   $<$0.3  &       &                  &          & $<$  20.61 &     ---   &            \\
                  NGC2759 &   137.1553 &    37.6216 & 7209 & -25.33 & DR2 &  0.09 &      1.6  &  P    &  9.7 $\pm$ 1.3   &          &  21.32     &  $<$1.7   &   $>$ -0.03\\
                  NGC2783 &   138.4145 &    29.9928 & 7037 & -25.75 & DR2 &  0.10 &     47.5  & E ?   & 12.4 $\pm$ 0.1   &  32 ( 15)&  22.78     &   25.6    &        0.28\\
                  NGC2789 &   138.7485 &    29.7303 & 6639 & -25.10 & DR2 &  0.09 &     91.3  & E ?   & 14.4 $\pm$ 0.1   &  43 ( 18)&  23.01     &   15.2    &        0.80\\
                  NGC2832 &   139.9453 &    33.7498 & 7237 & -26.43 & DR2 &  0.09 &     15.3  &FRI?   & 26.1 $\pm$ 1.1   &  45 ( 22)&  22.31     &  $<$0.7   &   $>$  1.37\\
                 UGC04972 &   140.4645 &    33.4019 & 7366 & -25.06 & DR2 &  0.10 &    425.0  & FRI   & 32.5 $\pm$ 0.1   & 140 ( 70)&  23.77     &  132.0    &        0.52\\
                 UGC04974 &   140.5431 &    33.8486 & 7314 & -25.05 & DR2 &  0.07 &   $<$0.4  &       &                  &          & $<$  20.71 &  $<$0.7   &            \\
                  NGC2892 &   143.2205 &    67.6174 & 7077 & -25.75 & DR2 &  0.07 &    801.0  & FRI   & 53.9 $\pm$ 0.1   & 105 ( 50)&  24.01     &  229.0$^N$&        0.56\\
                  NGC2918 &   143.9334 &    31.7054 & 7158 & -25.54 & DR2 &  0.07 &      0.7  &  P    &  8.0 $\pm$ 1.3   &          &  20.96     &  $<$0.6   &   $>$  0.03\\
                  NGC3158 &   153.4604 &    38.7649 & 7346 & -26.37 & DR2 &  0.09 &     27.5  & FRI   & 27.1 $\pm$ 1.1   &  63 ( 31)&  22.58     &    3.1$^N$&        0.98\\
                  NGC3222 &   155.6437 &    19.8871 & 6017 & -25.11 &   S &  0.10 &   $<$0.5  &       &                  &          & $<$  20.68 &  $<$0.7   &            \\
                  NGC3234 &   156.2472 &    28.0239 & 6827 & -25.00 & DR2 &  0.14 &     28.2  &  P    &  8.5 $\pm$ 0.1   &          &  22.53     &   13.8    &        0.32\\
                  NGC3598 &   168.7986 &    17.2627 & 6658 & -25.39 &   S &  0.25 &   $<$1.2  &       &                  &          & $<$  21.15 &  $<$0.7   &            \\
                  NGC3615 &   169.5277 &    23.3973 & 7178 & -25.66 &   S &  0.13 &      2.8  &  P    & 10.2 $\pm$ 1.2   &          &  21.56     &    2.1    &        0.13\\
                  NGC3665 &   171.1818 &    38.7628 & 2566 & -25.20 & DR2 &  0.07 &    512.0  & FRI   & 38.6 $\pm$ 0.1   & 315 ( 55)&  22.94     &  131.0$^N$&        0.61\\
                  NGC3710 &   172.7789 &    22.7680 & 7005 & -25.03 &   S &  0.57 &   $<$2.8  &       &                  &          & $<$  21.55 &  $<$0.7   &            \\
                  NGC3713 &   172.9251 &    28.1536 & 7483 & -25.25 & DR2 &  0.20 &   $<$1.0  &       &                  &          & $<$  21.15 &  $<$0.7   &            \\
                  NGC3805 &   175.1736 &    20.3430 & 7163 & -25.81 &   S &  0.27 &      7.8  &  P*   & 20.4 $\pm$ 0.6   &          &  22.01     &    3.6    &        0.35\\
                  NGC3816 &   175.4502 &    20.1036 & 6306 & -25.24 &   S &  0.28 &   $<$1.4  &       &                  &          & $<$  21.16 &  $<$0.7   &            \\
                  NGC3837 &   175.9851 &    19.8946 & 6672 & -25.00 &   S &  0.20 &      2.6  &  P    & 10.6 $\pm$ 0.3   &          &  21.48     &  $<$1.2   &   $>$  0.37\\
                  NGC3842 &   176.0090 &    19.9498 & 6857 & -25.93 &   S &  0.26 &   1030.0  & E ?   & 50.9 $\pm$ 0.1   & 340 (158)&  24.09     &   13.1    &        1.95\\
                  NGC3862 &   176.2708 &    19.6063 & 7053 & -25.59 &   S &  0.83 &  39600.0  & FRI   & 64.2 $\pm$ 0.2   & 460 (221)&  25.70     & 5320.0$^N$&        0.90\\
                  NGC3886 &   176.7733 &    19.8372 & 6442 & -25.16 &   S &  0.16 &   $<$0.8  &       &                  &          & $<$  20.94 &  $<$0.8   &            \\
                  NGC3894 &   177.2099 &    59.4156 & 3653 & -25.10 & DR2 &  0.06 &    319.0  & FRI   & 10.5 $\pm$ 0.0   &  67 ( 16)&  23.04     &  472.0    &       -0.18\\
                  NGC3919 &   177.6730 &    20.0151 & 6744 & -25.19 &   S &  0.13 &      2.5  &  P*   & 20.7 $\pm$ 0.8   &          &  21.46     &    1.4    &        0.26\\
                 UGC06846 &   178.1530 &    23.5825 & 7377 & -25.06 &   S &  0.15 &   $<$0.7  &       &                  &          & $<$  21.01 &  $<$0.7   &            \\
                  NGC3937 &   178.1776 &    20.6313 & 7214 & -25.71 &   S &  0.16 &      3.8  &  P    &  9.9 $\pm$ 1.2   &          &  21.71     &    4.2    &       -0.04\\
                  NGC3940 &   178.1935 &    20.9893 & 6967 & -25.30 &   S &  0.15 &   $<$0.8  &       &                  &          & $<$  20.97 &  $<$0.7   &            \\
                  NGC3971 &   178.9017 &    29.9959 & 7258 & -25.10 & DR2 &  0.08 &      1.3  &  P    &  9.2 $\pm$ 1.5   &          &  21.26     &  $<$0.7   &   $>$  0.31\\
                  NGC4008 &   179.5710 &    28.1925 & 4151 & -25.10 & DR2 &  0.12 &     38.2  &  P    &  8.8 $\pm$ 0.1   &          &  22.23     &    8.8    &        0.66\\
                  NGC4065 &   181.0257 &    20.2351 & 6888 & -25.34 &   S &  0.76 &     21.1  &  P    & 10.1 $\pm$ 3.4   &          &  22.41     &    5.1    &        0.63\\
                 UGC07115 &   182.0232 &    25.2373 & 7323 & -25.05 &   S &  0.14 &   2570.0  & FRI   & 27.6 $\pm$ 0.1   & 670 (335)&  24.55     &  629.0$^N$&        0.63\\
                  NGC4125 &   182.0251 &    65.1741 & 7323 & -25.28 & DR2 &  0.05 &      9.2  &  P    & 10.5 $\pm$ 0.4   &          &  22.10     &     ---   &   $>$  0.68\\
                 UGC07132 &   182.2909 &    31.5695 & 7318 & -25.17 &   S &  0.11 &      4.1  &  P    &  8.9 $\pm$ 0.8   &          &  21.75     &    2.6    &        0.21\\
                  NGC4213 &   183.9064 &    23.9819 & 7299 & -25.55 &   S &  0.17 &      4.2  &  P    &  9.3 $\pm$ 0.7   &          &  21.76     &  $<$0.7   &   $>$  0.79\\
                  NGC4227 &   184.1404 &    33.5220 & 6975 & -25.54 & DR2 &  0.08 &      2.1  &  P    & 10.5 $\pm$ 1.2   &          &  21.41     &  $<$0.7   &   $>$  0.49\\
                  NGC4229 &   184.1616 &    33.5609 & 7226 & -25.01 & DR2 &  0.09 &      0.4  &  P    &  8.1 $\pm$ 1.5   &          &  20.77     &  $<$0.7   &   $>$ -0.22\\
                   IC0780 &   184.9932 &    25.7717 & 7310 & -25.07 &   S &  0.16 &     33.6  & FRI   & 16.0 $\pm$ 0.4   &  38 ( 19)&  22.66     &   12.8    &        0.43\\
                  NGC4335 &   185.7579 &    58.4446 & 5049 & -25.26 & DR2 &  0.07 &    645.0  & FRI   & 63.5 $\pm$ 0.1   & 530 (182)&  23.63     & 1630.0$^N$&       -0.41\\
                  NGC4555 &   188.9216 &    26.5230 & 7251 & -25.97 &   S &  0.10 &     31.3  &  P    &  8.7 $\pm$ 0.3   &          &  22.63     &    2.0    &        1.24\\
                  NGC4583 &   189.5189 &    33.4589 & 7458 & -25.22 & DR2 &  0.06 &      2.6  &  P    &  8.0 $\pm$ 0.3   &          &  21.57     &    0.9    &        0.48\\
                      M60 &   190.9167 &    11.5526 & 1013 & -25.13 &   S &  0.27 &     99.3  & FRI   &  9.3 $\pm$ 0.9   &  65 (  4)&  21.42     &   29.9$^N$&        0.54\\
                  NGC4673 &   191.3945 &    27.0608 & 7424 & -25.32 & DR2 &  0.11 &      3.4  &  P    &  8.0 $\pm$ 0.5   &          &  21.68     &  $<$0.7   &   $>$  0.72\\
                  NGC4715 &   192.4911 &    27.8223 & 7463 & -25.13 & DR2 &  0.12 &   $<$0.6  &       &                  &          & $<$  20.93 &  $<$0.7   &            \\
                  NGC4816 &   194.0506 &    27.7455 & 7491 & -25.49 & DR2 &  0.09 &      0.4  &  P    &  7.0 $\pm$ 0.7   &          &  20.80     &  $<$0.8   &   $>$ -0.25\\
                 NGC4841A &   194.3832 &    28.4769 & 7350 & -25.94 & DR2 &  0.09 &      2.0  &  P    & 12.4 $\pm$ 1.9   &          &  21.45     &  $<$0.7   &   $>$  0.46\\
                  NGC4886 &   195.0338 &    27.9770 & 7076 & -26.67 & DR2 &  0.08 &      5.7  &  P    &  8.8 $\pm$ 0.2   &          &  21.87     &  $<$0.8   &   $>$  0.87\\
                  NGC4914 &   195.1789 &    37.3153 & 5216 & -25.77 & DR2 &  0.07 &   $<$0.3  &       &                  &          & $<$  20.37 &  $<$0.7   &            \\
                  NGC4952 &   196.2433 &    29.1223 & 6551 & -25.16 & DR2 &  0.07 &      0.7  &  P    &  6.7 $\pm$ 0.9   &          &  20.86     &  $<$0.5   &   $>$  0.13\\
                  NGC4957 &   196.3015 &    27.5698 & 7492 & -25.22 & DR2 &  0.09 &   $<$0.4  &       &                  &          & $<$  20.80 &  $<$0.7   &            \\
                  NGC4978 &   196.9606 &    18.4155 & 7091 & -25.13 &   S &  0.16 &      2.2  &  P    &  6.9 $\pm$ 1.6   &          &  21.46     &    1.3    &        0.24\\
                   IC0858 &   198.7164 &    17.2267 & 7494 & -25.29 &   S &  0.24 &   $<$1.2  &       &                  &          & $<$  21.24 &  $<$0.7   &            \\
                   IC0885 &   200.6287 &    21.3164 & 7461 & -25.13 &   S &  0.52 &   $<$2.6  &       &                  &          & $<$  21.57 &  $<$0.7   &            \\
                  NGC5127 &   200.9375 &    31.5658 & 5450 & -25.15 & DR2 &  0.18 &   8200.0  & FRI   & 39.1 $\pm$ 5.1   & 660 (245)&  24.80     & 1980.0$^N$&        0.64\\
                  NGC5141 &   201.2144 &    36.3785 & 5777 & -25.00 & DR2 &  0.08 &   3920.0  & FRII  & 34.5 $\pm$ 0.1   &  67 ( 26)&  24.53     &  822.0    &        0.70\\
                  NGC5322 &   207.3133 &    60.1904 & 2285 & -25.47 & DR2 &  0.07 &    312.0  & FRI   & 34.2 $\pm$ 0.1   & 170 ( 26)&  22.62     &   57.6    &        0.76\\
                  NGC5332 &   208.0331 &    16.9698 & 7365 & -25.33 &   S &  0.15 &      1.7  &  P    &  7.2 $\pm$ 3.9   &          &  21.37     &  $<$0.7   &   $>$  0.38\\
                  NGC5444 &   210.8506 &    35.1321 & 4536 & -25.28 & DR2 &  0.12 &   $<$0.6  &       &                  &          & $<$  20.49 &  $<$0.7   &            \\
                  NGC5490 &   212.4888 &    17.5455 & 5501 & -25.62 &   S &  0.46 &   3270.0  & FRI   & 24.9 $\pm$ 0.1   & 430 (161)&  24.40     &  786.0$^N$&        0.64\\
                  NGC5513 &   213.2860 &    20.4163 & 5633 & -25.21 &   S &  0.38 &   $<$1.9  &       &                  &          & $<$  21.19 &  $<$0.7   &            \\
                  NGC5525 &   213.9135 &    14.2826 & 6213 & -25.36 &   S &  0.16 &      2.7  &  P*   & 16.3 $\pm$ 1.4   &          &  21.43     &  $<$0.7   &   $>$  0.58\\
                  NGC5557 &   214.6071 &    36.4936 & 3802 & -25.65 & DR2 &  0.09 &   $<$0.4  &       &                  &          & $<$  20.20 &  $<$0.7   &            \\
                  NGC5784 &   223.5686 &    42.5578 & 5917 & -25.42 & DR2 &  0.06 &     23.8  &  P    & 12.9 $\pm$ 0.1   &          &  22.33     &    4.5    &        0.75\\
                   IC4562 &   233.9875 &    43.4932 & 6188 & -25.26 & DR2 &  0.06 &      5.8  &  P    &  9.2 $\pm$ 0.2   &          &  21.76     &    2.8    &        0.32\\
                  NGC5982 &   234.6658 &    59.3559 & 3494 & -25.40 & DR2 &  0.06 &      4.6  &  P    & 10.6 $\pm$ 0.5   &          &  21.16     &  $<$0.7   &   $>$  0.84\\
                 UGC10097 &   238.9303 &    47.8673 & 6453 & -25.50 & DR2 &  0.06 &    120.0  & FRI   & 26.4 $\pm$ 0.1   &  64 ( 56)&  23.11     &   41.2    &        0.48\\
                   IC1153 &   239.2625 &    48.1684 & 6408 & -25.19 & DR2 &  0.06 &      5.1  &  P    &  8.1 $\pm$ 0.2   &          &  21.73     &    2.3    &        0.36\\
                   IC1211 &   244.2164 &    53.0060 & 6073 & -25.07 & DR2 &  0.11 &   $<$0.6  &       &                  &          & $<$  20.72 &  $<$0.7   &            \\
                  NGC6125 &   244.7981 &    57.9841 & 5268 & -25.31 & DR2 &  0.06 &      2.6  &  P    &  8.0 $\pm$ 0.3   &          &  21.27     &  $<$0.7   &   $>$  0.58\\
                  NGC6364 &   261.1139 &    29.3902 & 7371 & -25.44 & DR2 &  0.11 &      2.0  &  P    &  7.2 $\pm$ 0.5   &          &  21.45     &     ---   &   $>$  0.01\\
                  NGC6375 &   262.3411 &    16.2068 & 6704 & -25.59 &   S &  0.65 &   $<$3.3  &       &                  &          & $<$  21.57 &     ---   &            \\
                 UGC10918 &   264.3892 &    11.1217 & 7016 & -25.81 &   S &  0.54 &    136.0  &  P    &  7.9 $\pm$ 0.2   &          &  23.23     &   85.1$^N$&        0.21\\
                  NGC6515 &   269.3549 &    50.7281 & 7228 & -25.29 &   S &  0.11 &      1.5  &  P    & 10.6 $\pm$ 2.0   &          &  21.29     &     ---   &   $>$ -0.14\\
                  NGC6524 &   269.8113 &    45.8871 & 6108 & -25.07 & DR2 &  0.08 &    147.0  & E ?   & 14.1 $\pm$ 0.0   &  38 ( 15)&  23.15     &   32.5$^N$&        0.68\\
                  NGC6575 &   272.7395 &    31.1162 & 7422 & -25.64 & DR2 &  0.19 &     18.4  &  P    &  6.8 $\pm$ 0.1   &          &  22.42     &   14.6$^N$&        0.10\\
                  NGC6619 &   274.7314 &    23.6556 & 5503 & -25.48 &   S &  0.31 &   $<$1.5  &       &                  &          & $<$  21.08 &     ---   &            \\
                  NGC6623 &   274.9286 &    23.7096 & 5307 & -25.28 &   S &  0.27 &   $<$1.3  &       &                  &          & $<$  20.98 &     ---   &            \\
                  NGC6628 &   275.5913 &    23.4785 & 4910 & -25.02 &   S &  0.29 &     16.1  &  P*   & 22.4 $\pm$ 0.4   &          &  22.00     &    7.7$^N$&        0.33\\
                  NGC6688 &   280.1672 &    36.2896 & 5858 & -25.64 & DR2 &  0.28 &   $<$1.4  &       &                  &          & $<$  21.08 &    3.2$^N$&            \\
                  NGC6702 &   281.7399 &    45.7057 & 5102 & -25.22 & DR2 &  0.62 &   $<$3.1  &       &                  &          & $<$  21.31 &     ---   &            \\
                 UGC11465 &   295.4255 &    50.6325 & 7364 & -25.96 &   S &  0.20 &  13100.0  & FRI   & 50.1 $\pm$ 0.1   & 480 (241)&  25.26     & 3090.0$^N$&        0.65\\
                 UGC11892 &   330.9670 &    35.9906 & 5727 & -25.12 & DR2 &  0.10 &      5.3  &  P    &  8.5 $\pm$ 0.3   &          &  21.65     &     ---   &   $>$  0.44\\
                   IC5180 &   332.8001 &    38.9273 & 6059 & -25.20 &   S &  0.29 &    502.0  & FRI   & 10.9 $\pm$ 0.1   & 580 (232)&  23.67     &     $^a$  &            \\
                 UGC11950 &   333.1315 &    38.6828 & 6227 & -25.61 &   S &  0.28 &     12.7  &  P    &  9.6 $\pm$ 0.1   &          &  22.10     &     ---   &   $>$  0.83\\
                  NGC7242 &   333.9146 &    37.2987 & 5822 & -26.37 &   S &  0.13 &      6.5  &  P    &  9.4 $\pm$ 1.0   &          &  21.75     &     ---   &   $>$  0.53\\
                  NGC7248 &   334.2190 &    40.5046 & 4493 & -25.03 & DR2 &  0.17 &      0.9  &  P    &  7.8 $\pm$ 4.2   &          &  20.65     &     ---   &   $>$ -0.37\\
                  NGC7265 &   335.6145 &    36.2098 & 5159 & -25.74 & DR2 &  0.12 &      2.1  &  P    &  9.1 $\pm$ 1.1   &          &  21.16     &     ---   &   $>$  0.03\\
                  NGC7274 &   336.0462 &    36.1259 & 6174 & -25.58 & DR2 &  0.11 &     19.4  & E ?   & 15.2 $\pm$ 0.3   &  35 ( 14)&  22.28     &    3.1$^N$&        0.82\\
                  NGC7315 &   338.8821 &    34.8035 & 6341 & -25.14 & DR2 &  0.11 &   $<$0.6  &       &                  &          & $<$  20.76 &     ---   &            \\
                  NGC7318 &   338.9913 &    33.9656 & 6232 & -25.56 & DR2 &  0.12 &      2.2  &  P    & 11.3 $\pm$ 2.3   &          &  21.34     &     ---   &   $>$  0.04\\
                  NGC7330 &   339.2341 &    38.5481 & 5369 & -25.34 & DR2 &  0.78 &   $<$3.9  &       &                  &          & $<$  21.46 &     ---   &            \\
                  NGC7335 &   339.3307 &    34.4479 & 6341 & -25.63 & DR2 &  0.12 &     16.4  &  P    &  9.0 $\pm$ 0.2   &          &  22.23     &     ---   &   $>$  0.94\\
                 UGC12179 &   341.2664 &    33.9962 & 7000 & -25.25 & DR2 &  0.09 &      3.5  &  P    &  7.0 $\pm$ 0.3   &          &  21.64     &     ---   &   $>$  0.25\\
                  NGC7386 &   342.5089 &    11.6987 & 7289 & -25.75 &   S &  0.79 &    156.0  &  P    &  9.5 $\pm$ 0.1   &          &  23.33     &   44.9    &        0.56\\
                 UGC12214 &   342.7540 &    31.3749 & 6607 & -25.06 & DR2 &  0.09 &      0.9  &  P    &  8.2 $\pm$ 1.3   &          &  21.02     &     ---   &   $>$ -0.34\\
                 UGC12242 &   343.6063 &    32.4517 & 6755 & -25.15 & DR2 &  0.09 &      5.4  &  P    & 10.8 $\pm$ 0.6   &          &  21.80     &     ---   &   $>$  0.45\\
                  NGC7426 &   344.0119 &    36.3614 & 5351 & -25.70 & DR2 &  0.13 &      6.9  &  P    &  8.1 $\pm$ 0.3   &          &  21.70     &    6.0$^N$&        0.06\\
                 NGC7436B &   344.4897 &    26.1500 & 7341 & -26.18 & DR2 &  0.14 &    106.0  & E ?   &  8.3 $\pm$ 0.4   & 145 ( 72)&  23.17     &   26.2$^N$&        0.63\\
                  NGC7512 &   348.0872 &    31.1256 & 7016 & -25.10 & DR2 &  0.08 &      0.4  &  P    & 16.0 $\pm$ 7.7   &          &  20.72     &     ---   &   $>$ -0.70\\
                 UGC12444 &   348.6223 &    31.5489 & 6321 & -25.09 & DR2 &  0.09 &   $<$0.4  &       &                  &          & $<$  20.64 &     ---   &            \\
                  NGC7539 &   348.6227 &    23.6848 & 6014 & -25.19 & DR2 &  0.20 &   $<$1.0  &       &                  &          & $<$  20.96 &     ---   &            \\
                  NGC7550 &   348.8170 &    18.9614 & 5054 & -25.48 & DR2 &  0.12 &     30.0  &  P    &  7.5 $\pm$ 0.1   &          &  22.29     &   44.9$^N$&       -0.18\\
                 UGC12482 &   349.3906 &    29.0194 & 6921 & -25.35 & DR2 &  0.18 &   1250.0  &Compl. & 48.6 $\pm$ 1.5   & 400 (189)&  24.19     &  239.0$^N$&        0.74\\
                  NGC7660 &   351.4528 &    27.0299 & 5657 & -25.17 & DR2 &  0.09 &   $<$0.5  &       &                  &          & $<$  20.57 &     ---   &            \\
                  NGC7680 &   352.1462 &    32.4157 & 5116 & -25.34 & DR2 &  0.09 &      1.8  &  P    &  8.6 $\pm$ 0.9   &          &  21.08     &     ---   &   $>$ -0.05\\
                  NGC7681 &   352.2287 &    17.3096 & 6773 & -25.78 & DR2 &  0.12 &      6.6  &  P*   & 19.5 $\pm$ 1.9   &          &  21.89     &     ---   &   $>$  0.54\\
                  NGC7722 &   354.6717 &    15.9548 & 4004 & -25.04 & DR2 &  0.16 &     11.1  &  P*   & 19.2 $\pm$ 0.3   &          &  21.66     &    4.7$^N$&        0.38\\
                  NGC7777 &   358.3021 &    28.2834 & 6898 & -25.33 & DR2 &  0.10 &      2.7  &  P    &  7.6 $\pm$ 0.4   &          &  21.52     &     ---   &   $>$  0.14\\
                 UGC12839 &   358.6239 &    28.3089 & 6930 & -25.16 & DR2 &  0.10 &      1.6  &  P    &  9.1 $\pm$ 1.2   &          &  21.29     &     ---   &   $>$ -0.11\\
                  NGC7785 &   358.8293 &     5.9158 & 3768 & -25.28 &   S &  0.36 &    316.0  &Compl. & 50.1 $\pm$ 9.2   & 100 ( 25)&  23.06     &   32.6$^N$&        1.02\\
                                 \hline
\end{longtable}
%\end{center}

\smallskip
\small{Column description: (1) name; (2 and 3) right ascension and declination;
(4) recession velocity (\kms); (5) K band absolute magnitude; (6)
image type; (7) r.m.s. noise of the LOFAR image (mJy beam$^{-1}$); (8)
flux density at 150 MHz from LOFAR (mJy); (9)
morphological description (P=point-like,
P*=sources in which the 3$\sigma$ level isophote does not reach a
radius of 15\arcsec, but with a size $\gtrsim 15
\arcsec$) (10)
size of the central component in the LOFAR images; (11) largest
angular size in arcseconds (and kpc) for the extended sources; (12)
luminosity at 150 MHz (W Hz$^{-1}$); (13) 1.4 GHz flux density (mJy)
from FIRST, or, from NVSS when the FIRST measurement is not available,
or not reliable (a ``$^N$'' indicates a measurement from NVSS images,
while a ``---'' represents a source undetected in the NVSS at a
typical limit of 2 mJy); (14) radio spectral index between 150 MHz and
1.4 GHz. $^a$: sources for which the flux density at 1.4 GHz can not
be obtained because they are not covered (or resolved out in the FIRST
images) and contaminated by a nearby source in the NVSS images.}
\end{landscape}
\twocolumn

\clearpage
\newpage

  \section{Images and notes on the radio morphology of the extended and diffuse sources.}

In this appendix, we show the images of all the extended and diffuse
sources and a brief description of the most interesting or complex
cases.

\begin{figure*}
\includegraphics[scale=0.13]{NGC0071.jpg}
\includegraphics[scale=0.13]{NGC0080.jpg}
\includegraphics[scale=0.13]{NGC0315.jpg}
\includegraphics[scale=0.13]{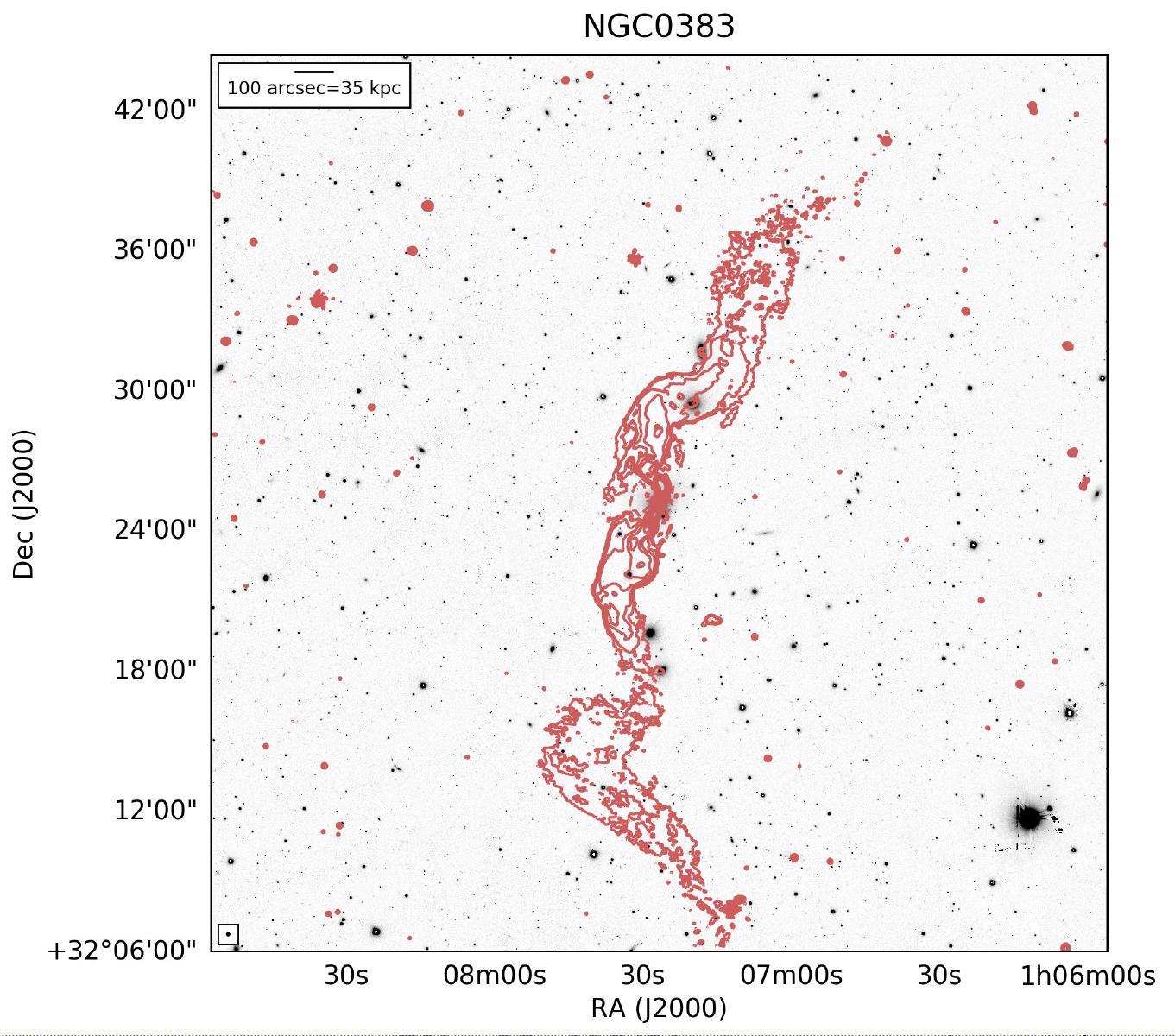}
\includegraphics[scale=0.13]{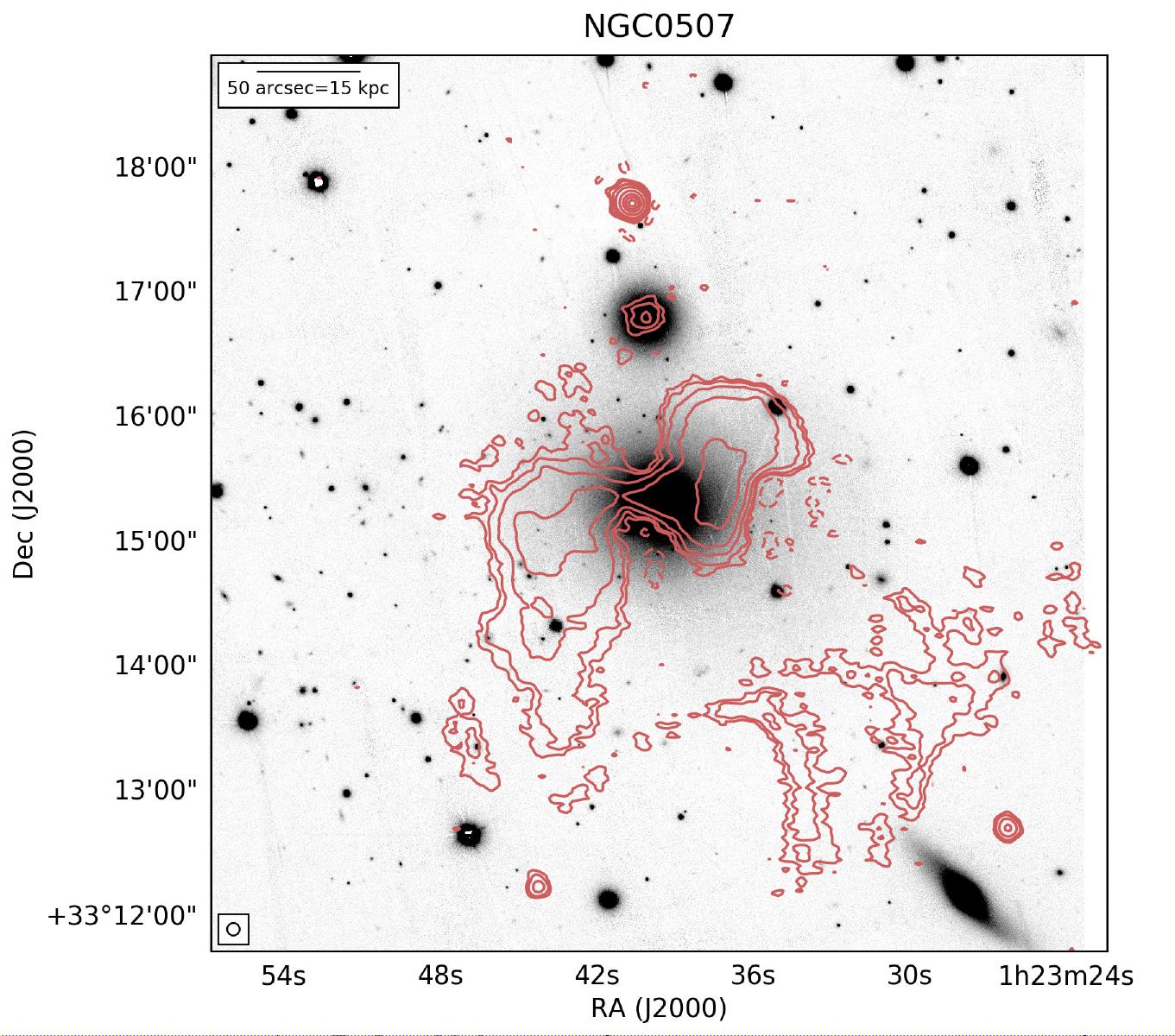}
\includegraphics[scale=0.13]{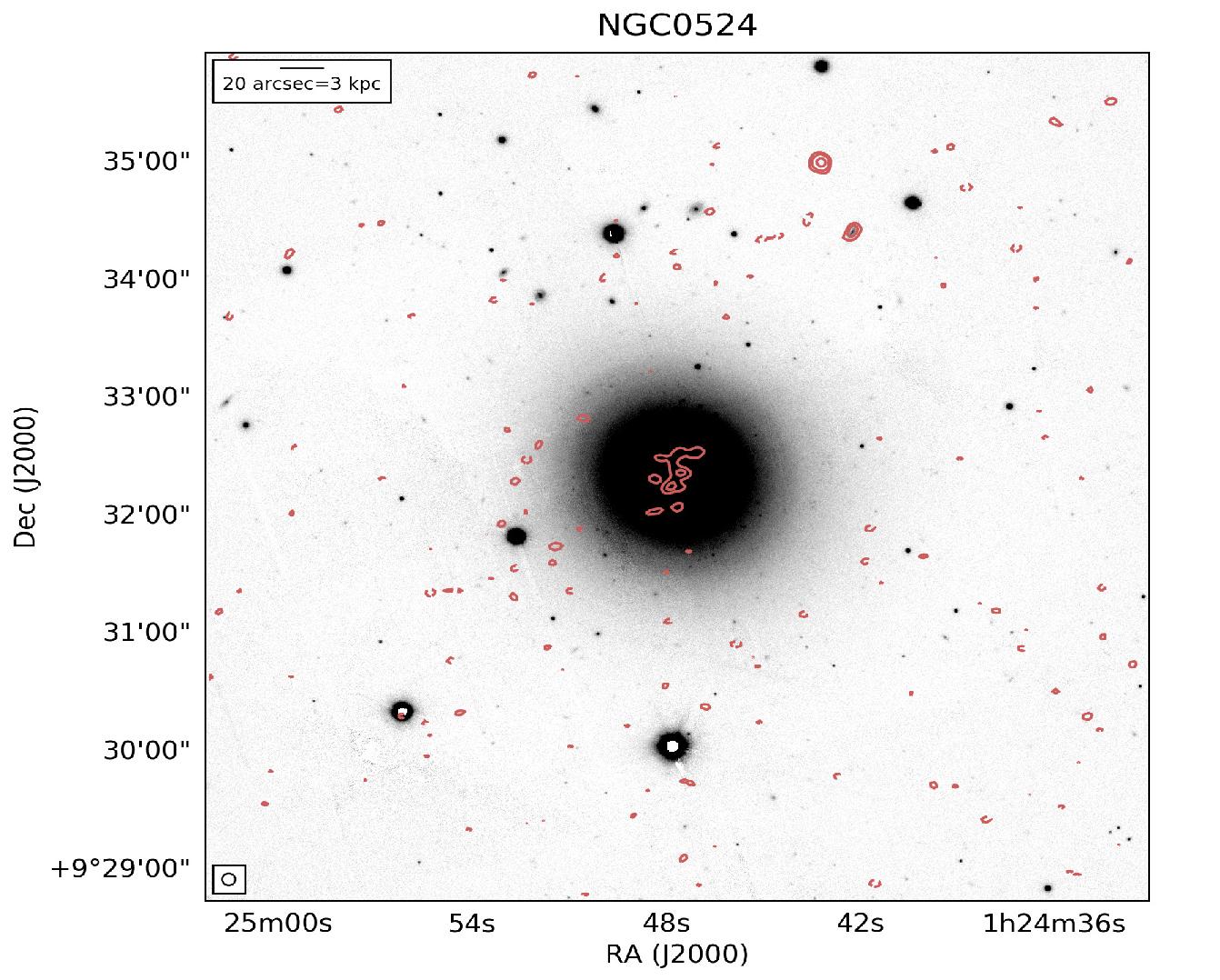}
\includegraphics[scale=0.13]{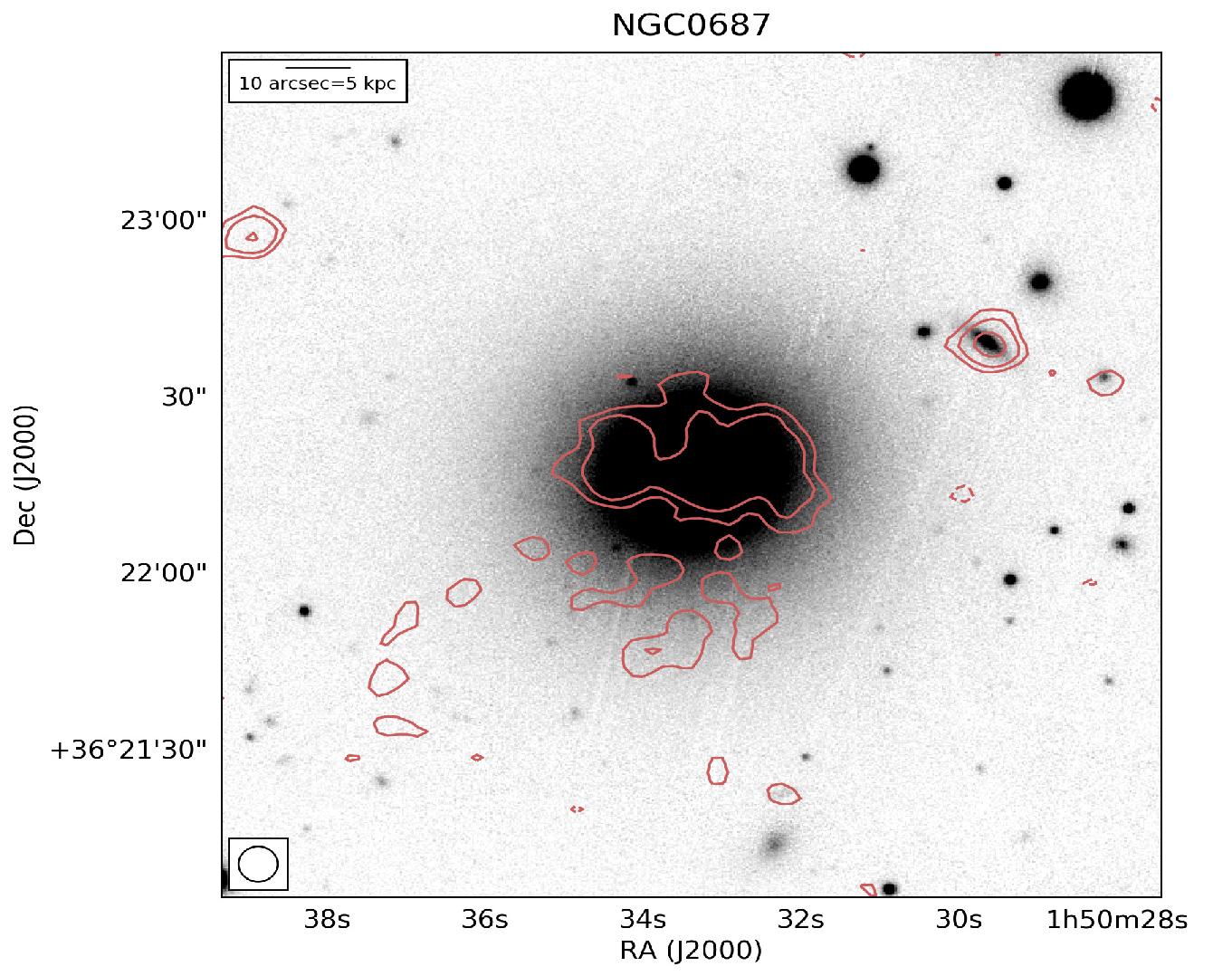}
\includegraphics[scale=0.13]{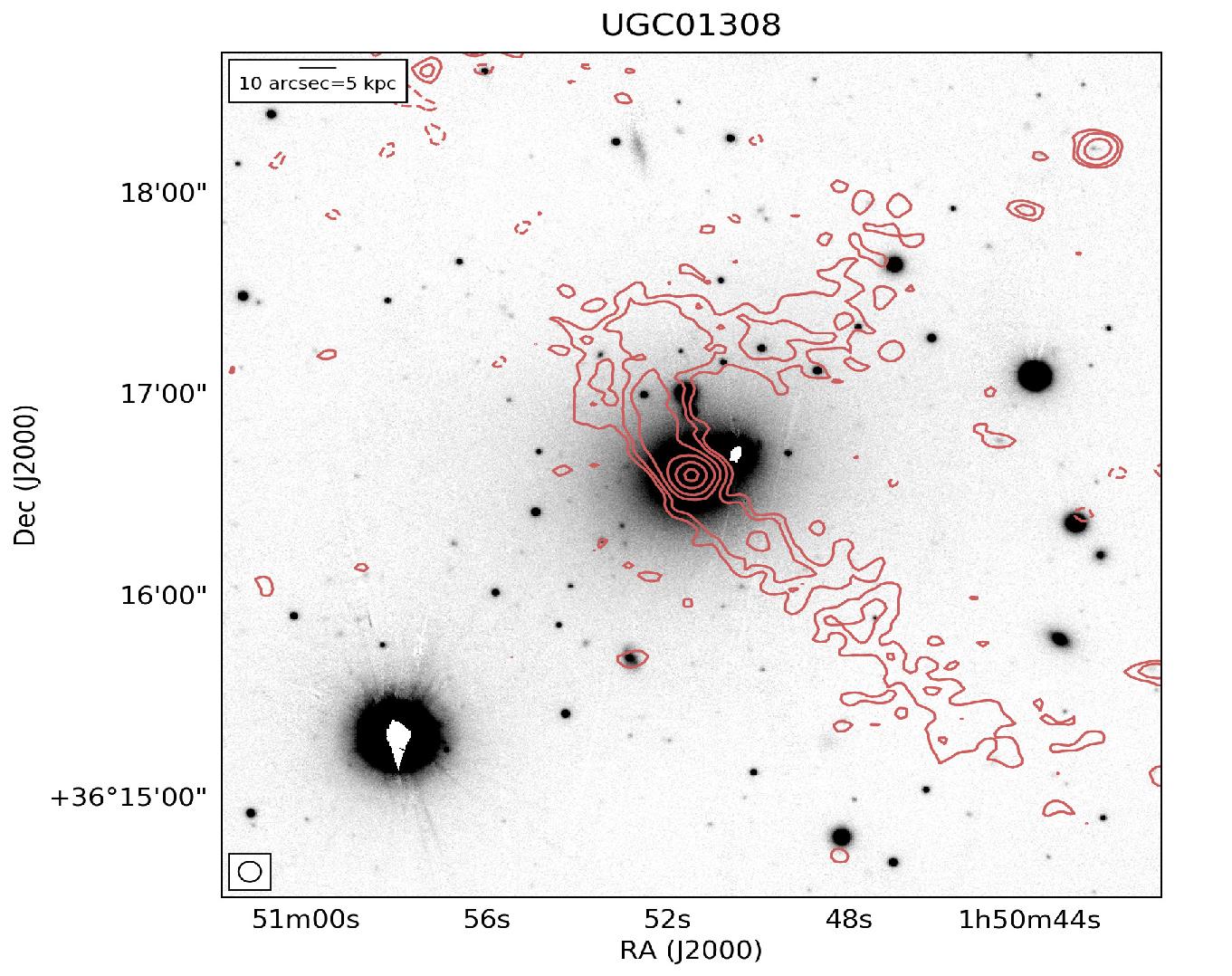}
\includegraphics[scale=0.13]{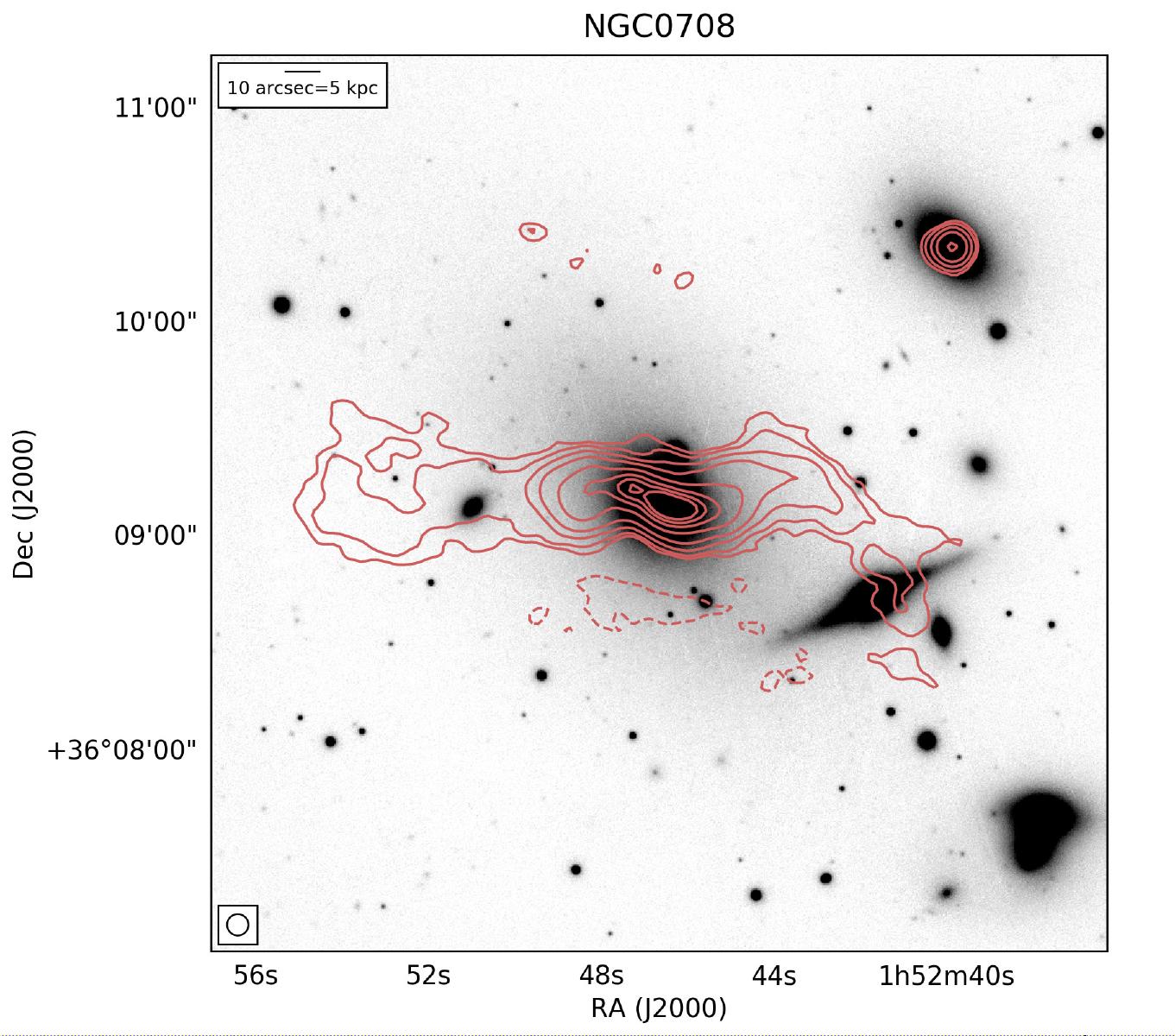}
\includegraphics[scale=0.13]{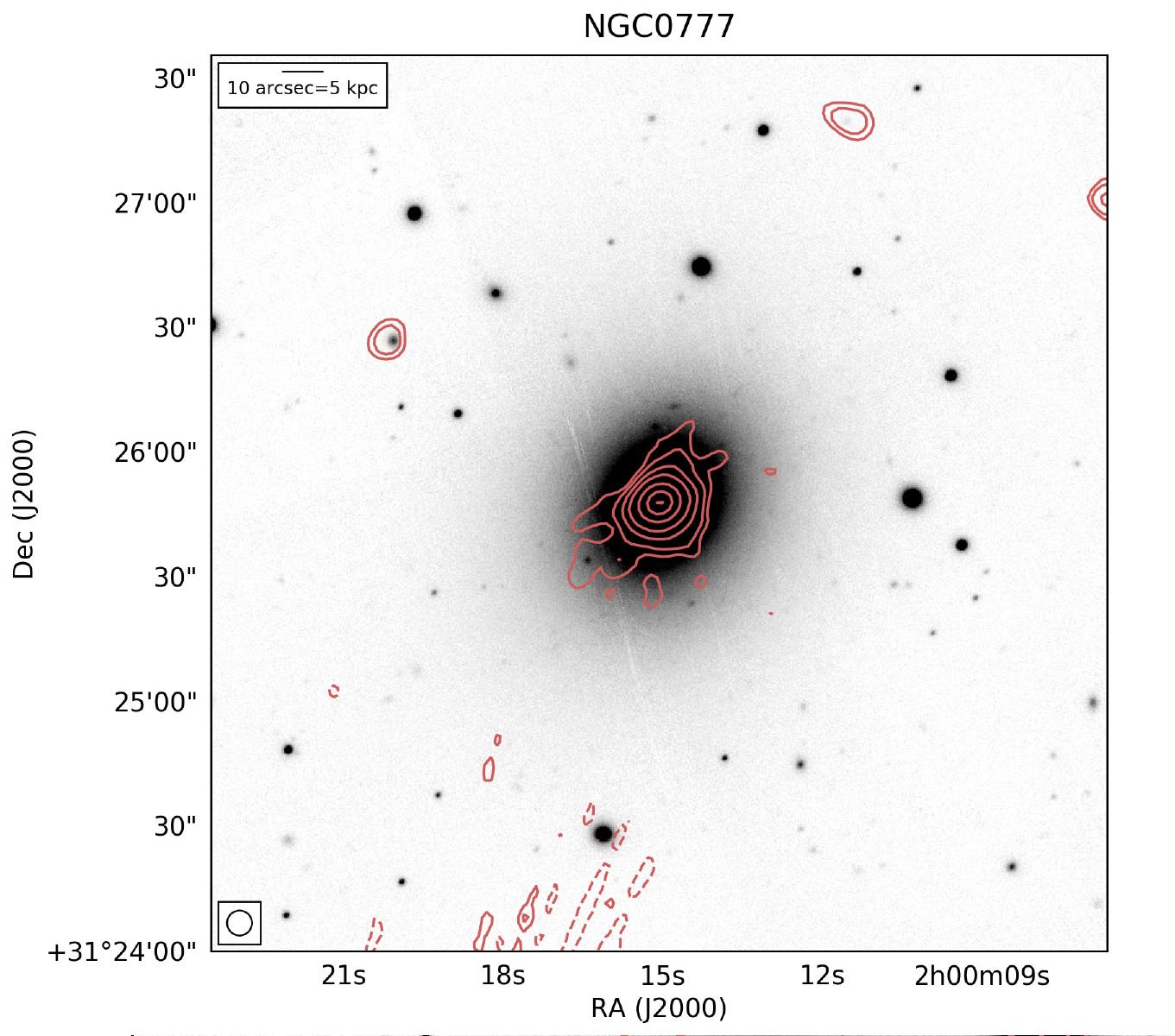}
\includegraphics[scale=0.13]{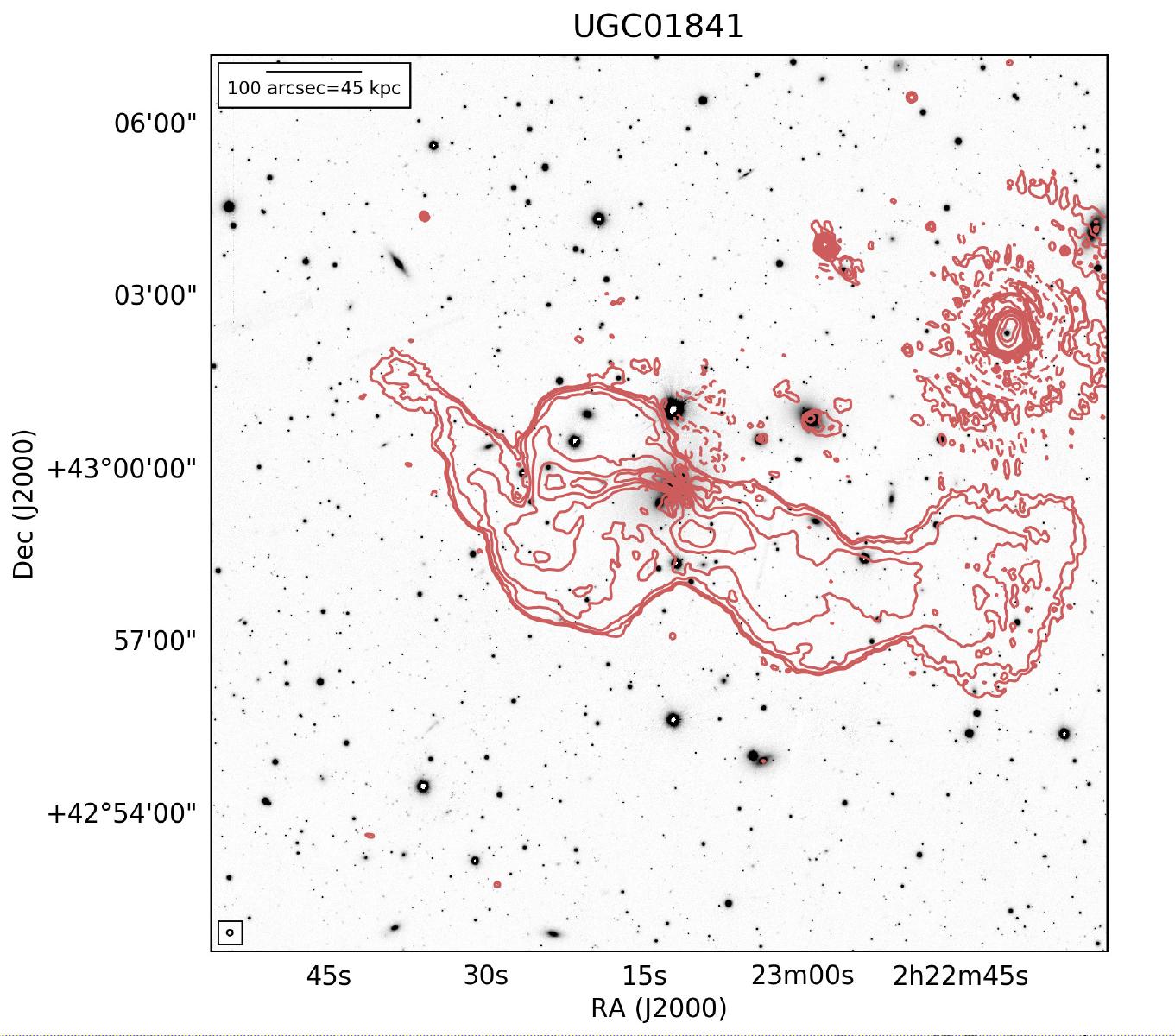}
\includegraphics[scale=0.13]{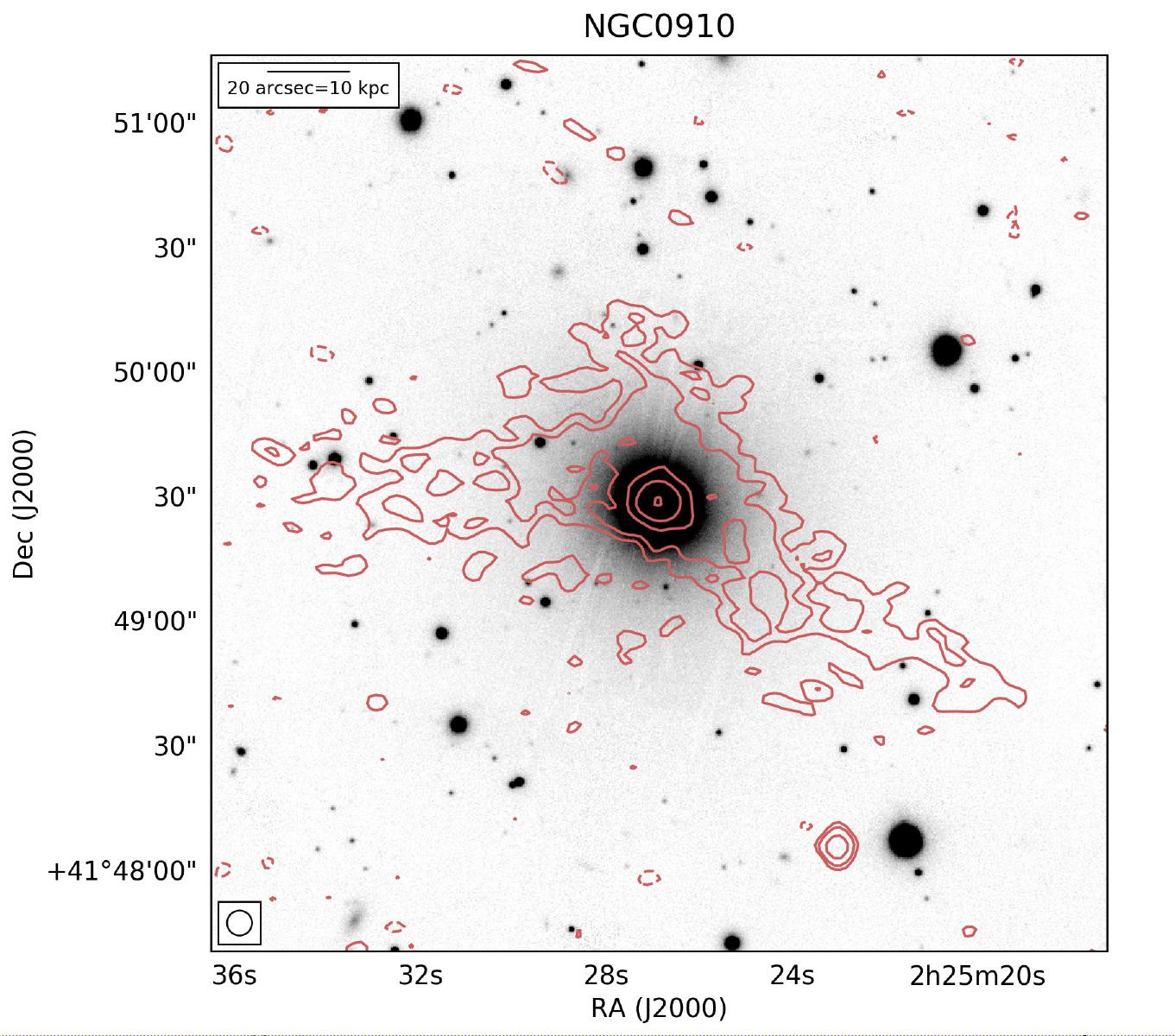}
\caption{LOFAR images at 150 MHz of the 46 galaxies showing extended
  emission. The lowest contour is drawn at three times the local
  r.m.s., as reported in Table \ref{tab}.}
\label{esteseall}
\end{figure*}

\addtocounter{figure}{-1}
\begin{figure*}
\includegraphics[scale=0.13]{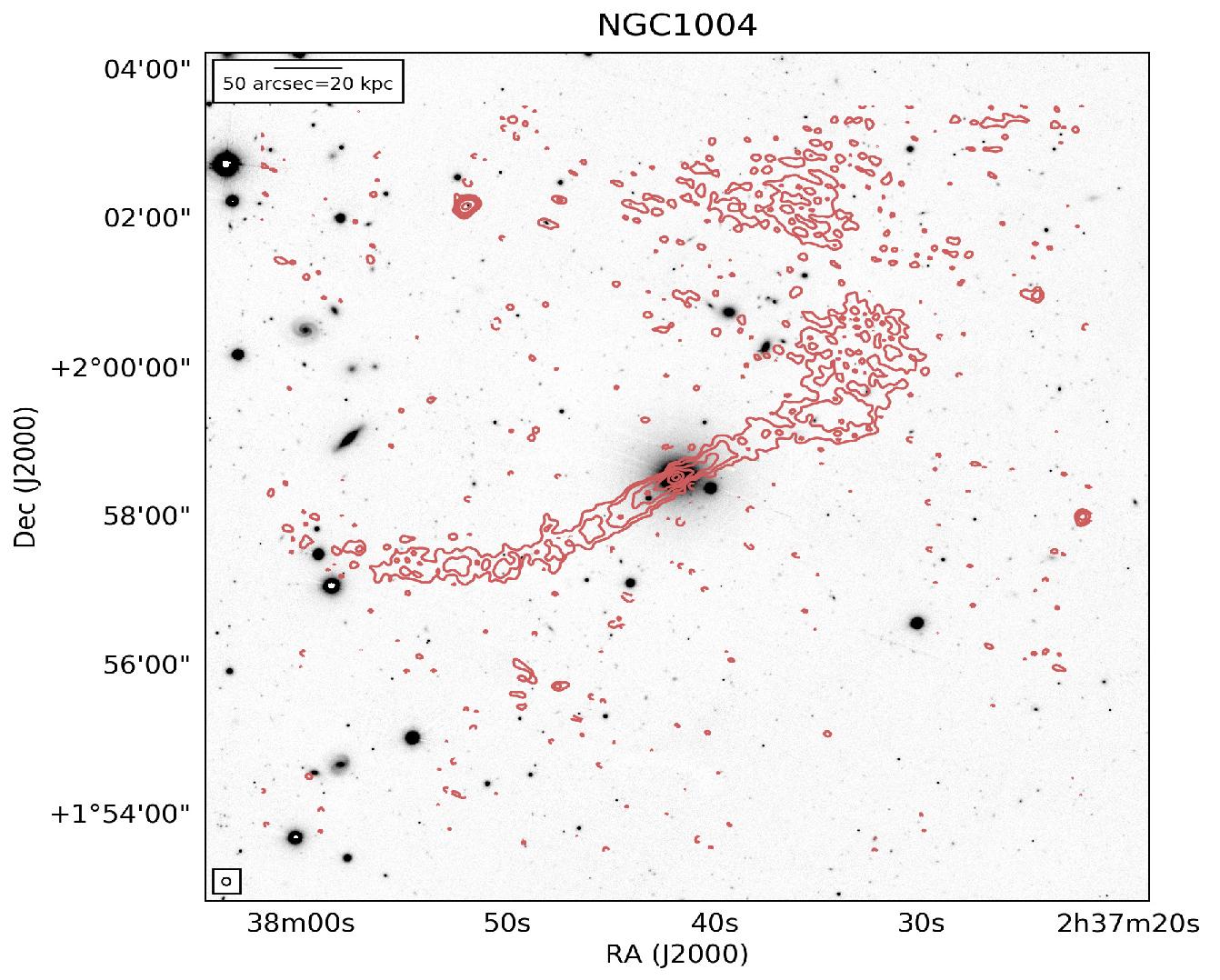}
\includegraphics[scale=0.13]{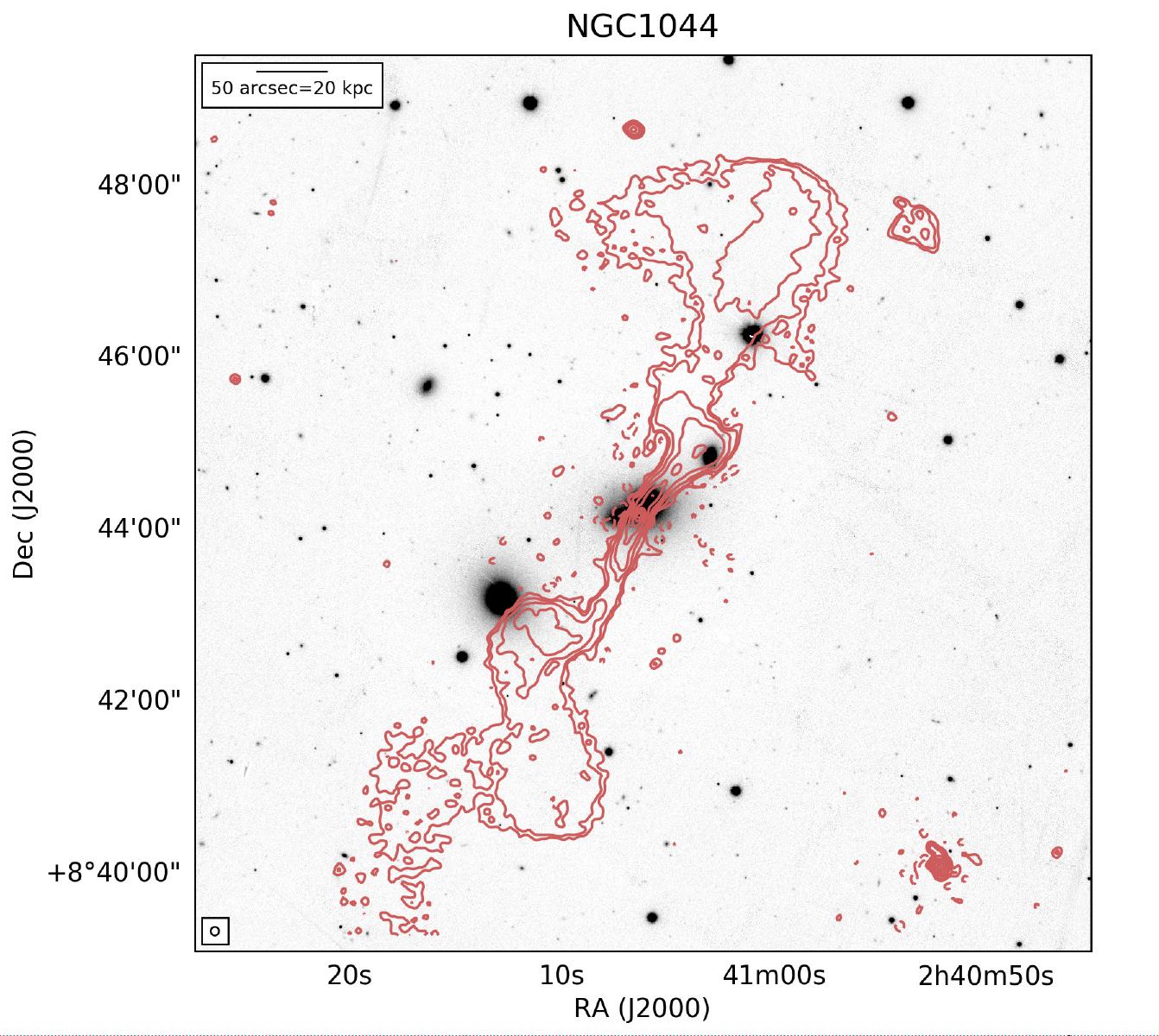}
\includegraphics[scale=0.13]{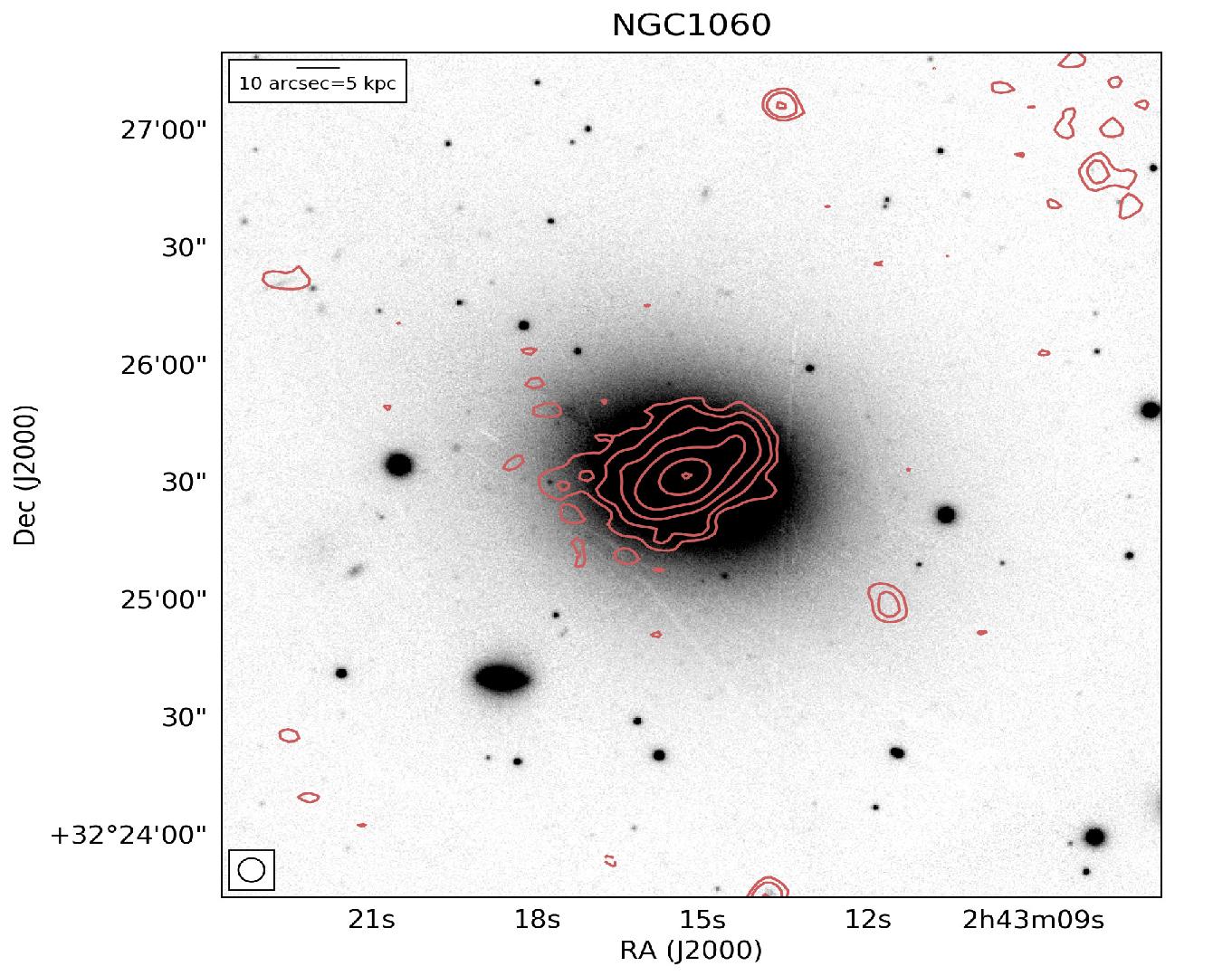}
\includegraphics[scale=0.13]{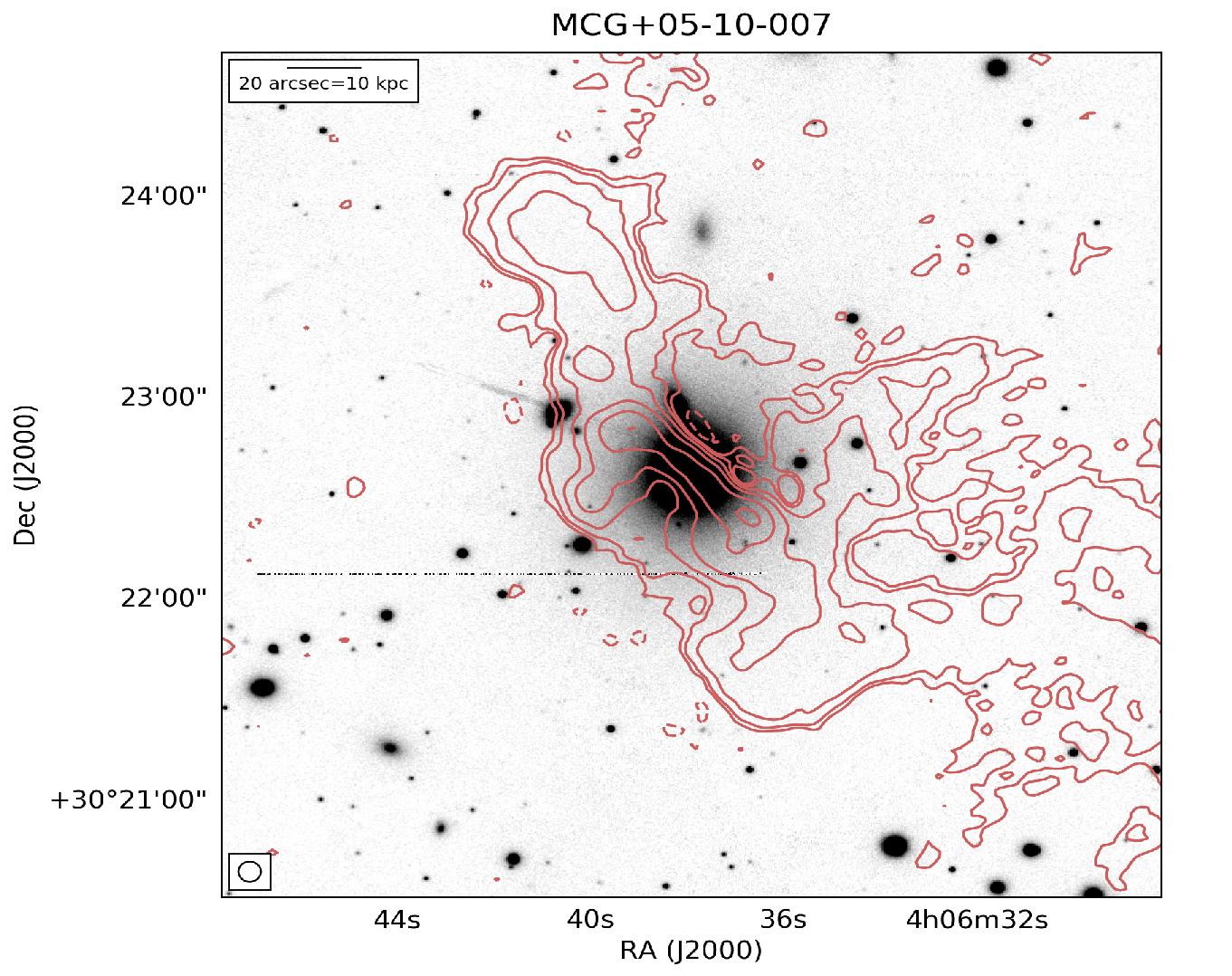}
\includegraphics[scale=0.13]{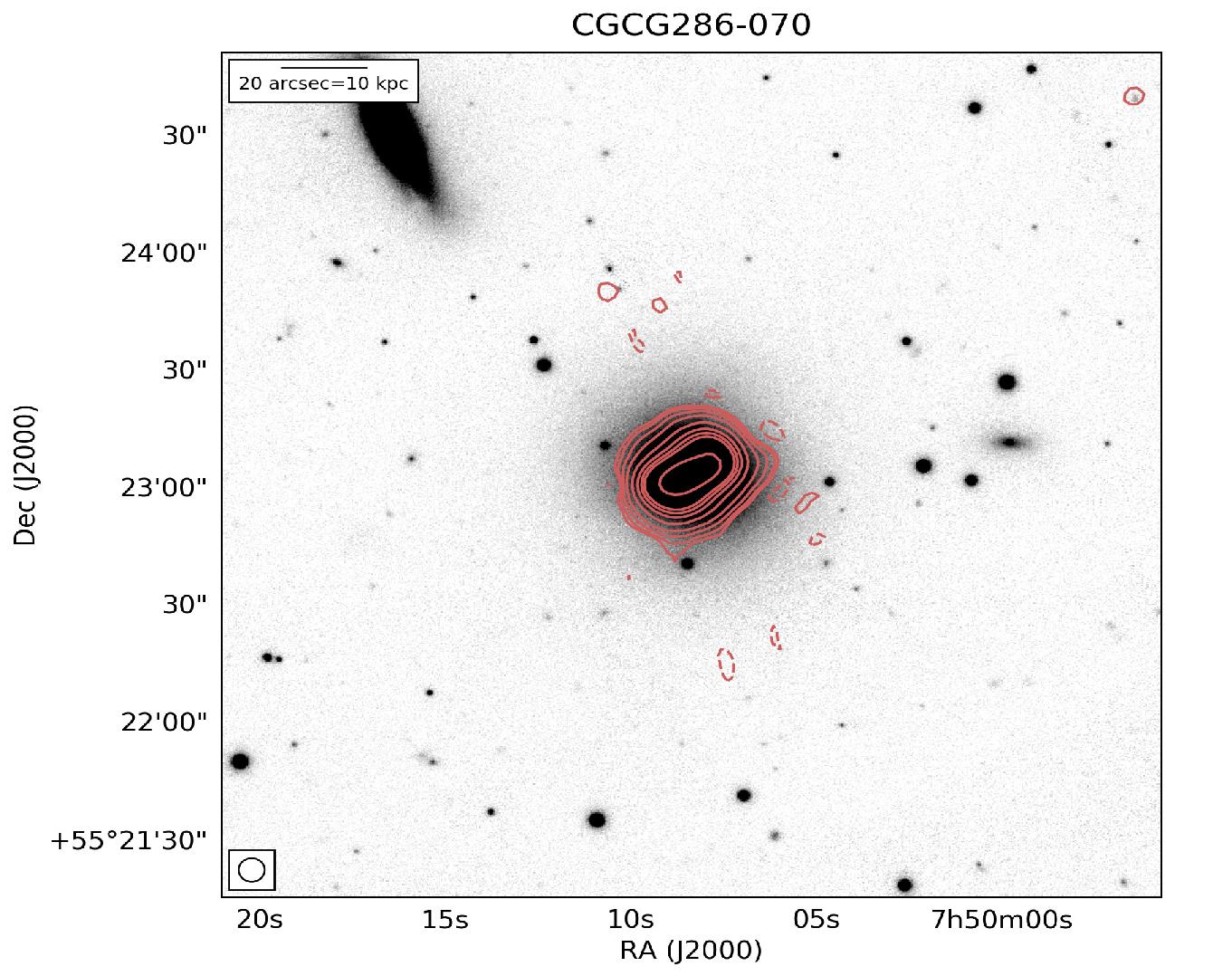}
\includegraphics[scale=0.13]{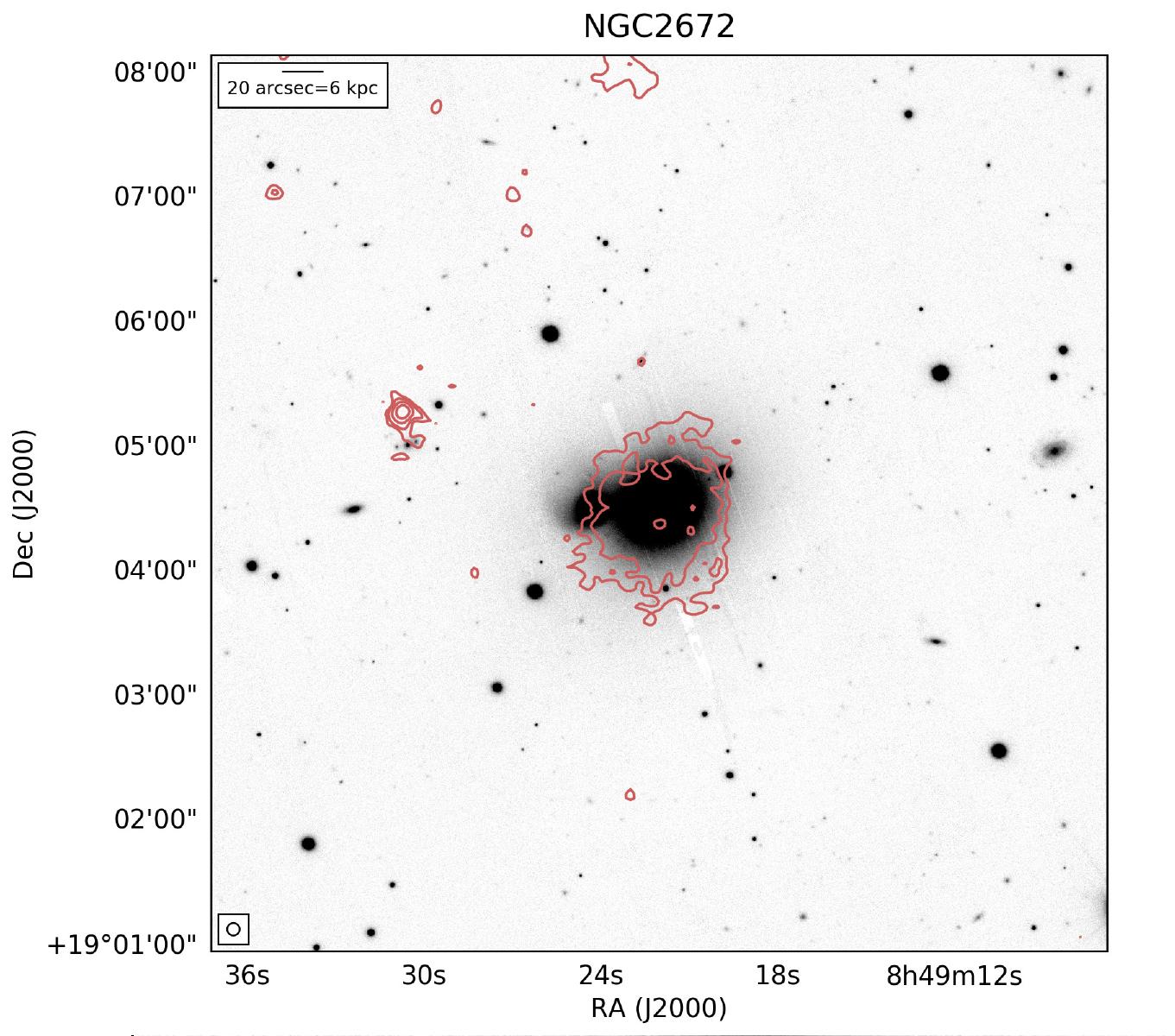}
\includegraphics[scale=0.13]{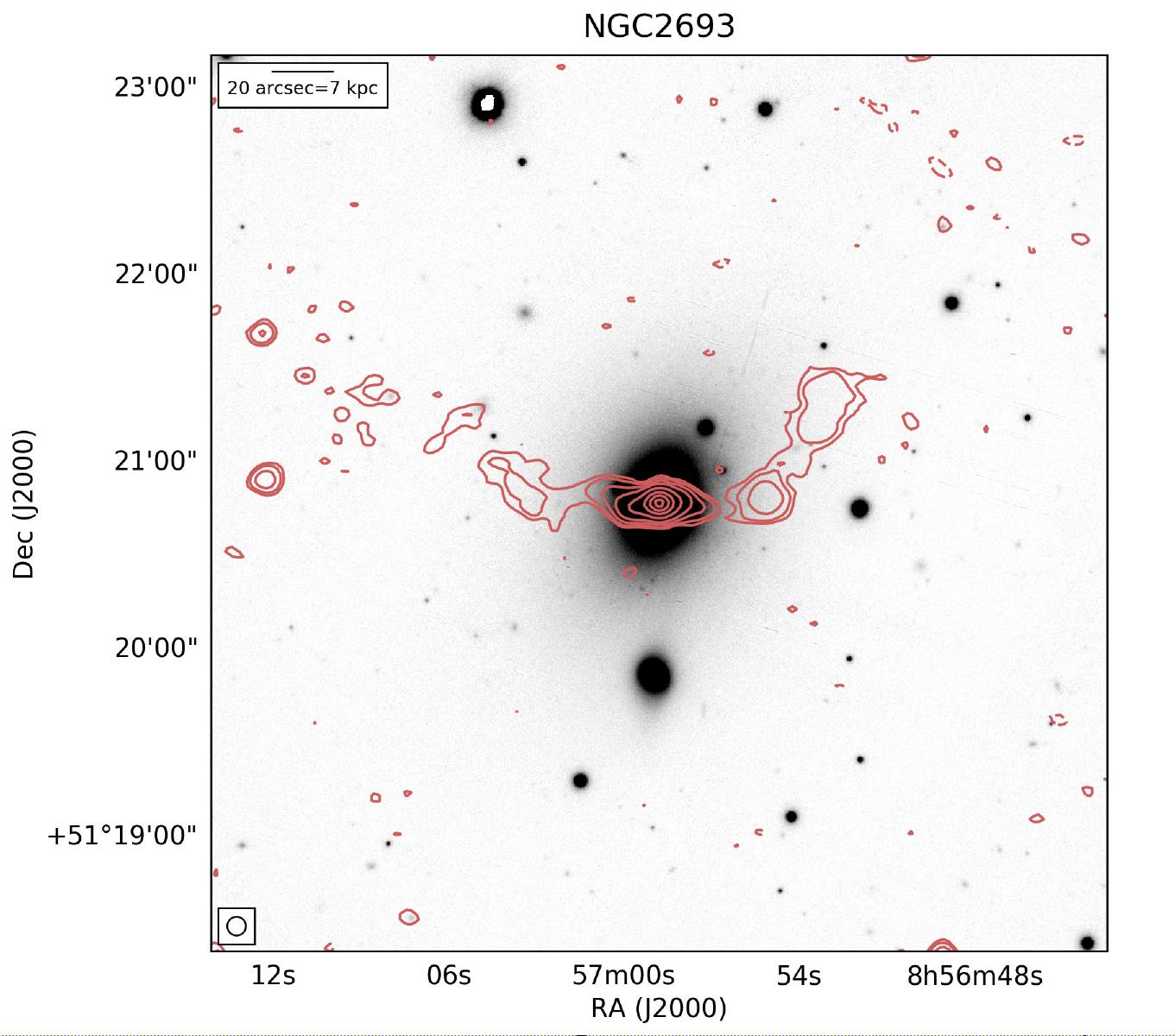}
\includegraphics[scale=0.13]{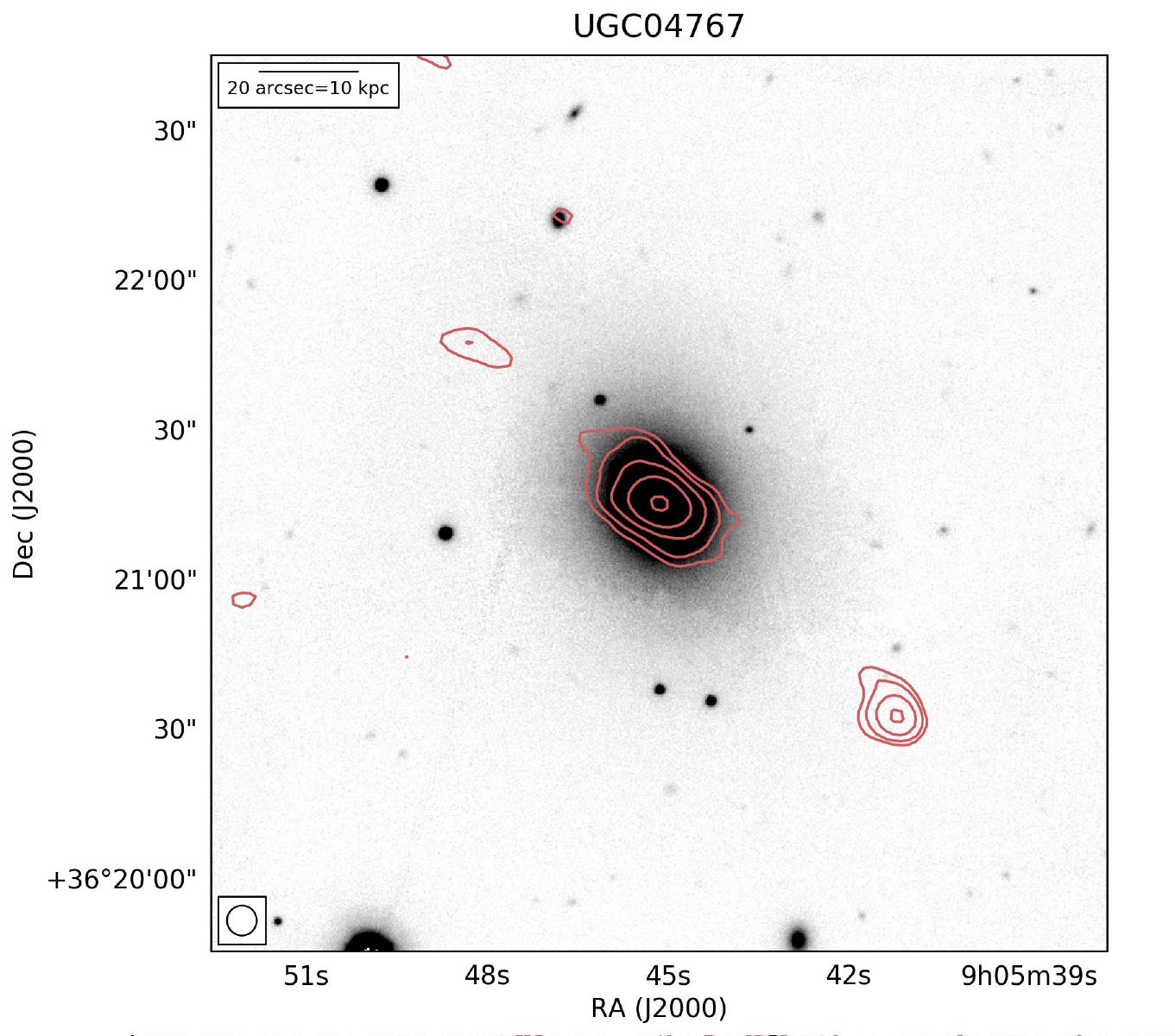}
\includegraphics[scale=0.13]{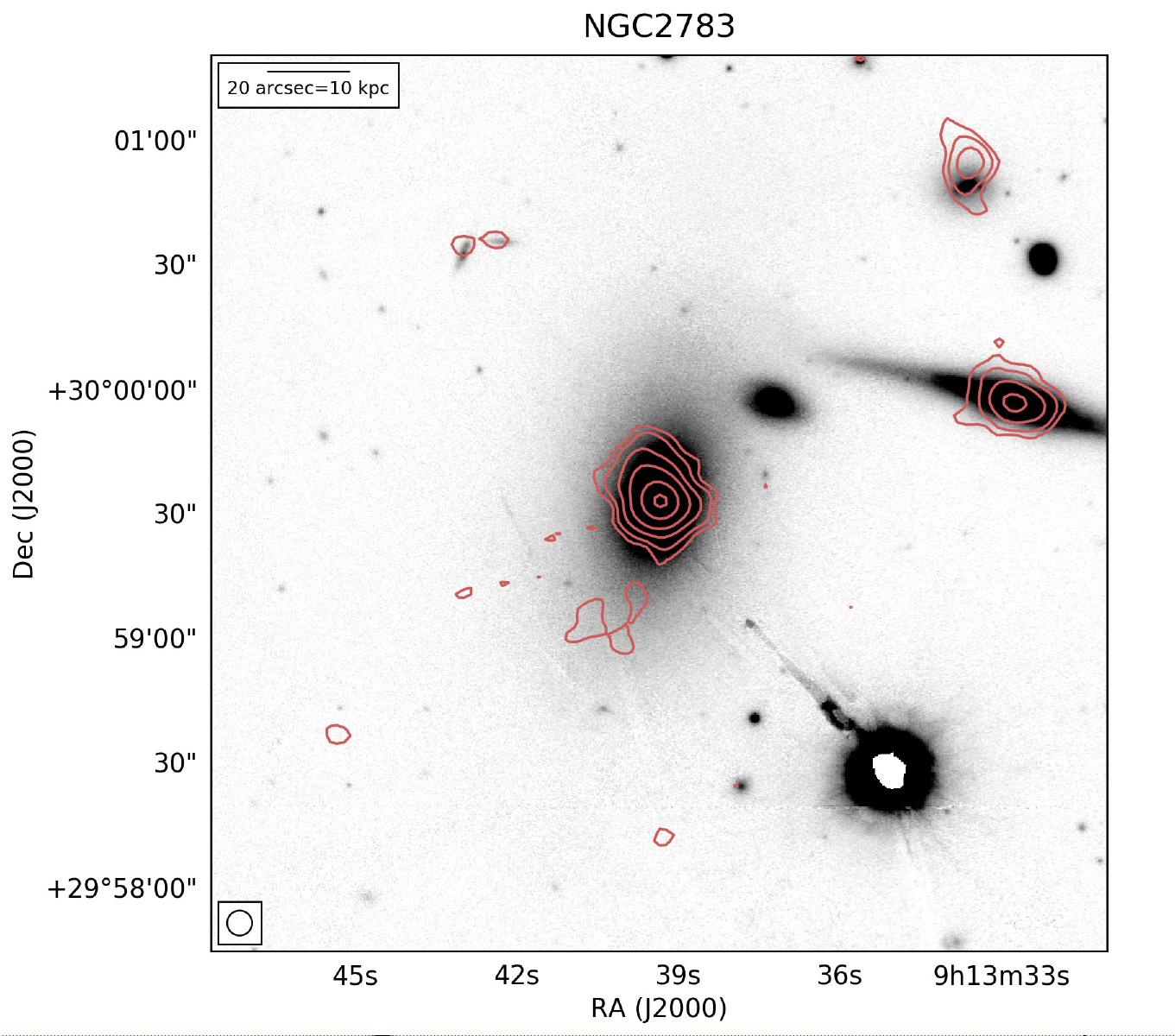}
\includegraphics[scale=0.13]{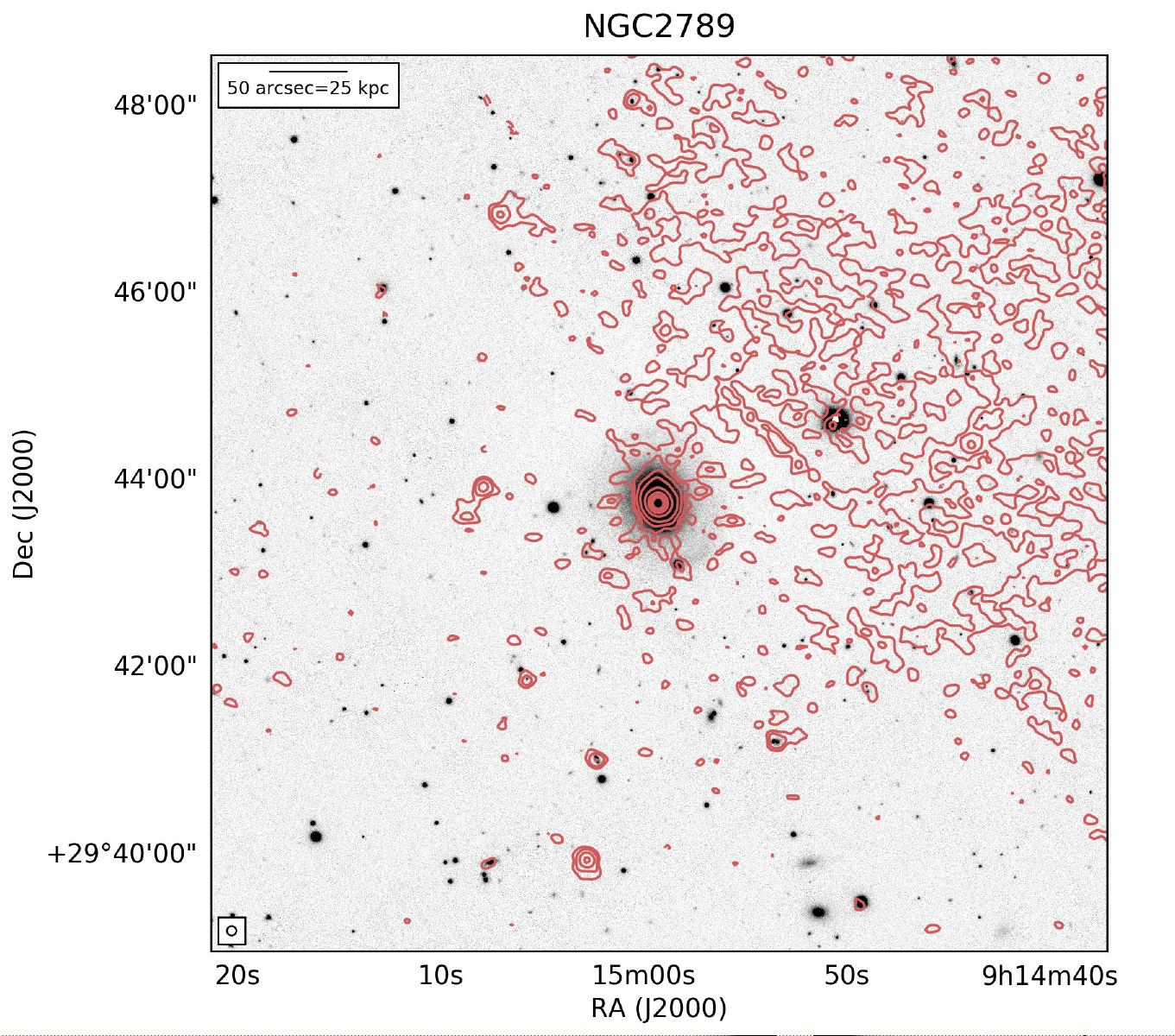}
\includegraphics[scale=0.13]{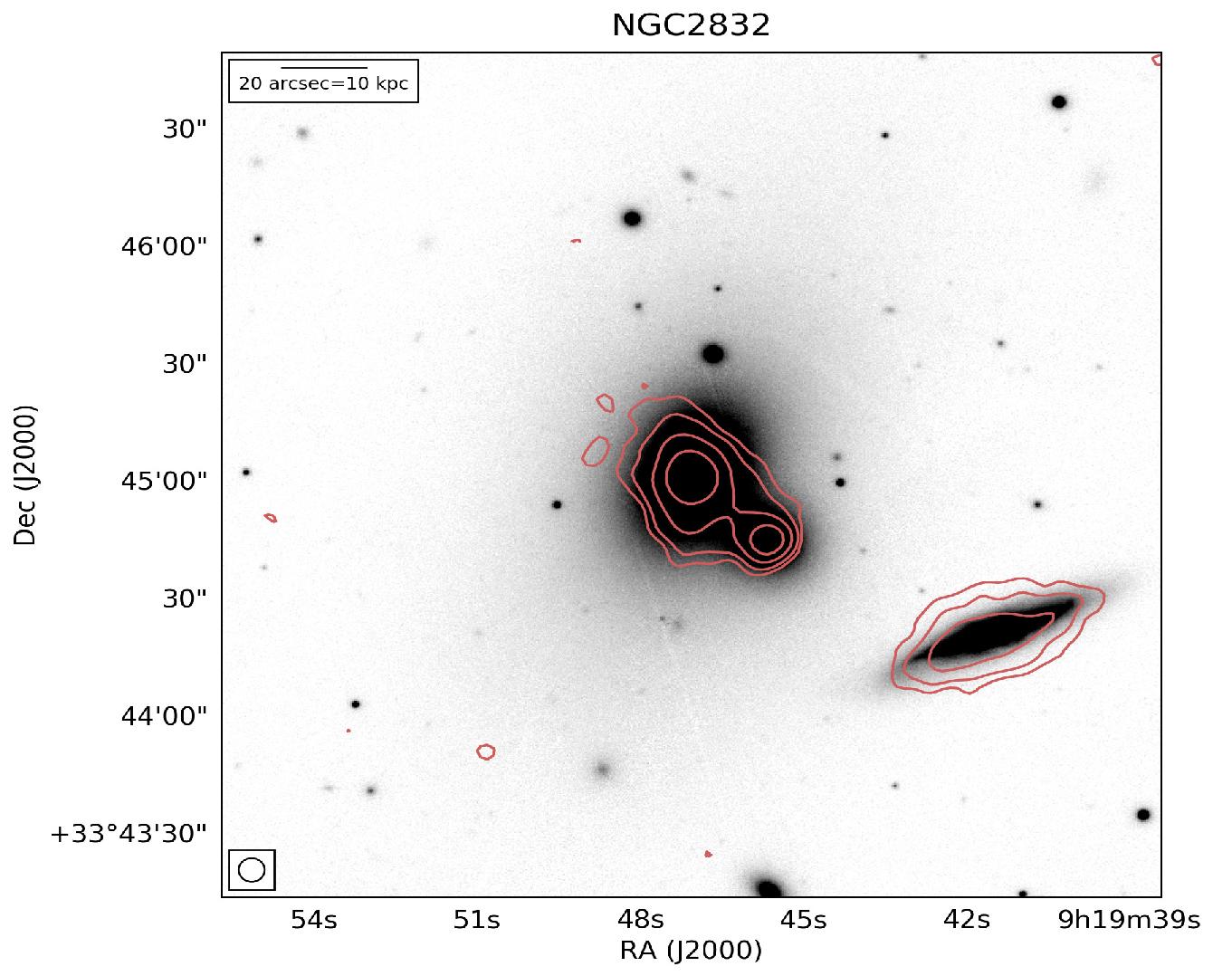}
\includegraphics[scale=0.13]{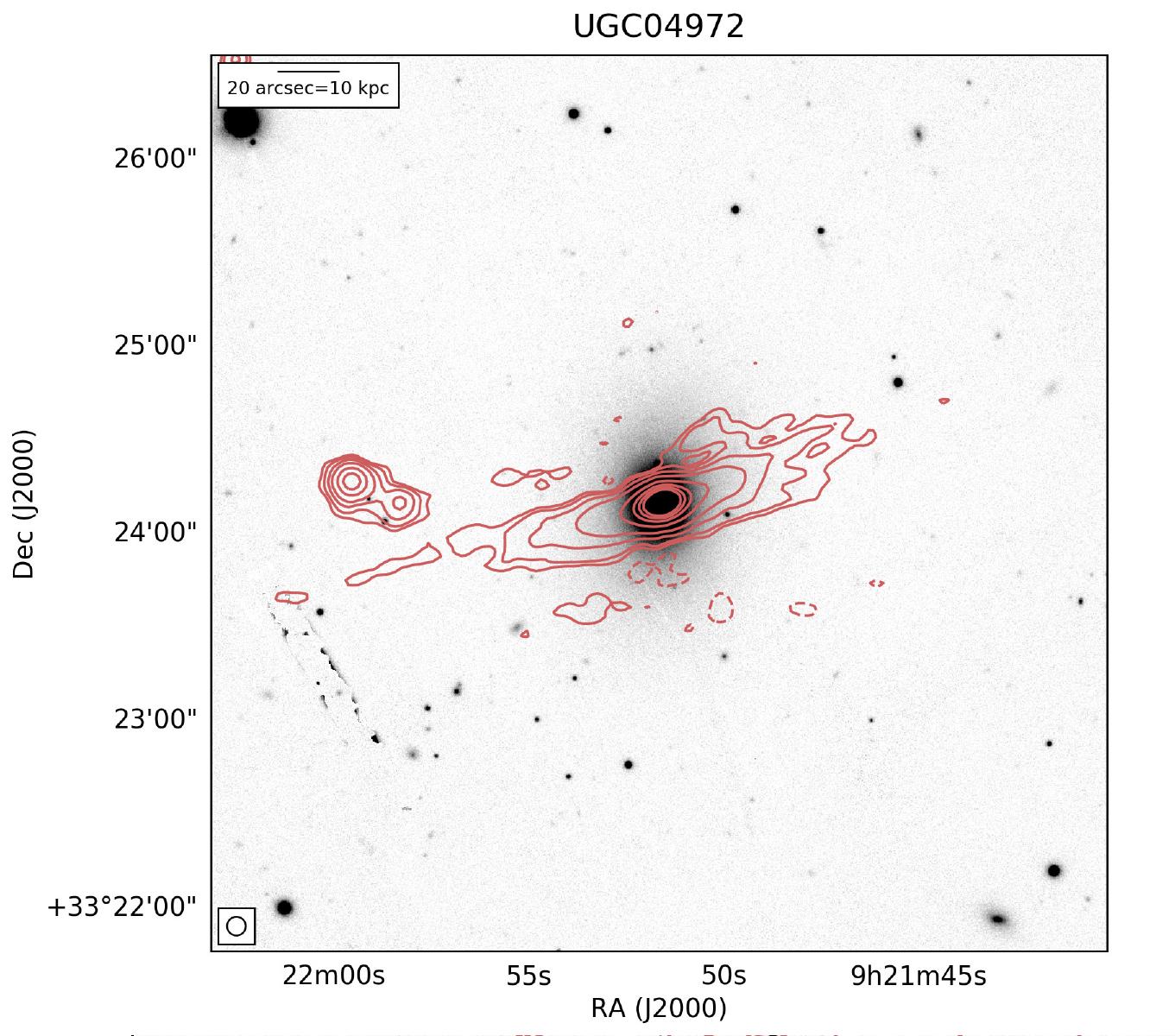}
\caption{(continued)}
\end{figure*}

\addtocounter{figure}{-1}
\begin{figure*}
\includegraphics[scale=0.13]{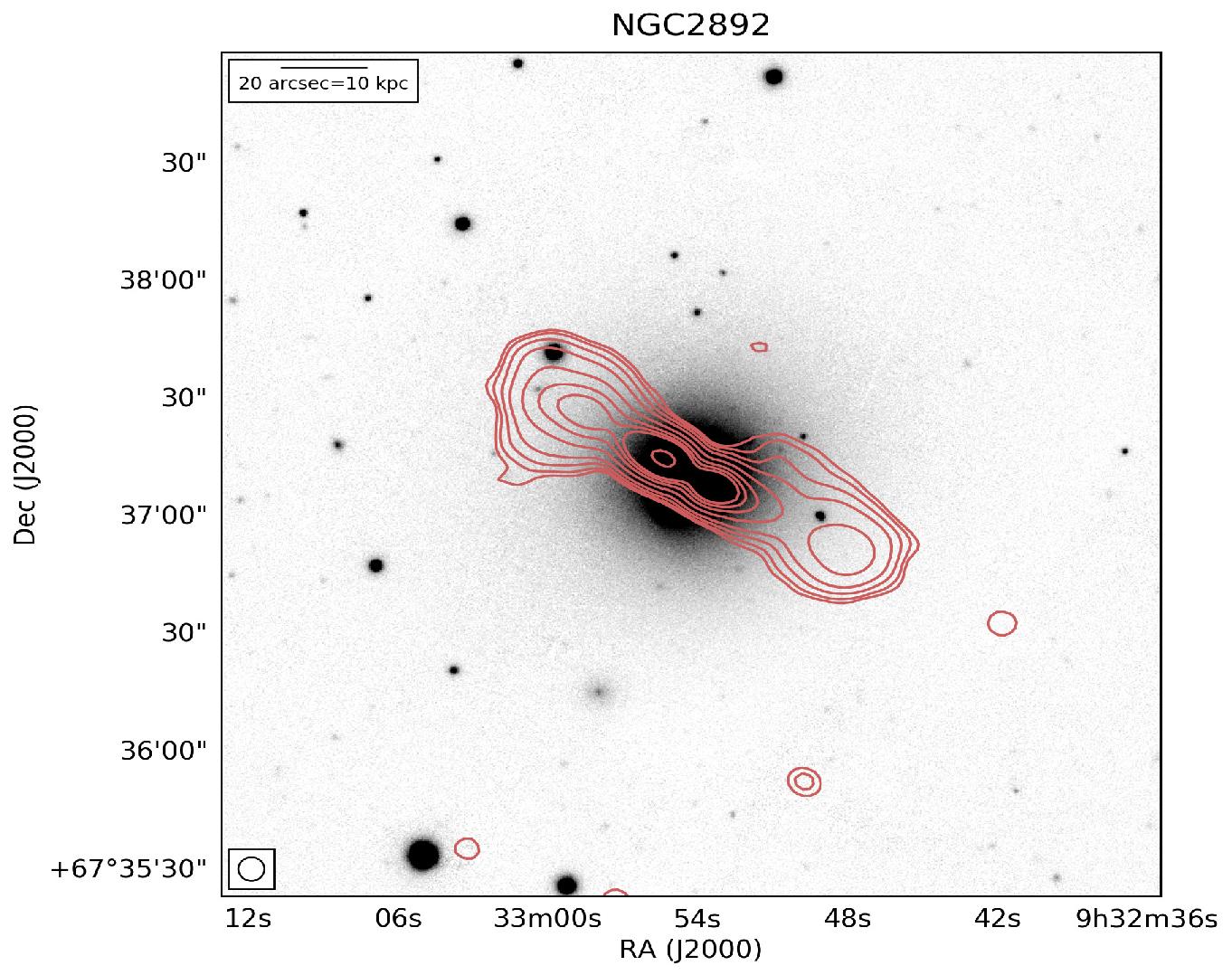}
\includegraphics[scale=0.13]{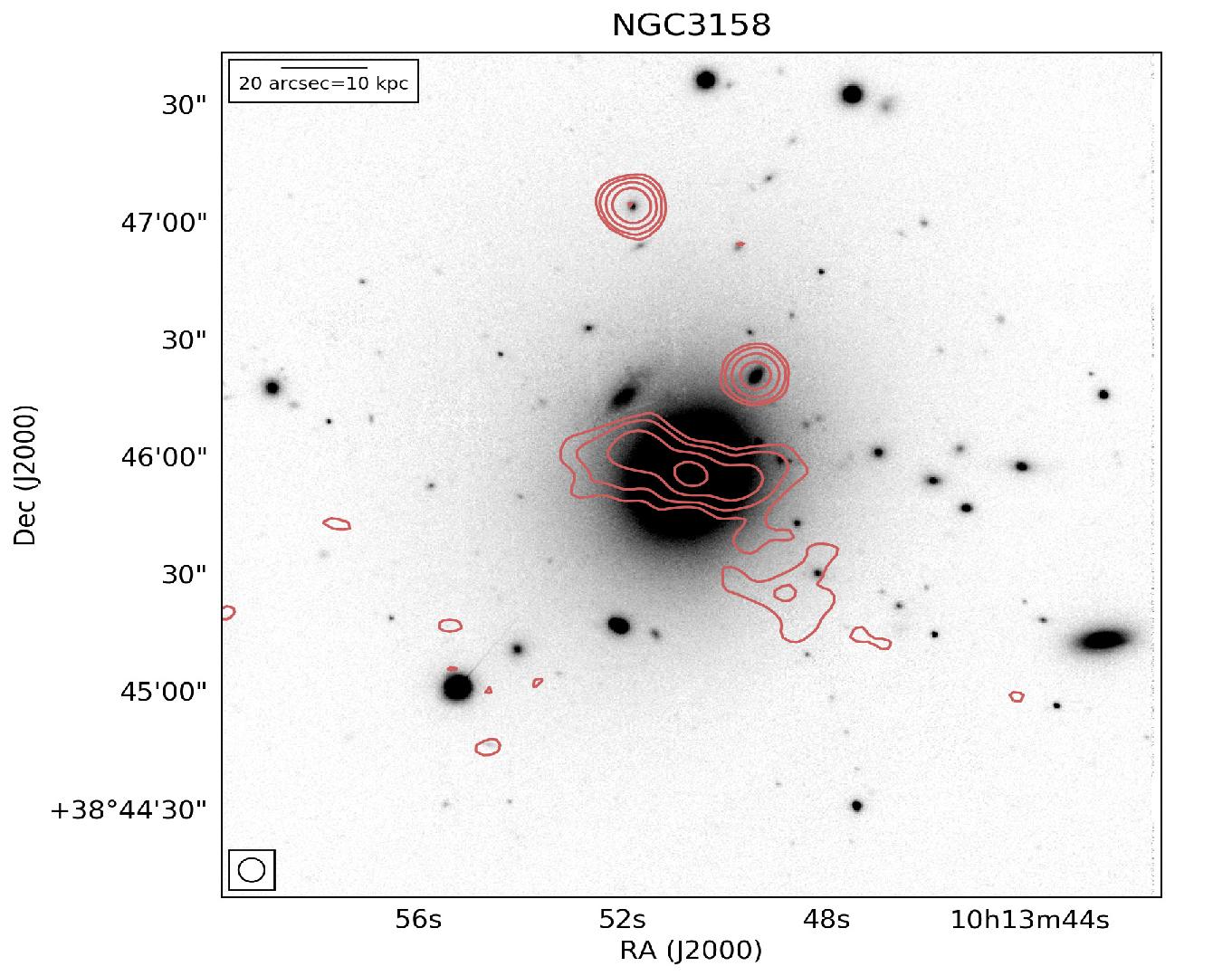}
\includegraphics[scale=0.13]{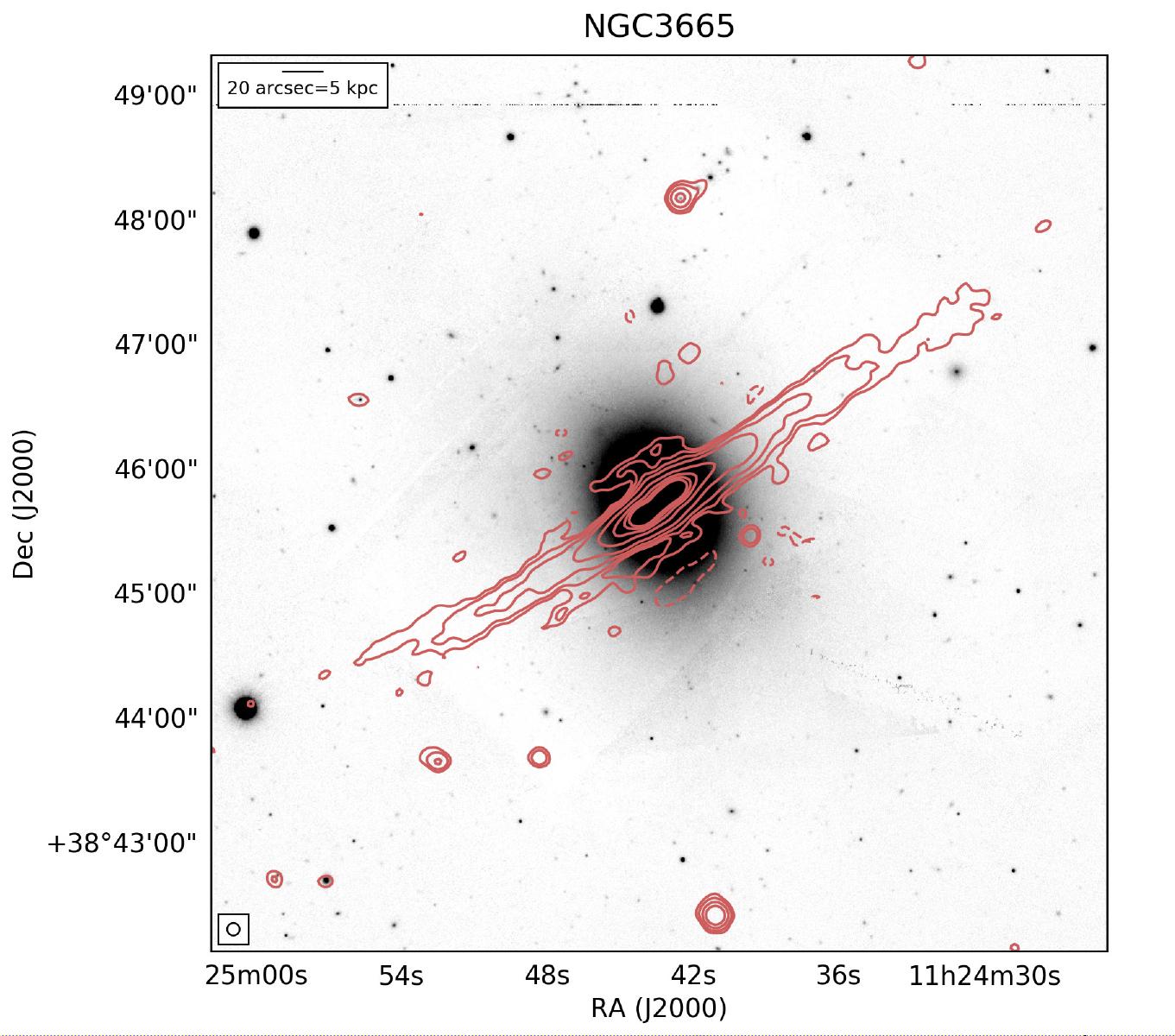}
\includegraphics[scale=0.13]{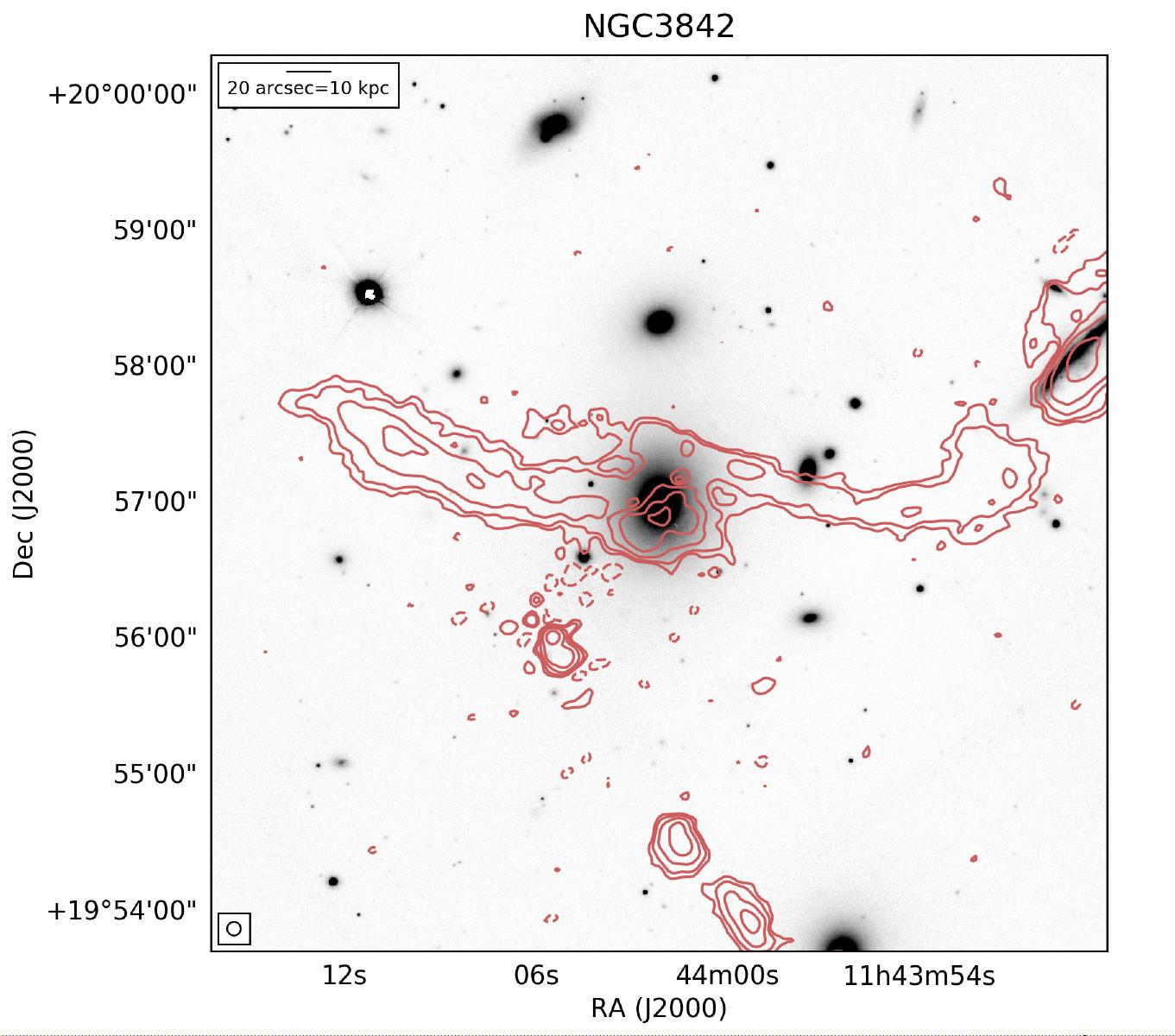}
\includegraphics[scale=0.13]{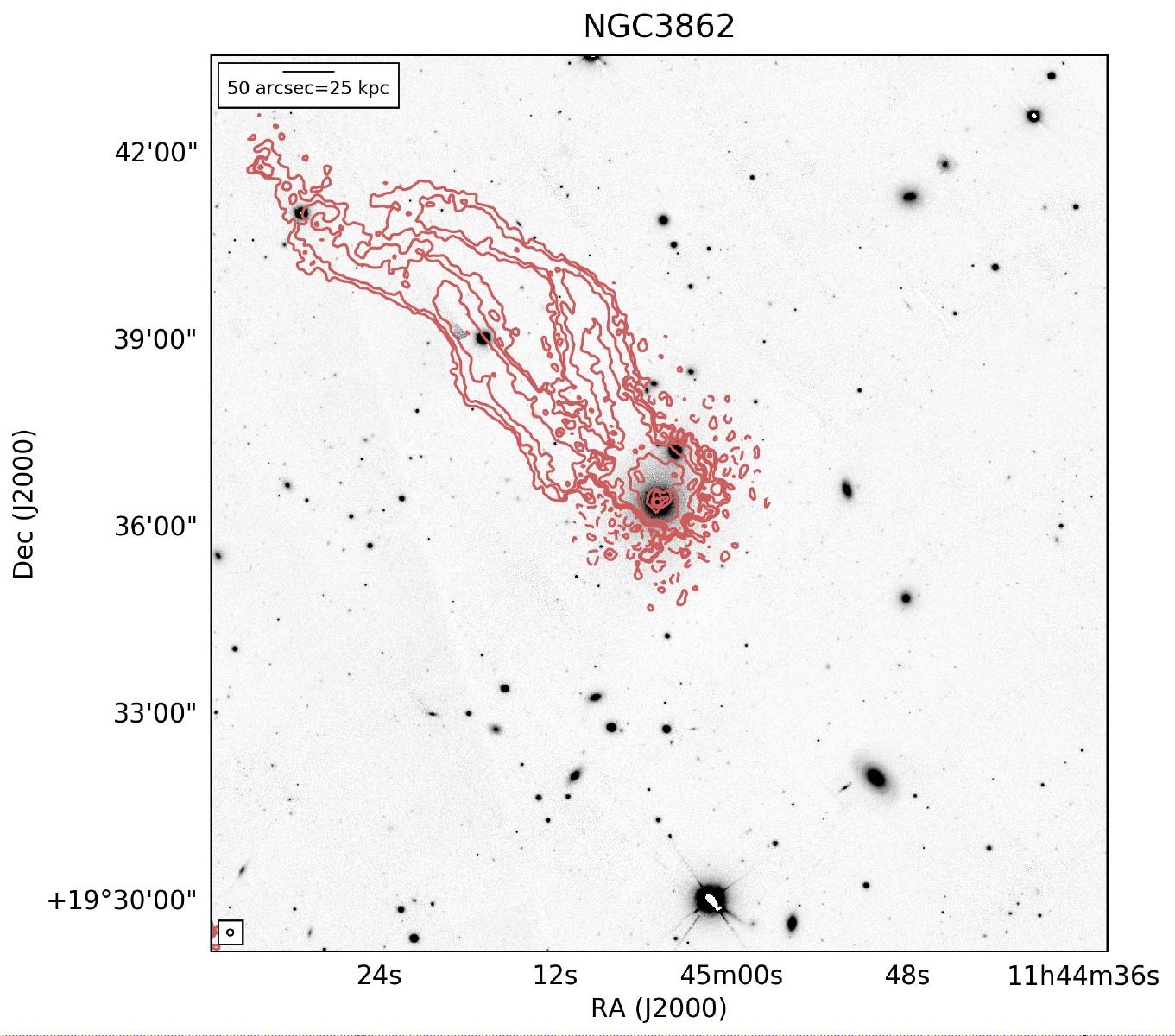}
\includegraphics[scale=0.13]{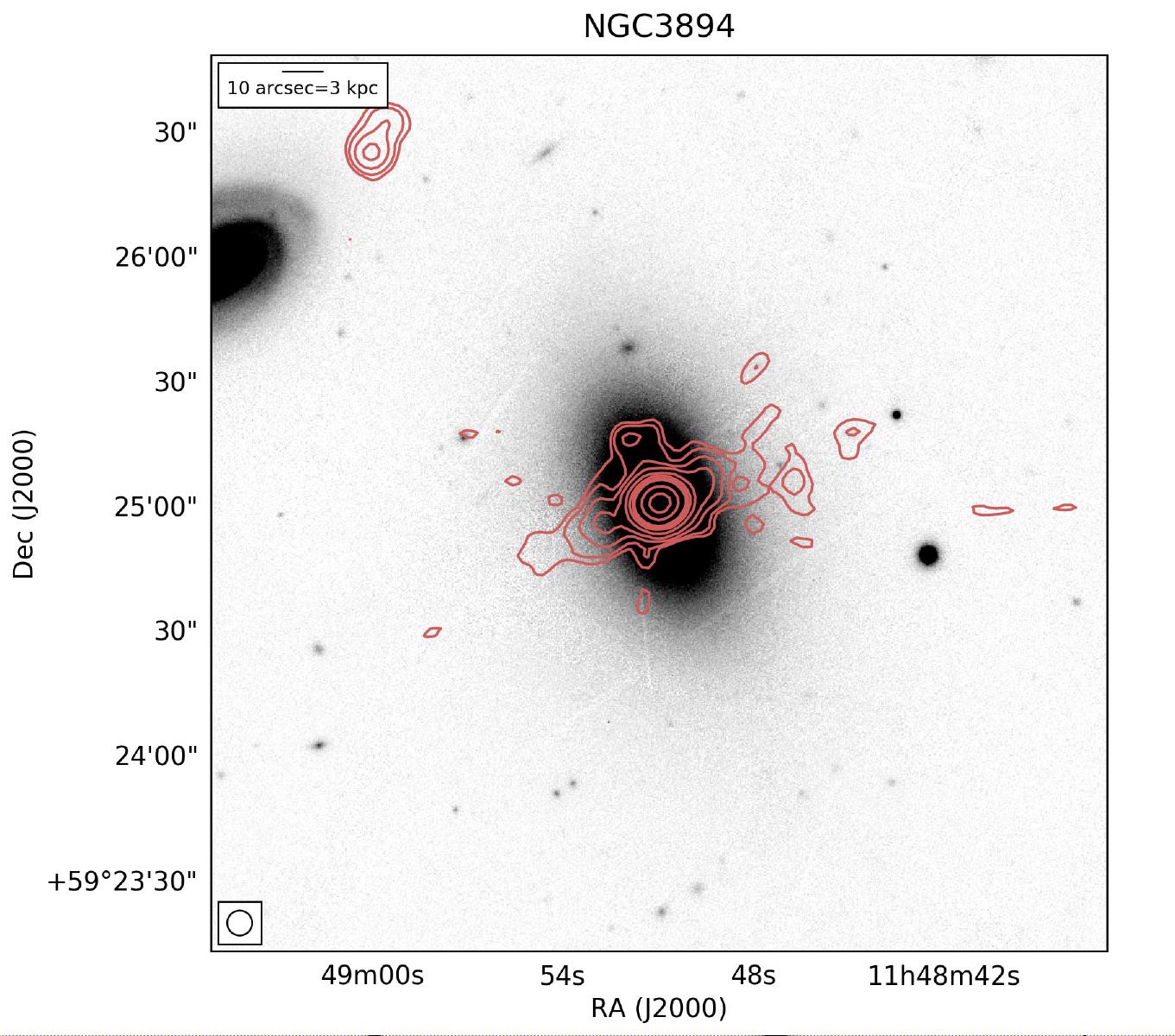}
\includegraphics[scale=0.13]{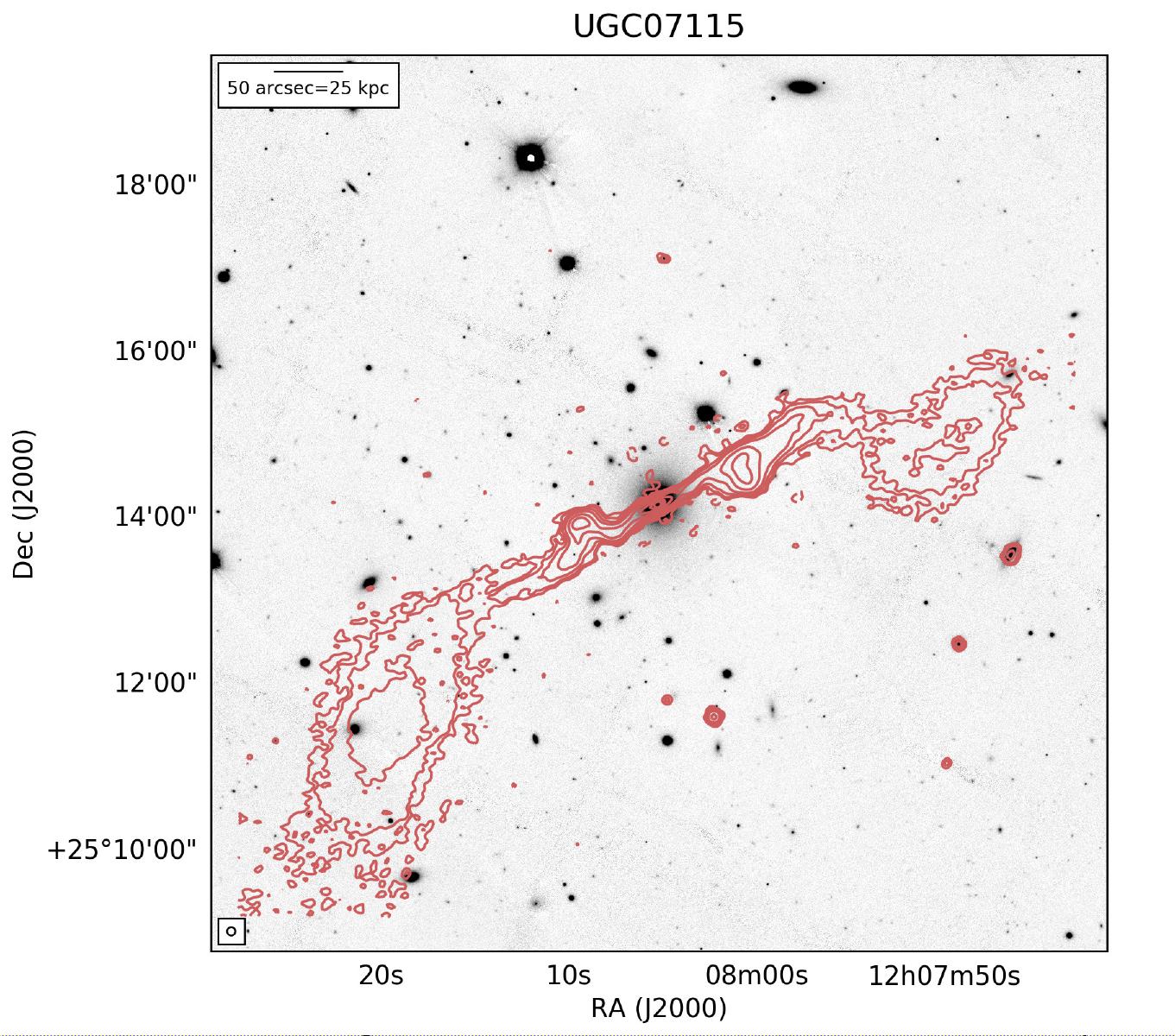}
\includegraphics[scale=0.13]{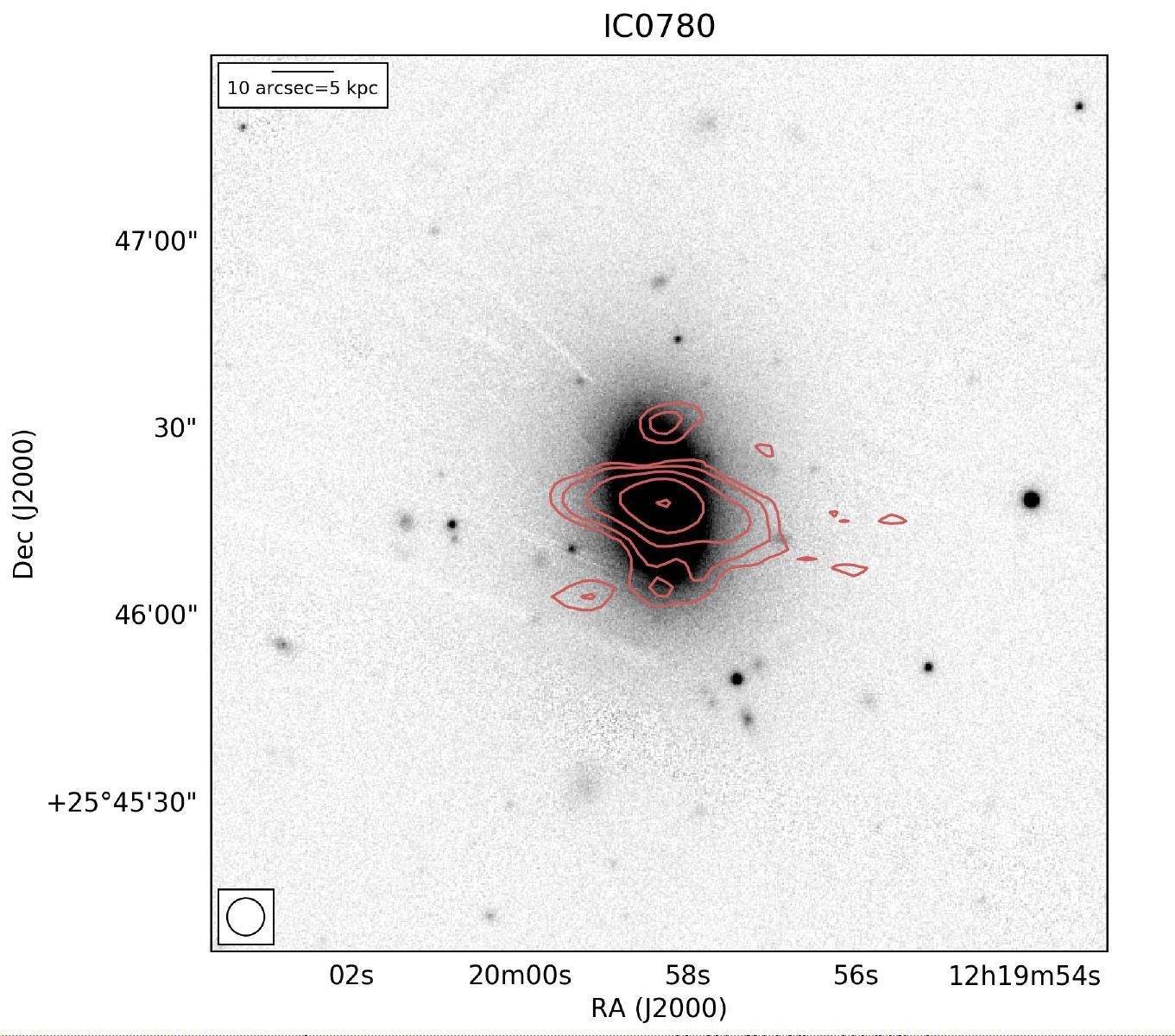}
\includegraphics[scale=0.13]{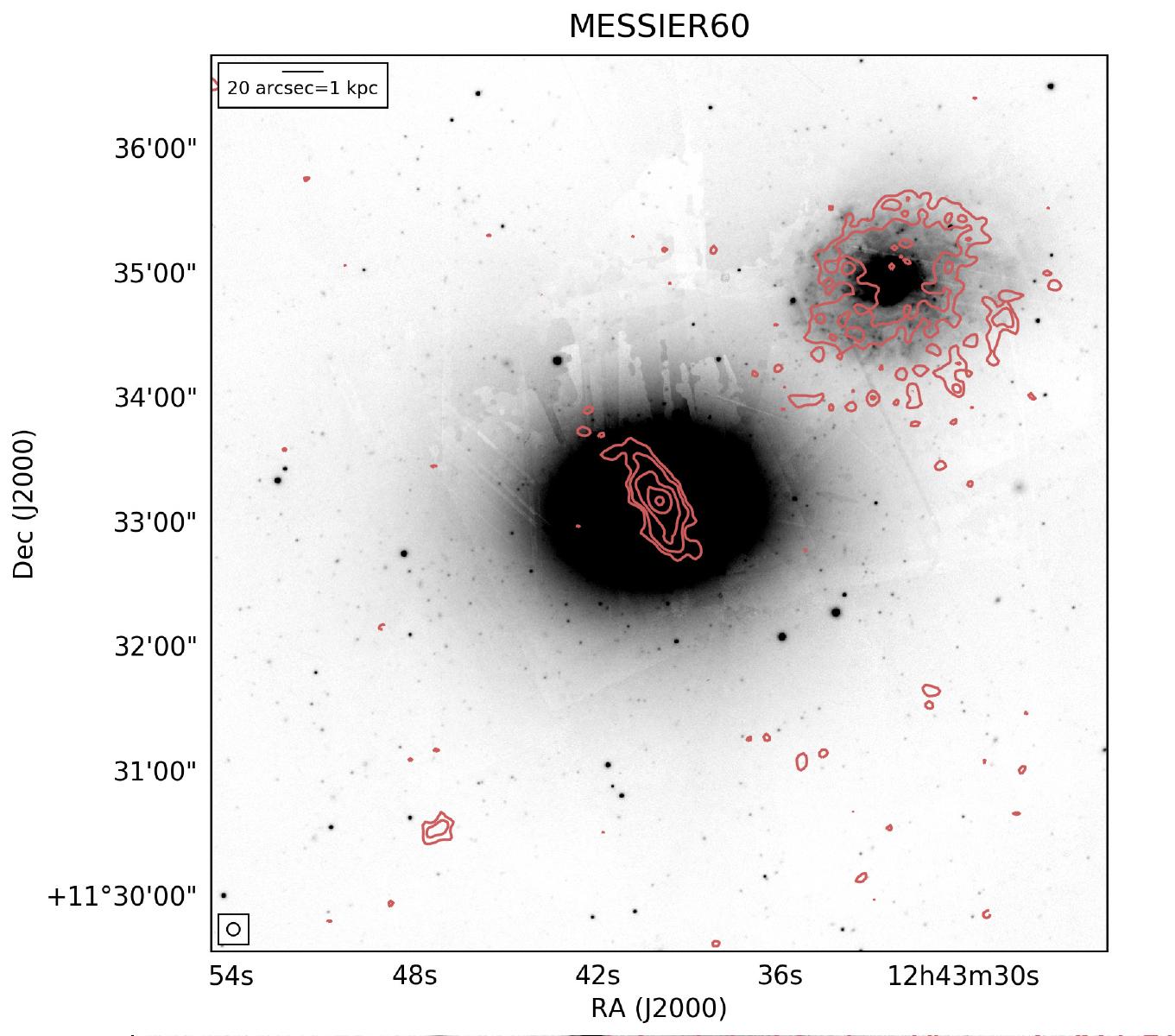}
\includegraphics[scale=0.13]{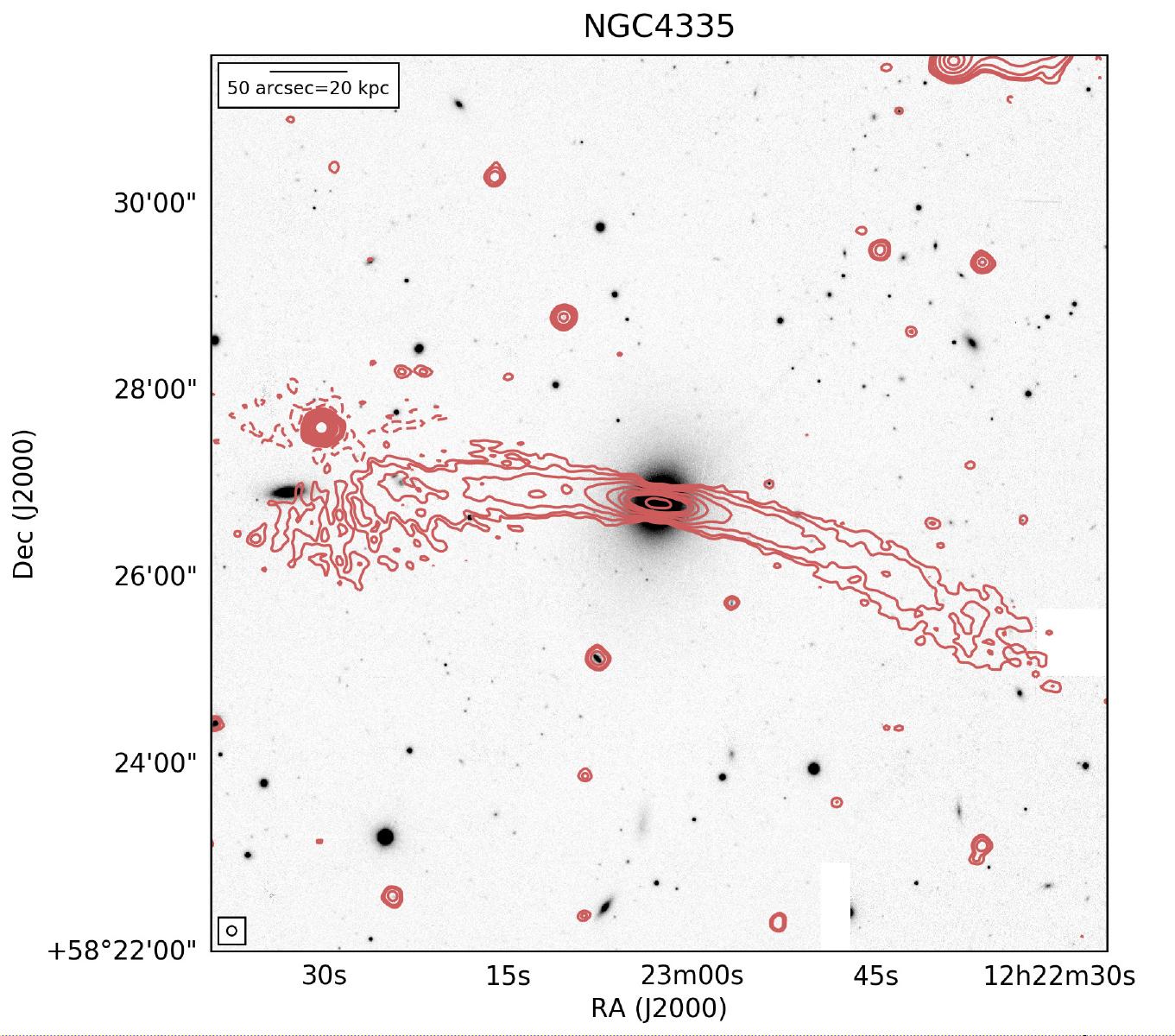}
\includegraphics[scale=0.13]{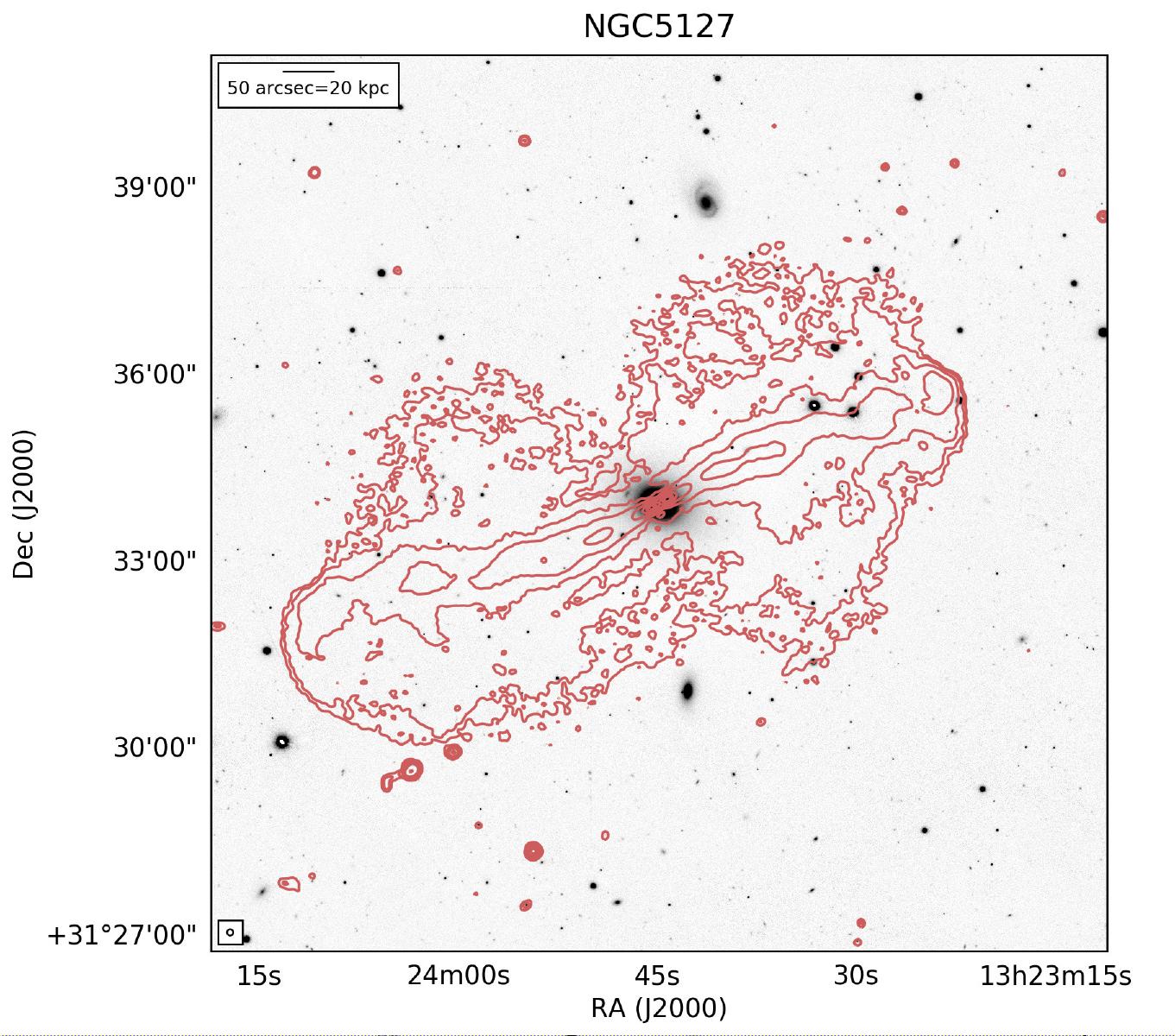}
\includegraphics[scale=0.13]{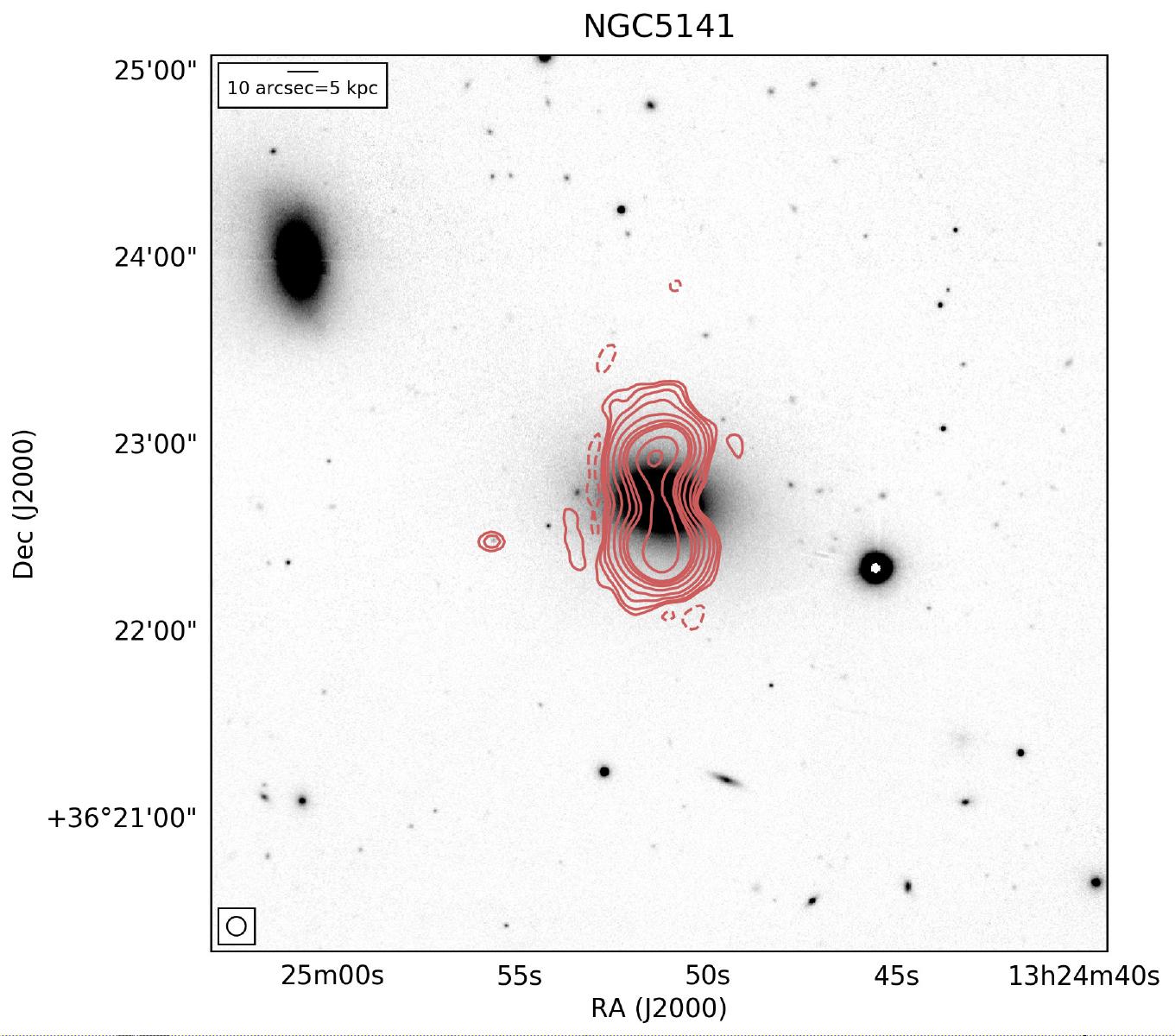}
\caption{(continued)}
\end{figure*}

\addtocounter{figure}{-1}
\begin{figure*}
\includegraphics[scale=0.13]{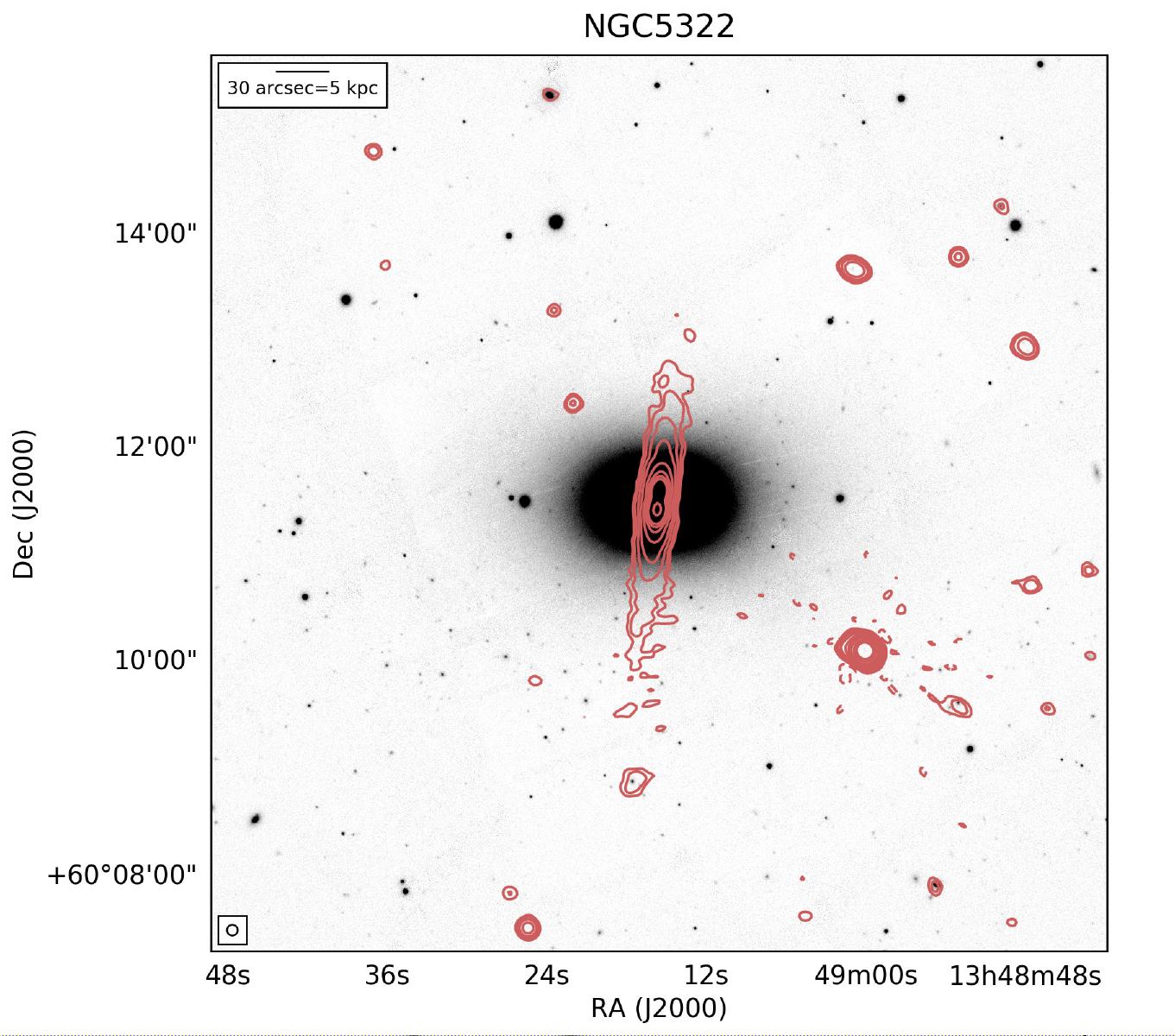}
\includegraphics[scale=0.13]{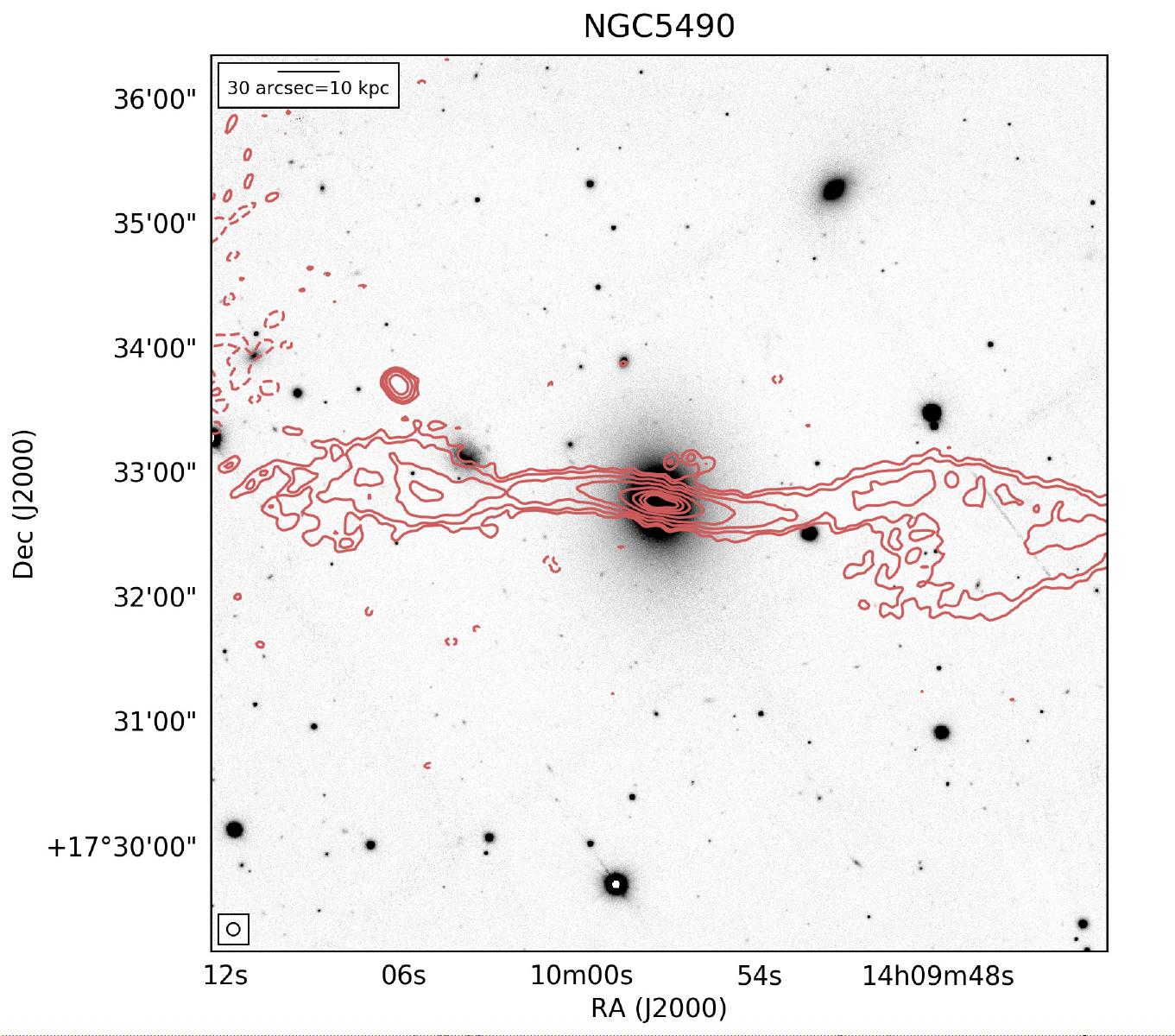}
\includegraphics[scale=0.13]{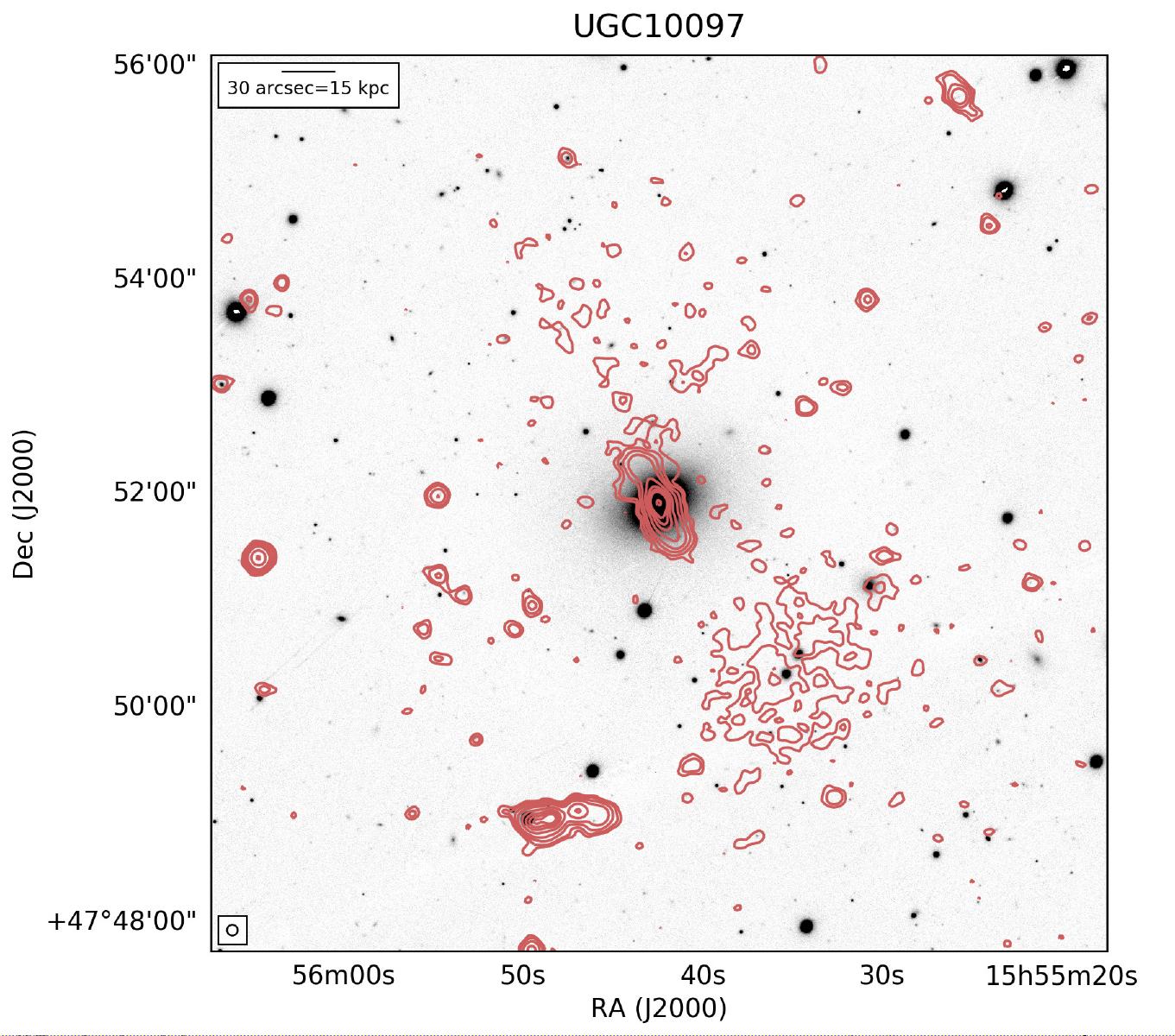}
\includegraphics[scale=0.13]{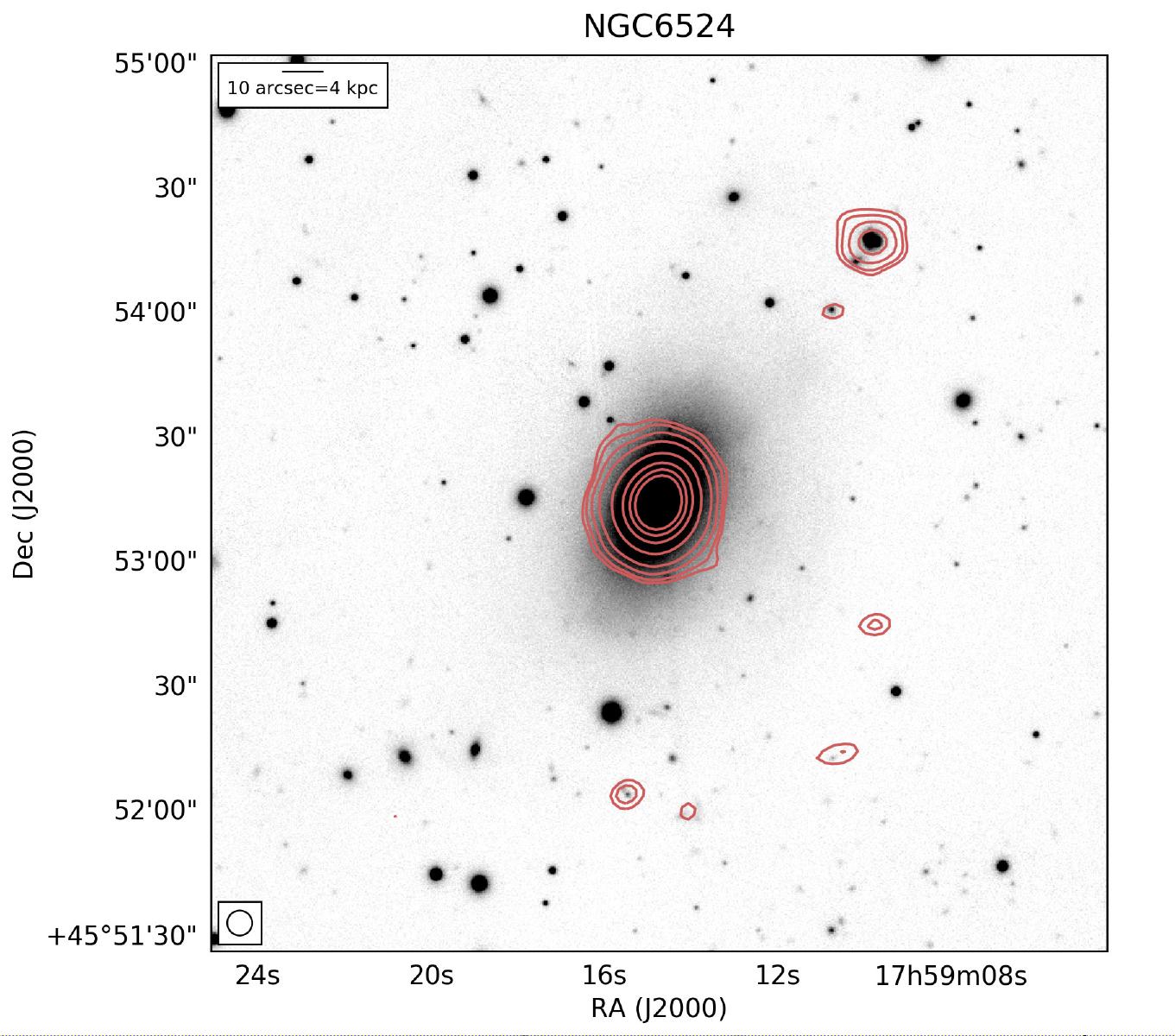}
\includegraphics[scale=0.13]{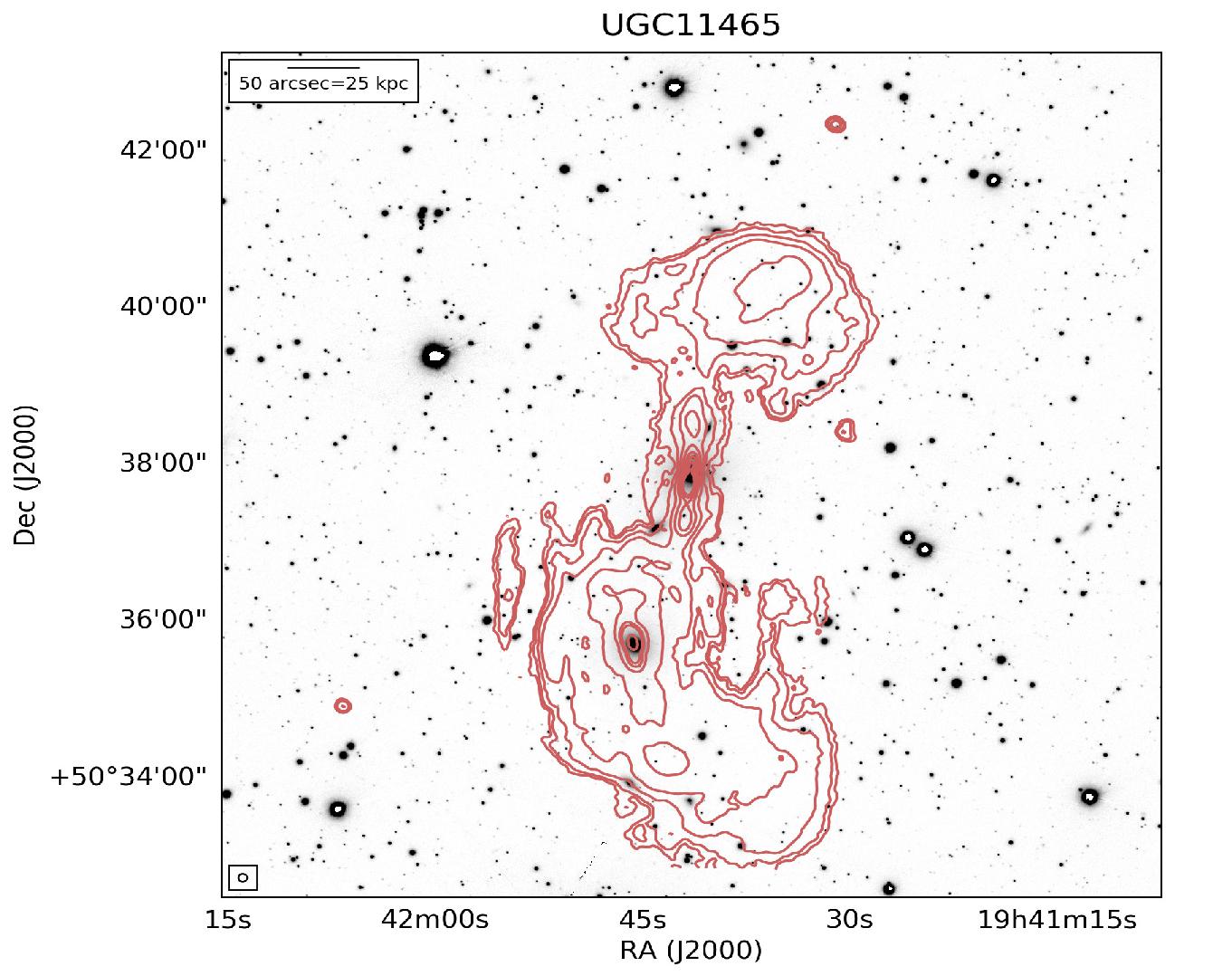}
\includegraphics[scale=0.13]{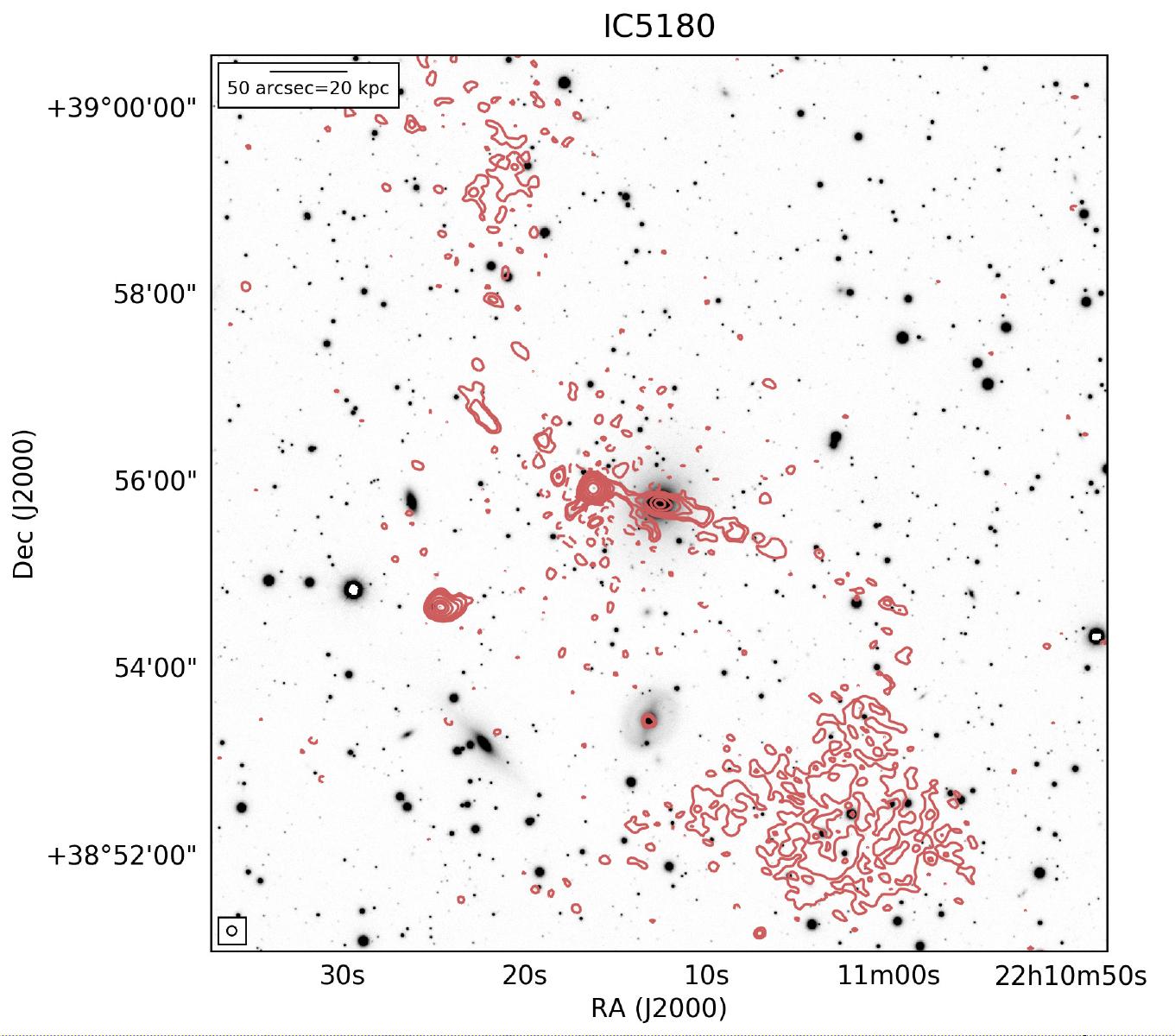}
\includegraphics[scale=0.13]{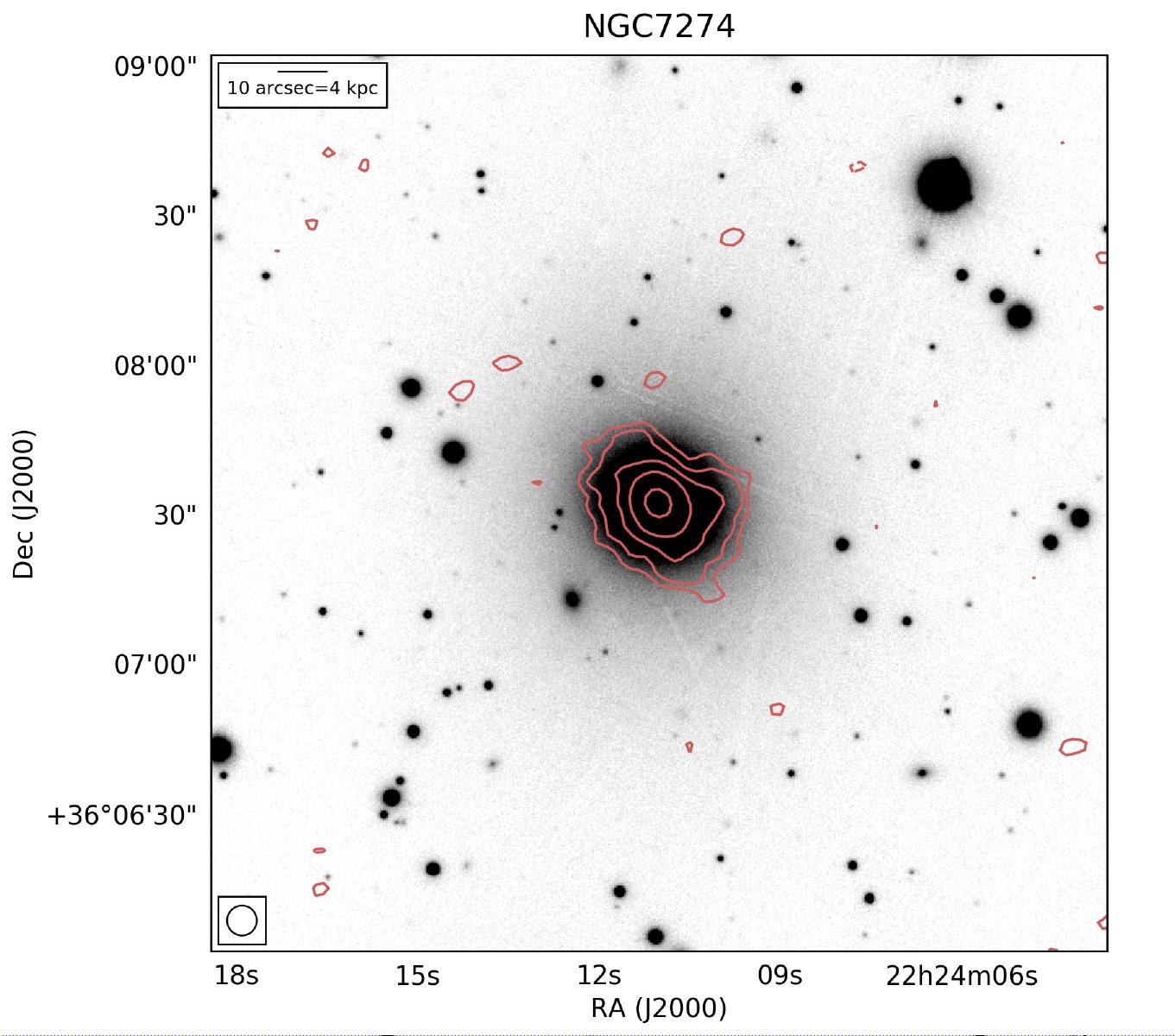}
\includegraphics[scale=0.13]{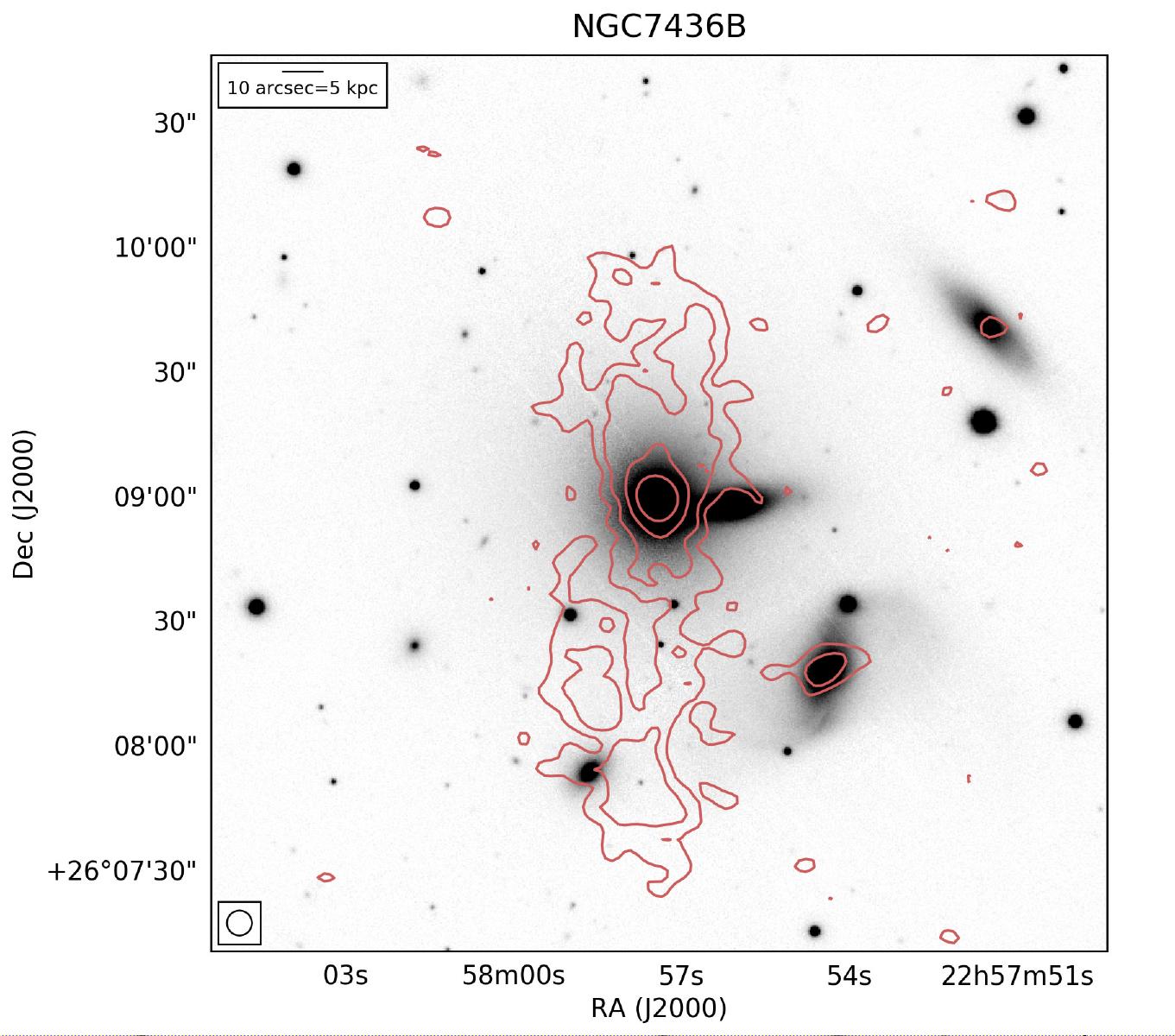}
\includegraphics[scale=0.13]{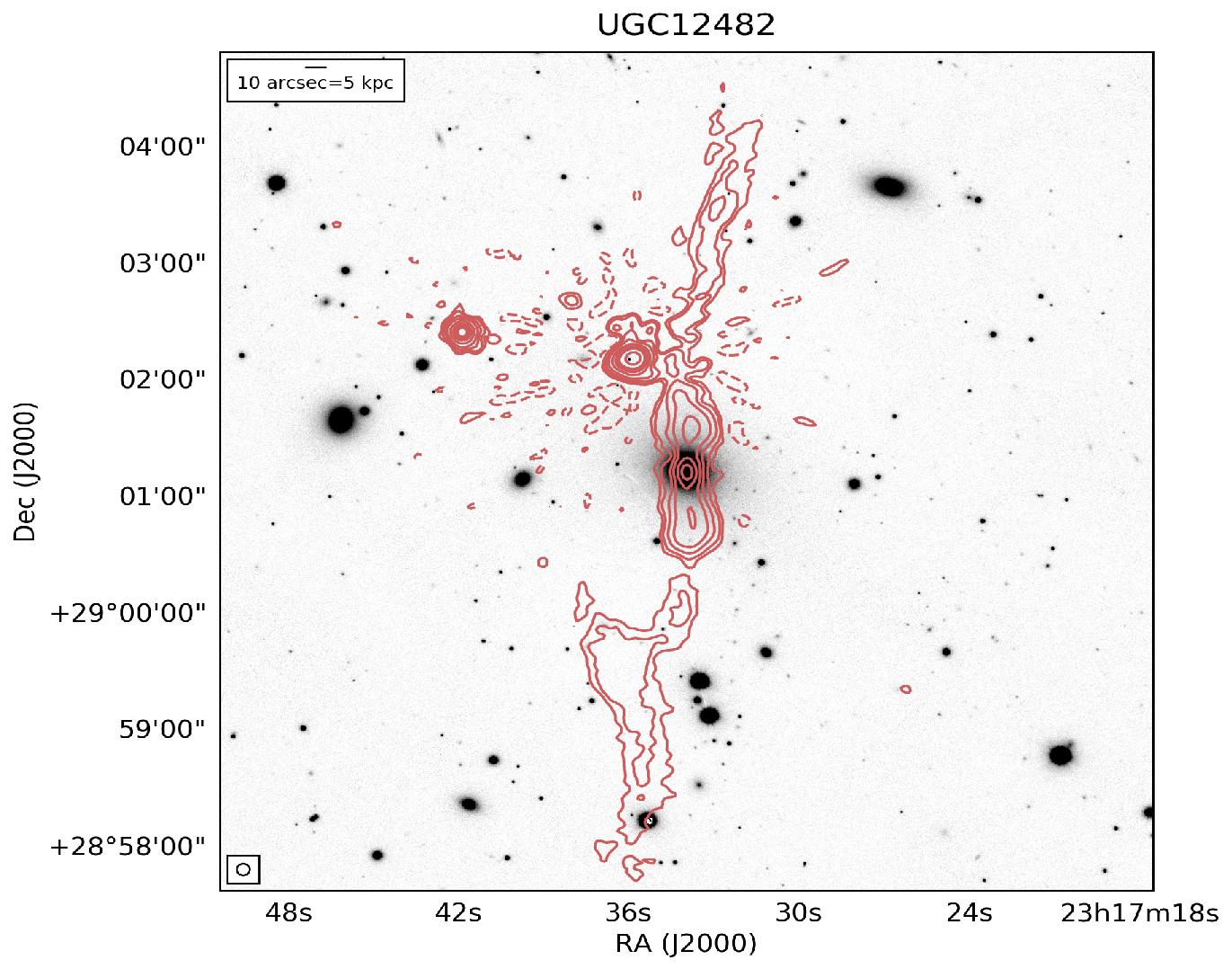}
\includegraphics[scale=0.13]{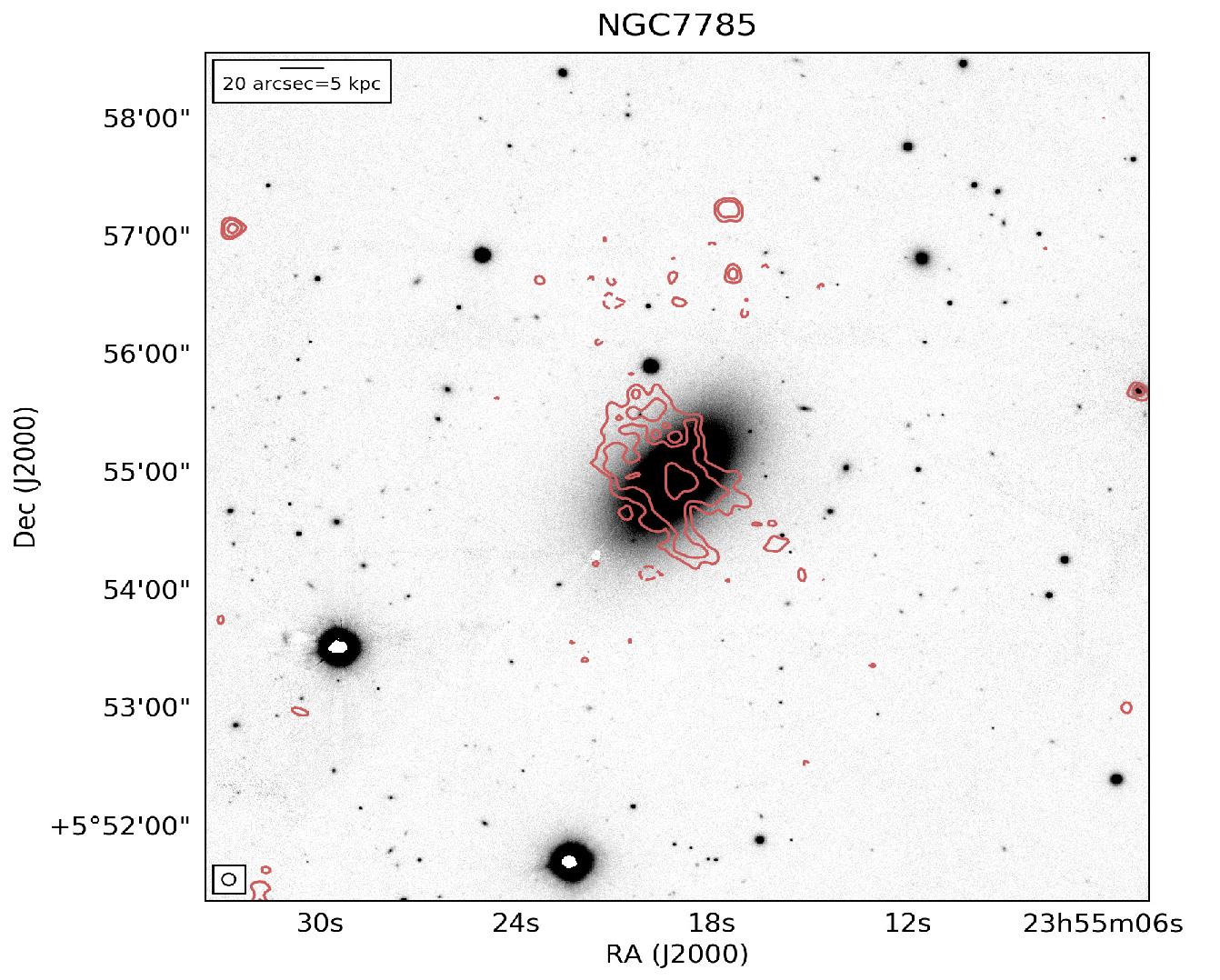}
\caption{(continued)}
\end{figure*}

\begin{figure*}
\includegraphics[scale=0.13]{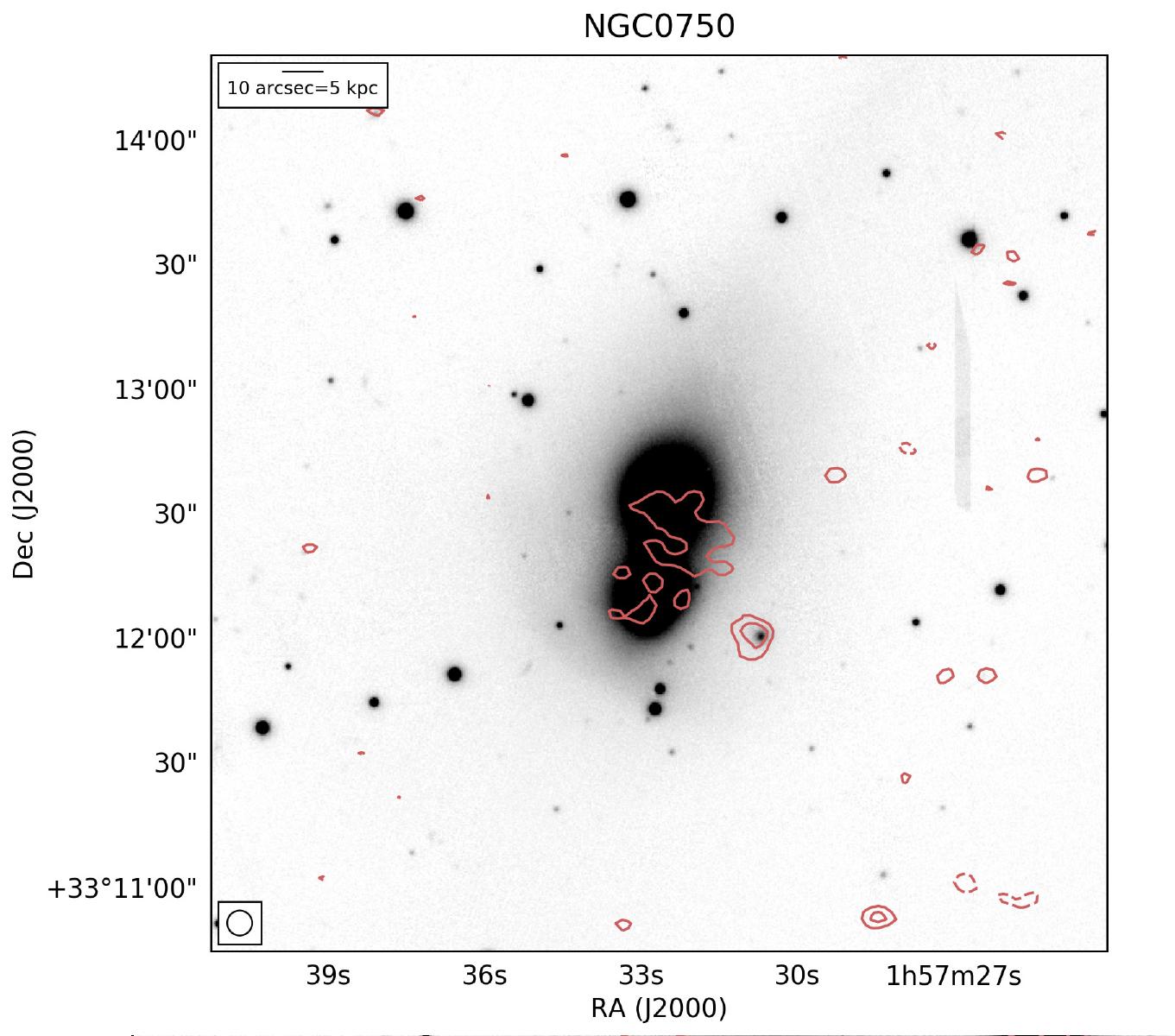}
\includegraphics[scale=0.13]{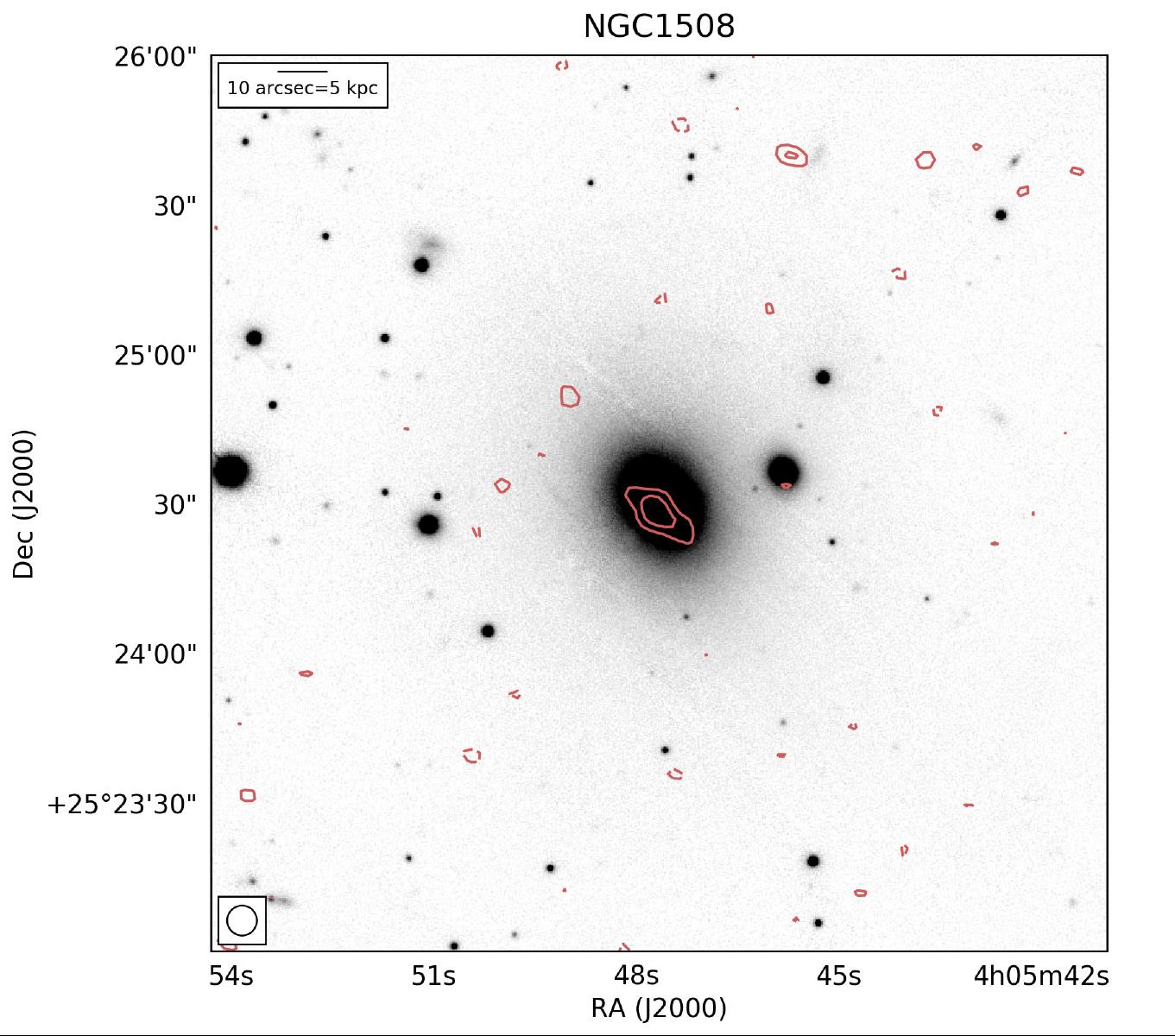}
\includegraphics[scale=0.13]{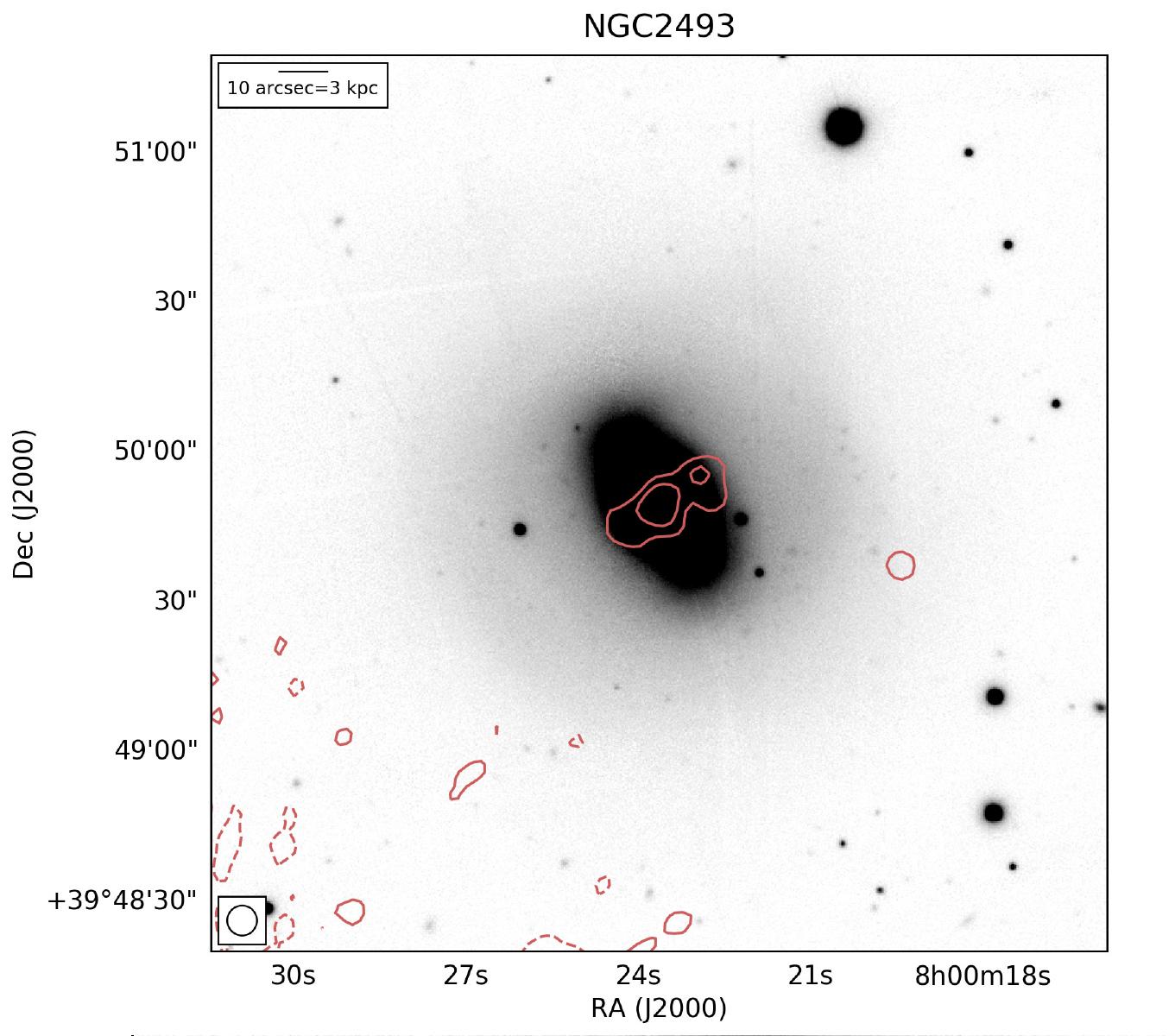}
\includegraphics[scale=0.13]{NGC3805.jpg}
\includegraphics[scale=0.13]{NGC3919.jpg}
\includegraphics[scale=0.13]{NGC5525.jpg}
\includegraphics[scale=0.13]{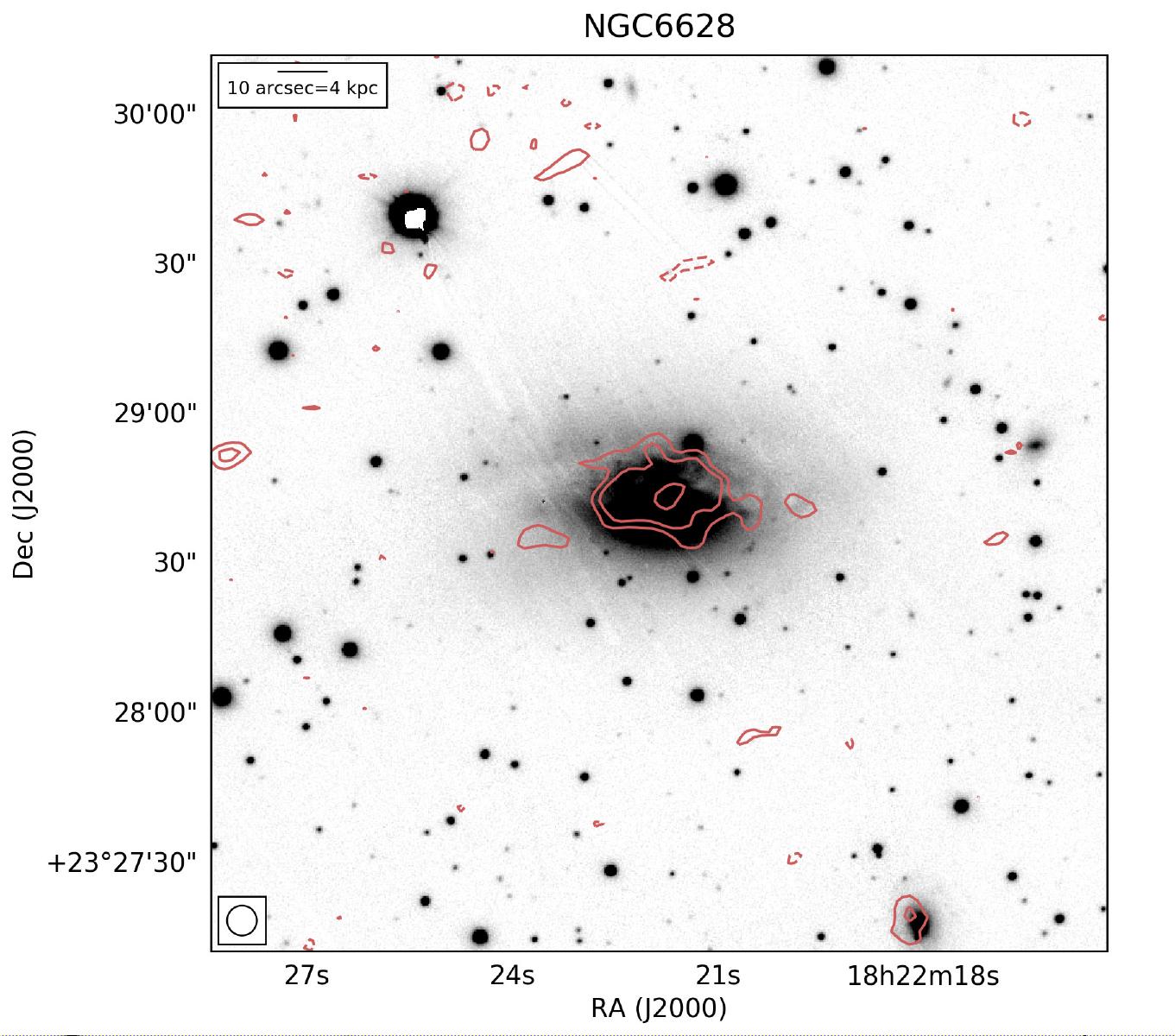}
\includegraphics[scale=0.13]{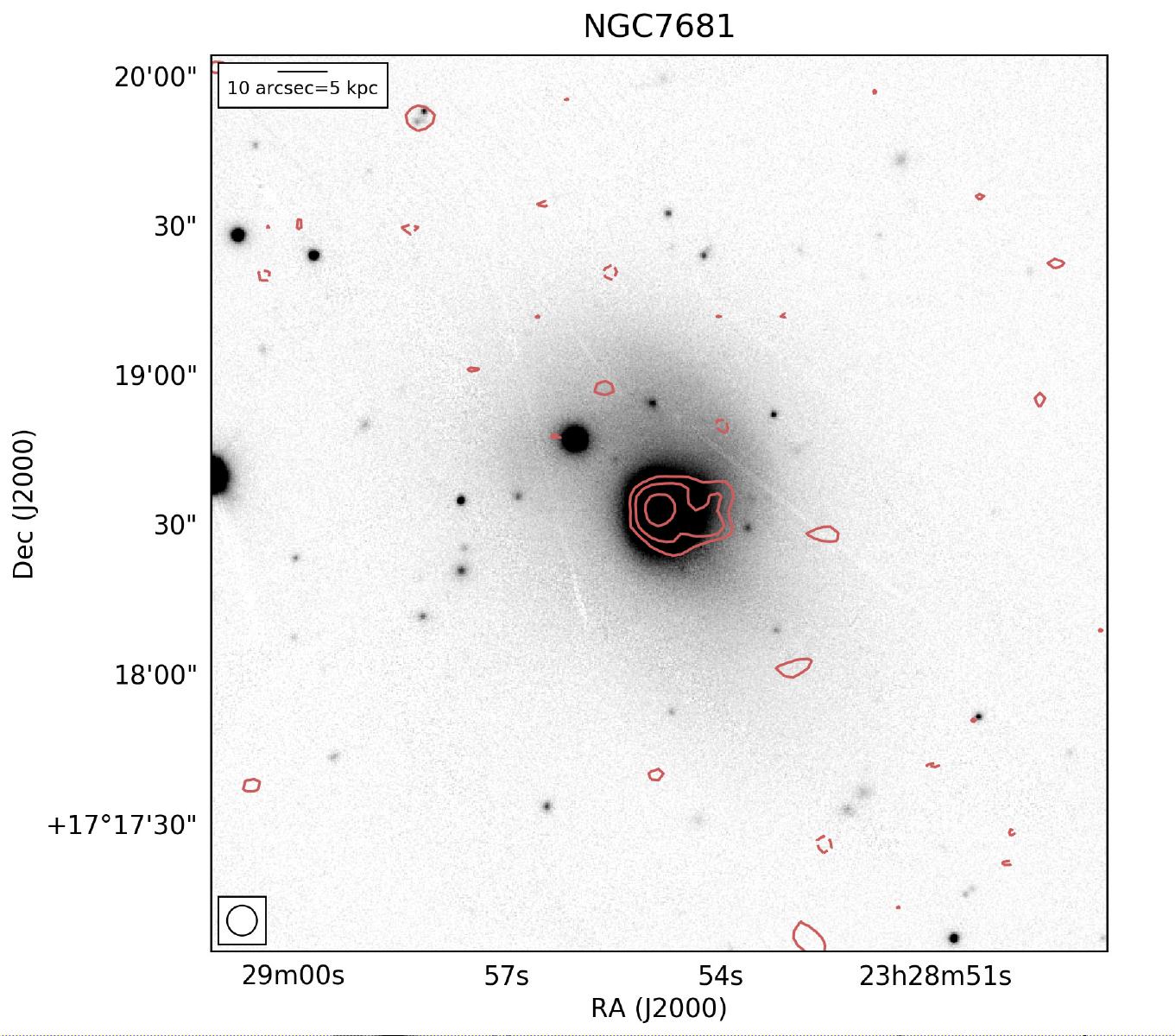}
\includegraphics[scale=0.13]{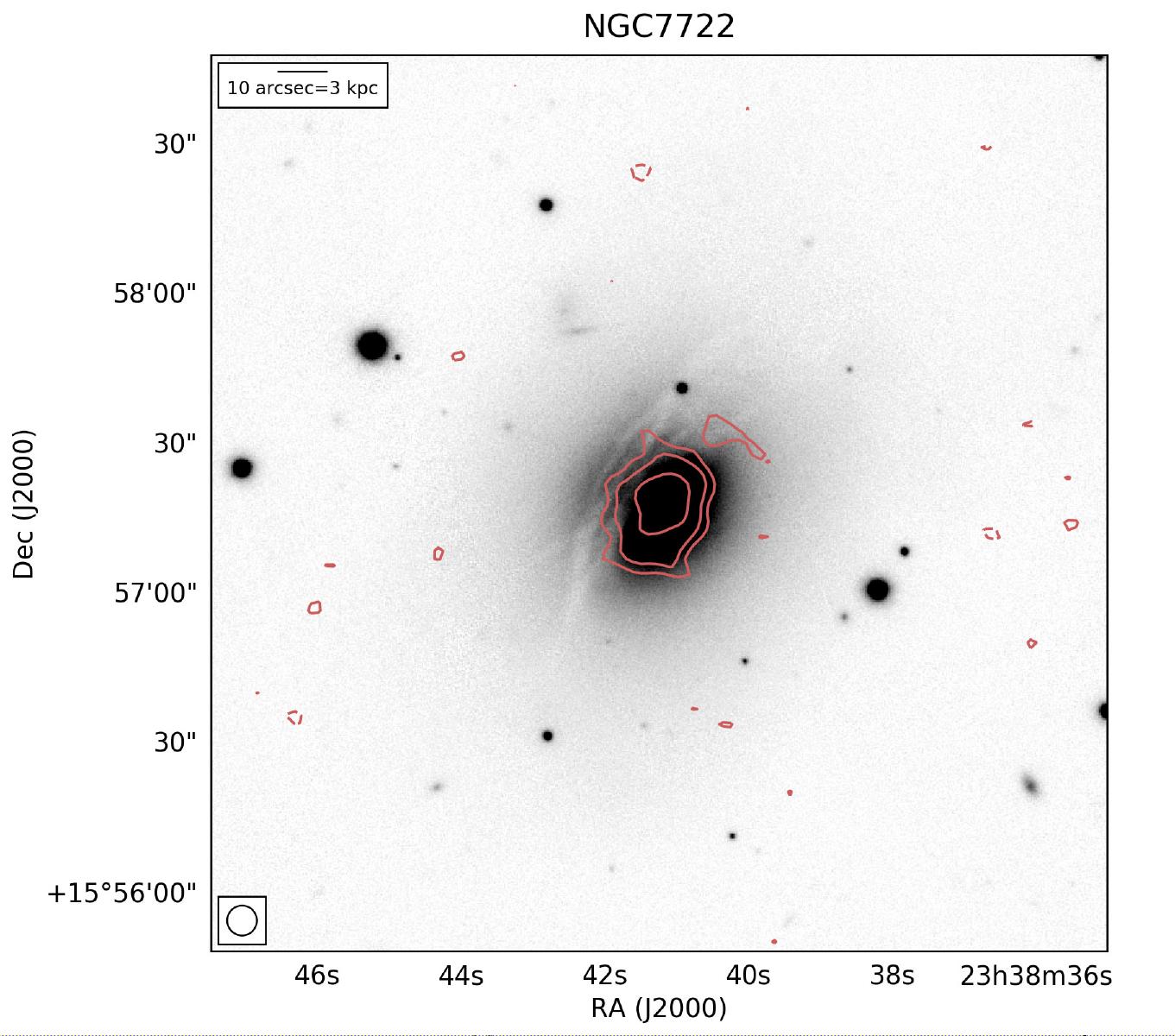}
\caption{LOFAR images at 150 MHz of the nine galaxies not
  classified as extended sources (that is, whose 3$\sigma$ level
  contours do not reach a radius of 15$\arcsec$ but whose FWHM,
  measured by fitting a 2D Gaussian to the central portion of the
  images, exceeds 15$\arcsec$.}
\label{diffuseall}
\end{figure*}

Here, we present the images and a description of the properties of the
individual sources with extended emission and report their
classification as in Table \ref{tab}.  We also briefly describe the
nine objects not included in the list of the extended sources (as they
do not reach a size of 15$\arcsec$ at the $3\sigma$ level) but with a
measured size $\gtrsim 15 \arcsec$ (the ``diffuse'' sources).

\noindent {\bf NGC~0071 (complex).} The radio emission to the north is
associated with a nearby spiral galaxy and, to a lesser extent, with a
smaller nearby elliptical. Diffuse, low-brightness emission to the east
is also present. This source is not covered by FIRST, and in the NVSS
it is confused with the nearby galaxies. No spectral index information
can be obtained.

\noindent {\bf NGC~0080 (complex).} The source is elongated in the NS
direction on small scales, but the more extended structure is oriented
along an EW axis. On the west side, the emission bends sharply toward
the south. This source is not detected at 1.4 GHz, resulting in a very
steep spectral index ($\alpha_{150}^{1400} > 1.76$).

\noindent {\bf NGC~0687 (diffuse).} This radio source is diffuse, and
it lacks a central compact source. It is not detected at 1.4 GHz,
resulting in a steep spectral index ($\alpha_{150}^{1400} > 1.23$).

\noindent {\bf NGC~0777 (FRI).} The source is dominated by a bright
compact core, from which emerge two small-scale (18 kpc of total
extent) jets.

\noindent {\bf NGC~0910 (complex).} Two elongated features emerge from
a compact central source. However, these are rather diffuse and do not
appear to be collimated jets. The overall spectral index is very steep
($\alpha_{150}^{1400} > 1.67$). 

\noindent {\bf CGCG~286-070 (FRI?).} The emission peak is located at the
center of the source, suggesting an FR~I morphology; yet, the small
angular size makes this classification uncertain.  

\noindent {\bf NGC~2672 (Diffuse).} Diffuse, ring-like structure with a
diameter of $\sim 30$ kpc. 

\noindent {\bf NGC~2783 (Ext. ?).} The asymmetry of the source suggests a
core-jet morphology, but the small angular size makes any
classification uncertain.

\noindent {\bf NGC~2789 (Ext. ?).} The central component of this
source is elongated in the NS direction, but the small angular size
makes any classification uncertain. However, two low brightness lobes are
visible in the low-resolution LOFAR image: the total size of the source exceeds 700 kpc.

\noindent {\bf NGC~2832 (FRI?).} The compact radio source on the SW
side is associated with a nearby galaxy. This source is elongated in
the NS direction, suggesting an FR~I morphology; yet, the small angular
size makes this classification uncertain. This source is not detected
at 1.4 GHz, resulting in a steep spectral index ($\alpha_{150}^{1400} > 1.37$).

\noindent {\bf NGC~3665 (FRI).} This galaxy presents two opposite
large-scale jets (55 kpc in total extent), for an overall FR~I
morphology. However, close to the nucleus ($\sim 3$ kpc) two bright
knots of higher surface brightness are seen, possibly indicating a
restarted source.

\noindent {\bf NGC~3842 (complex).} Two radio tails extend over $\sim$
160 kpc. However, the central regions are dominated by a small-scale,
H-shaped structure, $\sim$ 15 kpc wide, reminiscent of the
morphology seen in 3C~171 \citep{neff95}.  The steep spectral index
($\alpha_{150}^{1400}$ = 1.85) might indicate that this is a restarted source and
the tails are remnants of a previous phase of activity.

\noindent {\bf NGC~3894 (FRI).} The source is dominated by a bright
compact core, from which emerge two small-scale (16 kpc of total
extent) jets. The overall spectral shape is flat ($\alpha_{150}^{1400} = -0.18$), in
line with its high core dominance.

\noindent {\bf M~60 (FRI).} Small-scale ($\sim$ 4 kpc in length) FR~I. 

\noindent {\bf NGC~5141 (FRII).} The only gETG of the sample associated with
a radio source of FR~II morphology. In the radio-host luminosity
diagram of \citet{ledlow96}, this source is located close to the
boundary between FR~Is and FR~IIs.

\noindent {\bf UGC~10097 (FRI).} In addition to the central FR~I structure,
the low-resolution LOFAR images show two large-scale diffuse lobes,
reaching a radius of $\sim 200\arcsec$ ($\sim$500 kpc).

\noindent {\bf NGC~6524 (Ext. ?).} The angular size of this source is too small to explore its morphology.

\noindent {\bf IC~5180 (FRI).} In addition to the central FR~I structure,
the low resolution LOFAR images show two large-scale diffuse lobes.

\noindent {\bf NGC~7274 (Ext. ?).} The angular size of this source is too small to explore its morphology.

\noindent {\bf UGC~12482 (Complex).} The central region shows the presence of a
triple source, extending for $\sim 30$ kpc, while long symmetric tails
reach a distance of $\sim$ 90 kpc, suggesting that this might be a
restarted source and that the tails are remnants of a previous phase
of activity.

\medskip
Notes on the sources not classified as extended but with sizes larger
than $15 \arcsec$:

\noindent {\bf NGC~0750 (Diffuse).} The radio emission is diffuse and
it extends to include a southern companion galaxy.

\noindent {\bf NGC~1508 (FRI?).} This source is elongated in
the NE-SW direction, suggesting an FR~I morphology; however, the small angular
size makes this classification uncertain.

\noindent {\bf NGC~2493 (FRI?).} This source is elongated in
the NE-SW direction, suggesting an FR~I morphology; however, the small angular
size makes this classification uncertain. The spectral index is rather steep ($\alpha_{150}^{1400} > 1.05$).

\noindent {\bf NGC~3805 (Diffuse).} The radio emission is diffuse and
it extends mostly on the southern side of the host.

\noindent {\bf NGC~3919 (Ext. ?).} The radio emission is slightly
elongated in the NS direction, but the small angular size makes this
classification uncertain.

\noindent {\bf NGC~5525 (Ext. ?).} The radio emission is slightly
elongated in the NS direction, but the small angular size makes this
classification uncertain.

\noindent {\bf NGC~6628 (Ext. ?).} Radio source elongated in the EW
direction, approximately aligned with the optical axis.

\noindent {\bf NGC~7681 (Ext. ?).} The emission peak is located at the
host's center, with diffuse emission extending toward the west.

\noindent {\bf NGC~7722 (Diffuse).} Diffuse radio emission cospatial
with the optical emission.

\end{appendix}

\end{document}